\documentclass[journal]{IEEEtran}
\usepackage{amsmath,amssymb,amsfonts}
\usepackage{tabularx}
\usepackage{array}
\usepackage[caption=false,font=footnotesize,labelfont=rm,textfont=rm]{subfig}
\usepackage{textcomp}
\usepackage{stfloats}
\usepackage{url}
\usepackage{verbatim}
\usepackage{graphicx}
\usepackage{cite}
\usepackage[ruled,vlined,linesnumbered]{algorithm2e}
\SetAlFnt{\footnotesize}  
\SetAlCapFnt{\footnotesize}  
\SetAlCapNameFnt{\footnotesize}  

\usepackage{booktabs}
\usepackage{xspace}
\usepackage{balance}
\usepackage{hyperref}
\usepackage[frozencache,cachedir=.]{minted}
\usepackage[most]{tcolorbox}
\usepackage{comment}
\usepackage{epstopdf}

\usepackage[utf8]{inputenc}
\usepackage{pifont}
\usepackage{newunicodechar}
\newunicodechar{✓}{\ding{51}}
\usepackage{enumitem}
\usepackage{ulem}
\usepackage{multirow} 
\usepackage{multicol}
\usepackage{threeparttable}
\usepackage{makecell,rotating,diagbox}
\usepackage{setspace}
\usepackage[framemethod=tikz]{mdframed}
\graphicspath{{./Figure}}

\usepackage{xcolor}
\usepackage{listings}
\usepackage{inconsolata} 
\usepackage{orcidlink}
\usepackage{pifont}
\usepackage[T1]{fontenc}

\definecolor{functioncolor}{HTML}{4672C4}
\definecolor{classcolor}{HTML}{D79655}
\definecolor{keywordcolor}{HTML}{C678DD}
\definecolor{commentcolor}{HTML}{7F7F7F}
\definecolor{stringcolor}{HTML}{A2D27E}

\newcommand{\rev}[1]{\textcolor{black}{#1}} 
\newcommand{\tool}{\textsc{PCART}\xspace}
\newcommand{\benchmark}{\textsc{PCBench}\xspace}

\begin{document}

\title{PCART: Automated Repair of Python API Parameter Compatibility Issues}



\author{Shuai Zhang, 
Guanping Xiao, 
Jun Wang, 
Huashan Lei, 
Gangqiang He, 
Yepang Liu, 
and Zheng Zheng

\thanks{This work was supported in part by the National Natural Science Foundation of China under Grants 62002163, 61932021, and 62372021, the National Key R\&D Program of China under Grant 2024YFB3311503, and the Open Research Fund of State Key Laboratory of Novel Software Technology under Grant KFKT2025B13. (\textit{Corresponding author: Guanping Xiao.})}
\thanks{Shuai Zhang, Guanping Xiao, Jun Wang, Huashan Lei, and Gangqiang He are with the College of Computer Science and Technology and the Key Laboratory for Safety-critical Software Development and Verification, Nanjing University of Aeronautics and Astronautics, Nanjing, China. (email: \{shuaizhang, gpxiao, junwang, leihuashan, gangqiang.he\}@nuaa.edu.cn)}

\thanks{Guanping Xiao is also with the State Key Lab. for Novel Software Technology, Nanjing University, P.R. China.}

\thanks{Yepang Liu is with the Department of Computer Science and Engineering, Southern University of Science and Technology, Shenzhen, China. (email: liuyp1@sustech.edu.cn)}

\thanks{Zheng Zheng is with the School of Automation Science and Electrical Engineering, Beihang University, Beijing, China. (email: zhengz@buaa.edu.cn)}



}


\markboth{Journal of \LaTeX\ Class Files,~Vol.~14, No.~8, August~2021}%
{Shell \MakeLowercase{\textit{et al.}}: A Sample Article Using IEEEtran.cls for IEEE Journals}


\maketitle

\begin{abstract}

In modern software development, Python third-party libraries play a critical role, especially in fields like deep learning and scientific computing. However, API parameters in these libraries often change during evolution, leading to compatibility issues for client applications reliant on specific versions. Python's flexible parameter-passing mechanism further complicates this, as different passing methods can result in different API compatibility. Currently, no tool can automatically detect and repair Python API parameter compatibility issues. To fill this gap, we introduce \tool, the first solution to fully automate the process of API extraction, code instrumentation, API mapping establishment, compatibility assessment, repair, and validation. \tool handles various types of Python API parameter compatibility issues, including parameter addition, removal, renaming, reordering, and the conversion of positional to keyword parameters. To evaluate \tool, we construct \benchmark, a large-scale benchmark comprising 47,478 test cases mutated from 844 parameter-changed APIs across 33 popular Python libraries. Evaluation results demonstrate that \tool is both effective and efficient, significantly outperforming existing tools (MLCatchUp and Relancer) and the large language model ChatGPT (GPT-4o), achieving an F1-score of 96.51\% in detecting API parameter compatibility issues and a repair precision of 91.97\%. Further evaluation on 30 real-world Python projects from GitHub confirms \tool's practicality. We believe \tool can significantly reduce the time programmers spend maintaining Python API updates and advance the automation of Python API compatibility issue repair.
\end{abstract}


\section{Introduction} 


\IEEEPARstart{S}{oftware} libraries evolve continuously, and migrating client code to adapt to changing APIs is a long-standing challenge across programming languages~\cite{lamothe2021systematic}. When libraries introduce breaking changes such as modified method signatures, renamed classes, or altered parameters, client applications must update their API usages to remain compatible. This migration process is often tedious and error-prone, as even seemingly minor changes can lead to compilation errors or runtime failures if not handled correctly~\cite{zhong2024compiler}. A large body of prior work has explored migration in statically typed languages such as Java and Android, leveraging static type information and compiler feedback to detect and guide adaptations. For example, techniques mine API change patterns from documentation~\cite{xi2019migrating,huang2021repfinder,li2020cda,liu2023automatically} or learn edit scripts from repository histories~\cite{xing2007api, nguyen2010graph, yamaguchi2022two, kang2019semantic}. However, these approaches typically assume the presence of type checking and compile-time error reporting, assumptions that do not hold in dynamically typed languages. Moreover, many breaking changes are insufficiently documented (only about 20\% are explicitly documented~\cite{cossette2012seeking}), leaving developers with limited guidance. As a result, automated API migration remains a fundamental challenge rather than a simple engineering task.

\begin{figure}[!t]
\centering
\includegraphics[width=\linewidth]{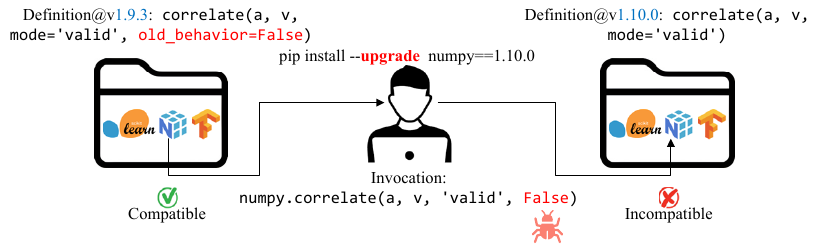} 
\vspace{-4mm}
\caption{An example of Python API parameter compatibility issues.} 
\label{Upgrading Library Versions}
\vspace{-4mm}
\end{figure}

Python exemplifies these challenges in unique and particularly demanding ways. It is currently the most widely used programming language (ranked first in the TIOBE index as of September 2025~\cite{TIOBE}) and hosts a rich ecosystem of over 650,000 third-party libraries~\cite{PyPI}. These libraries evolve rapidly, and refactoring, bug fixes, or feature additions often trigger parameter-level API changes such as adding, removing, renaming, or reordering parameters~\cite{zhang2020python}. Python's highly flexible parameter passing mechanisms (i.e., mixing positional and keyword arguments, optional and variadic forms) combined with the lack of compile-time type checking, mean that many incompatibilities surface only at runtime. Furthermore, whether an incompatibility manifests depends on how an API is invoked. For example, if a positional parameter is removed, positional passing is incompatible, but omission remains compatible. Fig.~\ref{Upgrading Library Versions} illustrates this: upgrading NumPy from version 1.9.3 to 1.10.0 breaks the call \mintinline{python}{numpy.correlate(a, v, 'valid', False)} with a \textit{TypeError}, because the parameter \mintinline{python}{old_behavior} was removed. Such issues directly undermine program reliability and stability. Addressing them requires reasoning not only about API definitions but also about dynamic invocation contexts. In addition, establishing precise API mappings and conducting automated validation, both of which require contextual runtime information, pose substantial technical challenges.


\begin{figure*}[!t]
\centering
\includegraphics[width=\linewidth]{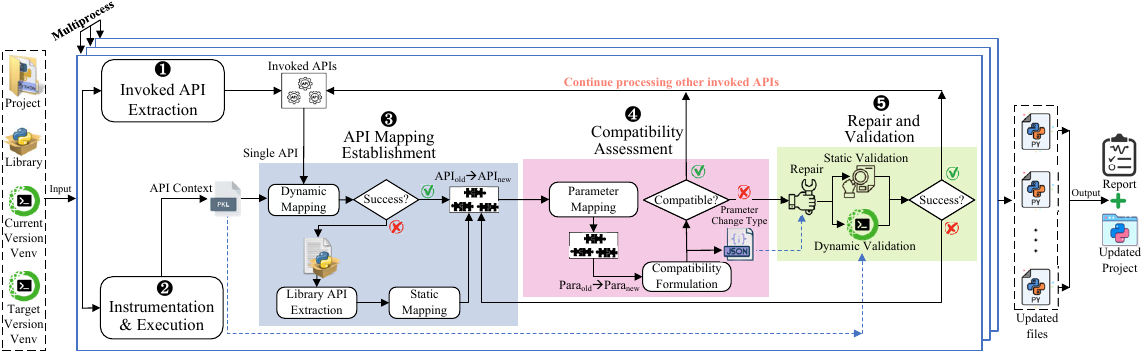}
\vspace{-4mm}
\caption{Overview of our \tool approach.} 
\label{overview}
\vspace{-4mm}
\end{figure*}


Existing Python-focused migration tools are insufficient to address these challenges. Most emphasize detecting deprecated APIs (e.g., PyCompat~\cite{zhang2020python}, DLocator~\cite{wang2020exploring}, APIScanner~\cite{vadlamani2021apiscanner}) or analyzing evolution patterns (e.g., AexPy~\cite{du2022aexpy}, pidiff~\cite{GitHub:pidff}), without supporting automated repair. Attempts at repair are limited: MLCatchUp~\cite{haryono2021mlcatchup} uses static, manually provided mappings to update machine learning APIs but cannot generalize, while Relancer~\cite{zhu2021restoring} relies on dynamic error-driven iteration in Jupyter notebooks. These approaches fall short in three ways: \textbf{(1)} they assess compatibility primarily at the definition level, overlooking the impact of parameter-passing methods on actual usage; \textbf{(2)} they lack automation, relying on manual mappings or weak validation (e.g., successful execution alone); and \textbf{(3)} they offer limited repair strategies, failing to handle the flexibility and diversity of Python call syntax.

To address these limitations, we present \tool, an \underline{a}utomated \underline{r}epair \underline{t}ool for \underline{P}ython API parameter \underline{c}ompatibility issues. \tool is, to our knowledge, the first approach to achieve end-to-end automation of parameter-level API migration in a dynamic language environment. As shown in our framework overview (Fig.~\ref{overview}), unlike prior work, \tool combines static and dynamic analysis to automatically (i) extract invoked APIs, (ii) instrument and execute code to capture runtime behavior, (iii) establish precise mappings between old and new API signatures, (iv) assess compatibility considering both parameter changes and invocation styles, and (v) generate and validate repairs. Repairs are inferred automatically without predefined templates, covering parameter addition, removal, renaming, reordering, and positional-to-keyword conversion. Validation is performed through a hybrid static–dynamic strategy, ensuring both formal consistency and runtime executability. This end-to-end pipeline, from detection to validated repair, requires no manual intervention, making it fundamentally more automated and reliable than prior tools in Python or other ecosystems.

\rev{A practical challenge underlying this workflow is how to obtain complete and reliable API signature information. Although third-party libraries do expose function definitions with parameter names and defaults, purely static extraction is often unreliable in practice due to aliasing, same-name APIs across modules, and mismatches between fully qualified source definitions and the symbols actually invoked in client code. To address this, \tool primarily leverages Python's reflection (\mintinline{python}{inspect}) to extract the effective signatures of the concrete objects reached at runtime, thereby aligning parameter definitions with the true call targets observed during execution. When reflection is not applicable, \tool falls back to static parsing of library sources.} 



To evaluate \tool, we construct a large-scale benchmark, \benchmark, comprising 47,478 test cases across 844 parameter-changed APIs from 33 widely used Python libraries. We compare \tool against state-of-the-art repair tools (MLCatchUp and Relancer), evaluate it on 30 real-world GitHub projects, and benchmark it against ChatGPT (GPT-4o)~\cite{ChatGPT}. Results show that \tool achieves an F1-score of 96.51\% for detection and a repair precision of 91.97\%, substantially outperforming existing methods, i.e., MLCatchUp (86.42\%/8.82\%) and Relancer (78.51\%/0.00\%). Finally, we analyze its efficiency, with an average processing time of 3,096 ms per API call. These results demonstrate that \tool not only advances the automation of Python API migration but also provides insights relevant to migration in other dynamic languages such as JavaScript, Ruby, and R.

In summary, we make the following contributions:

\begin{itemize}
    \item \textbf{Automated API Migration Approach.} We present \tool, the first fully automated approach for detecting, repairing, and validating Python API parameter compatibility issues. \tool achieves end-to-end automation without requiring a prior change database, bridging a key gap in dynamic language migration research. \tool is open source and available at \url{https://github.com/pcart-tools}. 

    \item \textbf{Benchmark for API Compatibility.} We introduce \benchmark, a large-scale benchmark with 47,478 test cases from 844 APIs across 33 libraries. It covers diverse parameter change types and passing methods, offering a rigorous evaluation foundation for migration techniques.

    \item \textbf{Experimental Evaluation.} We comprehensively evaluate \tool against state-of-the-art tools and real-world projects, demonstrating its superior detection accuracy, repair precision, and practical feasibility. Comparisons with ChatGPT further highlight its robustness.

\end{itemize}

\section{Background and Challenges}\label{sec:background}
\subsection{Characteristics of Python API Parameters}\label{sec:background-characteristics}

Python demonstrates exceptional flexibility in its API parameter passing mechanism, markedly contrasting with traditional programming languages such as C/C++ and Java~\cite{python}. This flexibility is reflected in several key aspects:

\textbf{(1) Supports Positional and Keyword Parameters.} Python employs a special syntax (i.e., \mintinline{python}{*}) in API definitions to distinguish positional and keyword parameters, the two fundamental types of parameters. Parameters located before ``\mintinline{python}{*}'' are positional parameters, which must be passed in order according to their positions if not accompanied by a parameter name; otherwise, they can be passed out of order if the parameter name is given. Parameters after ``\mintinline{python}{*}'' are keyword parameters, which require the inclusion of the parameter name when used; otherwise, a syntax error will occur.

As illustrated in Listing~\ref{pos_and_key}, when the API \mintinline{python}{Rule} of the Rich library is upgraded from version 2.3.1 to 3.0.0, the positional parameters \mintinline{python}{character} and \mintinline{python}{style} are transformed into keyword parameters. Consequently, in Rich version 2.3.1, if called without specifying parameter names, upgrading to version 3.0.0 would result in a compatibility issue. Conversely, if called with parameter names, the upgrade 
is compatible.

\begin{figure}[!t]
\centering
\begin{minipage}{3.4in}
\begin{lstlisting}[language=Python, label=pos_and_key, caption=Examples of positional/keyword parameters and different parameter passing methods.]
#API definition in library Rich 2.3.1
def Rule(title:Union[str,Text]='',character:str=None,style:Union[str,Style]='rule.line')

#API definition in library Rich 3.0.0
def Rule(title:Union[str,Text]='',*,character:str=None,style:Union[str,Style]='rule.line')

from rich.rule import Rule
#Incompatible calling from Rich 2.3.1 to 3.0.0
rule = Rule('', None, 'rule.line')

#Compatible calling from Rich 2.3.1 to 3.0.0
rule = Rule('', character=None, style='rule.line')
\end{lstlisting}
\end{minipage}
\vspace{-4mm}
\end{figure}

\textbf{(2) Supports Optional Parameters.} In Python API definitions, optional parameters (i.e., parameters with default values) are not necessarily passed during API calls. Listing~\ref{default parameter} shows that the API \mintinline{python}{Proxy} defined in HTTPX version 0.18.2 includes an optional parameter \mintinline{python}{mode}, which is removed in 0.19.0. If the removed optional parameter is used with the invocation of API \mintinline{python}{Proxy} in HTTPX 0.18.2, such a call would become incompatible upon upgrading to the new version 0.19.0. In contrast, if this optional parameter is not used, 
upgrading HTTPX from 0.18.2 to 0.19.0 remains compatible.

\begin{figure}[!t]
\centering
\begin{minipage}{3.4in}
\begin{lstlisting}[language=Python, label=default parameter, caption=Examples of optional parameter and different parameter passing methods.]
#API definition in library HTTPX 0.18.2
def Proxy(self,url:URLTypes,*,headers:HeaderTypes=None,mode:str='DEFAULT')

#API definition in library HTTPX 0.19.0
def Proxy(self,url:URLTypes,*,headers:HeaderTypes=None)

import httpx
proxy_url = 'http://localhost:8080'
proxy_headers = {'Custom-Header': 'Value'}

#Incompatible calling from HTTPX 0.18.2 to 0.19.0
proxy = httpx.Proxy(proxy_url,headers=proxy_headers,mode='DEFAULT')

#Compatible calling from HTTPX 0.18.2 to 0.19.0
proxy = httpx.Proxy(proxy_url,headers=proxy_headers)
\end{lstlisting}
\end{minipage}
\vspace{-4mm}
\end{figure}

\textbf{(3) Supports Variadic Parameters.} Python introduces variadic parameters, i.e., \mintinline{python}{*args} and \mintinline{python}{**kwargs}, to permit APIs to accept an arbitrary number of positional and keyword arguments, respectively. 
As illustrated in Listing~\ref{variadic}, the API \mintinline{python}{pdist} defined in version 0.19.1 of the SciPy library accepts specific parameters: \mintinline{python}{p}, \mintinline{python}{w}, \mintinline{python}{V}, and \mintinline{python}{VI}. However, in version 1.0.0, these specific parameters are replaced with \mintinline{python}{*args} and \mintinline{python}{**kwargs} to allow the API to accept a broader range of arguments. Despite the removal of some parameters in version 1.0.0, using these removed parameters 
remains compatible.

\begin{figure}[!t]
\centering
\begin{minipage}{3.4in}
\begin{lstlisting}[language=Python, label=variadic, caption=Examples of variadic parameters.]
#API definition in library SciPy 0.19.1
def pdist(X, metric='euclidean', p=None, w=None, V=None, VI=None)

#API definition in library SciPy 1.0.0
def pdist(X, metric='euclidean', *args, **kwargs)

from scipy.spatial.distance import pdist
#Compatible calling from SciPy 0.19.1 to 1.0.0
pdist(X, 'euclidean', None, None, V=None, VI=None)
\end{lstlisting}
\end{minipage}
\vspace{-4mm}
\end{figure}

The aforementioned flexible parameter-passing mechanism of Python provides developers with a convenient and efficient programming experience, it also impacts the compatibility of APIs during the evolution of Python libraries. 
Through an in-depth analysis of six popular Python frameworks, Zhang \textit{et al.}~\cite{zhang2020python} found that 8 out of 14 common API change patterns directly involve parameter changes. This result underscores the prevalence and significance of parameter changes in Python API evolution. A recent large-scale empirical study on API breaking changes 
revealed that 
34 out of 61 cases are caused by parameter changes~\cite{du2022aexpy}, highlighting the prevalence and critical role of parameter changes in API breaking changes. 
Automating the resolution of API parameter compatibility issues can significantly reduce developers' time spent on manually maintaining client code affected by breaking API parameter changes.

\subsection{Challenges in Automated Detection and Repair of Python API Parameter Compatibility Issues}\label{sec:background-challenges}

To automatically detect and repair Python API parameter compatibility issues, we face the following challenges: 

\textbf{Challenge 1.} \textit{Compatibility Assessment.} How to precisely assess the compatibility of invoked APIs is challenging. First, 
the compatibility assessment 
depends not only on the changes in API definitions but also on the actual usage of parameter-passing methods. From Listings~\ref{pos_and_key} to \ref{variadic}, we can observe that breaking parameter changes does not necessarily mean calling the API would cause compatibility issues. Parameter-passing methods used in projects greatly impact API compatibility.

Second, a runnable API invocation 
does not imply it is truly compatible, as not all API parameter compatibility issues would result in a program crash. As the example shown in Listing~\ref{run but incmp}, the \mintinline{python}{maxcardinality} parameter of the API \mintinline{python}{min_weight_matching} is removed in NetworkX version 3.0. Since this parameter is passed by its position 
when calling API \mintinline{python}{min_weight_matching} in version 2.8.8, upon upgrading to the new version, its value would erroneously be assigned to the \mintinline{python}{weight} parameter, due to the removal of the parameter \mintinline{python}{maxcardinality}. The code snippet is executable without any syntax errors. However, the semantics of the API have been changed in the program. 

\begin{figure}[!t]
\centering
\begin{minipage}{3.4in}
\begin{lstlisting}[language=Python, label=run but incmp, caption=An example of runnable but incompatible code snippet.]
#API definition in library NetworkX 2.8.8
def min_weight_matching(G,maxcardinality=None,weight='weight')

#API definition in library NetworkX 3.0
def min_weight_matching(G,weight='weight')

import networkx as nx
G = nx.Graph()
#Runnable but incompatible calling from Networkx 2.8.8 to 3.0
matching = nx.min_weight_matching(G, None)
\end{lstlisting}
\end{minipage}
\vspace{-4mm}
\end{figure}


\textbf{Challenge 2.} \textit{Automated Establishment of API Mappings.}  
Establishing API mappings automatically is crucial for implementing a fully automated detection and repair tool. Existing tools, such as MLCatchUp~\cite{haryono2021mlcatchup} and Relancer~\cite{zhu2021restoring}, require users to manually provide the old and new signatures of the updated APIs or to manually pre-build a database of breaking API changes. 
For detecting and repairing API parameter compatibility issues, we need to establish two types of API mappings: 
(1) mappings of API signatures between the old and the new library versions: $API_{old} \rightarrow API_{new}$; (2) mappings of parameters between the old and the new API signatures: $Parameter_{old} \rightarrow Parameter_{new}$. In the following, we discuss the challenges in automated establishing these two types of API mapping relationships. 

\textit{(1) Establishing API Signature Mappings.} To establish $API_{old} \rightarrow API_{new}$ mappings, one solution is to extract the definition of the invoked API from the library source code. Existing tools, such as DLocator~\cite{wang2020exploring} and PyCompat~\cite{zhang2020python}, mainly employ manual or semi-automated approaches to extract API definitions 
by comparing the call paths of APIs invoked in the project against their real paths in the library source code. However, the effectiveness of such a solution is significantly impacted by the same-name APIs, API aliases, and 
API overloading in the source code of Python libraries.

\begin{figure}[!t]
\centering
\includegraphics[width=2.25in]{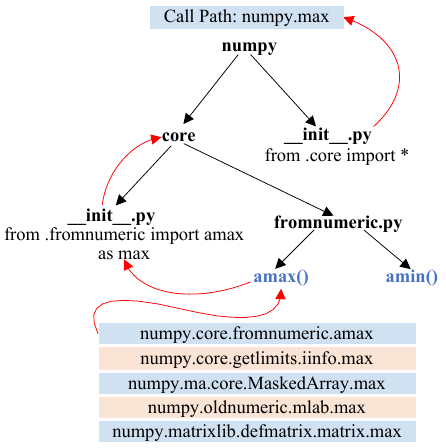} 
\vspace{-4mm}
\caption{An example of alias and import mechanism of Python APIs.}
\label{import and init}
\vspace{-4mm}
\end{figure}

On the one hand, the same-name APIs and API aliases may generate multiple uncertain matching results. 
Developers of Python libraries often use the ``\mintinline{python}{__init__.py}'' file and the \mintinline{python}{import} mechanism to create API aliases, 
aiming at shortening the call paths of public APIs available for user usage~\cite{Import}. However, API aliases can lead to inconsistencies between an API's call path in the project and its real path in the library source code. 
As shown in Fig.~\ref{import and init}, in the NumPy source code (version 1.10.0), the real path of API \mintinline{python}{max} is 
\mintinline{python}{numpy.core.fromnumeric.amax}. Due to the alias and \mintinline{python}{import} mechanism, the call path provided for user invocation is \mintinline{python}{numpy.max}. If the terms ``numpy'' and ``max'' are used for comparison, the matching result is uncertain, because 
there are several other APIs with the same name \mintinline{python}{max}, such as \mintinline{python}{numpy.core.getlimits.iinfo.max} and \mintinline{python}{numpy.ma.core.MaskedArray.max}. 
Our preliminary statistics of the same-name APIs on 33 popular Python libraries (described in Section~\ref{sec:benchmark}) find that the proportion of the same-name APIs against the total APIs ranges from 3.90\% to 24.38\%  per version across different libraries.

On the other hand, API overloading also poses challenges in precisely establishing $API_{old} \rightarrow API_{new}$ mappings. 
Although Python does not support function overloading, many third-party libraries, such as PyTorch, TensorFlow, and NumPy, have implemented function overloading through C/C++ extensions. These overloads can automatically select and invoke the correct C/C++ function version based on the types and numbers of arguments passed during API invocation. For example, in PyTorch 1.5.0~\cite{PyTorch}, the API  \mintinline{python}{torch.max} has three overloading forms: \mintinline{python}{torch.max(input)}, \mintinline{python}{torch.max(input, dim, keepdim=False, out=None)}, and \mintinline{python}{torch.max(input, other, out=None)}. Due to multiple overloading forms, it is difficult to identify which one is called in the project, even if the definition is correctly extracted from the library source code. This may result in establishing wrong  $API_{old} \rightarrow API_{new}$ mappings. 

\textit{(2) Establishing Parameter Mappings.} After establishing API signature mappings, it is necessary to establish parameter mappings (i.e., $Parameter_{old} \rightarrow Parameter_{new}$) for analyzing parameter changes. Intuitively, the establishment of parameter mappings relies on the name of parameters for matching. Once the mapping is determined, further analysis of parameter changes, such as position change or type change, can be conducted. However, establishing correct mappings becomes more complicated when parameter renaming or removal occurs. As shown in Listing~\ref{rename and removal}, given the signatures of TensorFlow API \mintinline{python}{DispatchServer} between 2.3.4 and 2.4.0 versions, 
the parameter \mintinline{python}{start} can be mapped explicitly based on its name between the two versions. However, the relationship between \mintinline{python}{port} and \mintinline{python}{protocol} parameters in 2.3.4 and the \mintinline{python}{config} parameter in the new version 2.4.0 is difficult to determine. 
Through checking API documents, the changes are removal (\mintinline{python}{port} and \mintinline{python}{protocol}) and addition (\mintinline{python}{config}). 
Similarly, in the \mintinline{python}{drop} API of the Polars library, the parameters \mintinline{python}{name} and \mintinline{python}{columns} from versions 0.14.17 to 0.14.18 might not seem like renaming, 
yet they actually are. 

\begin{figure}[!t]
\centering
\begin{minipage}{3.4in}
\begin{lstlisting}[language=Python,label=rename and removal,caption=Examples of parameter renaming and removal.]
#API definition in library Tensorflow 2.3.4
def DispatchServer(port, protocol=None, start=True)

#API definition in library Tensorflow 2.4.0
def DispatchServer(config=None, start=True)


#API definition in library polars 0.14.17
def drop(name: 'str | list[str]')

#API definition in library polars 0.14.18
def drop(columns: 'str | list[str]')
\end{lstlisting}
\end{minipage}
\vspace{-3mm}
\end{figure}


\textbf{Challenge 3.} \textit{Automated Repair and Validation.} 
Automated repair and validation constitute a critical component in addressing Python API parameter compatibility issues. 
MLCatchUp~\cite{haryono2021mlcatchup} repairs compatibility issues based on the manual given API signatures. The static method 
requires manual effort to validate the repair results, which is error-prone and time-consuming. Relancer~\cite{zhu2021restoring}, on the other hand, repairs by dynamically executing each line of code in Jupyter notebooks, based on runtime error messages, but it is susceptible to the impact of multiple compatibility issues within a single file or across several source files. For example, 
if the code snippet shown in Listing~\ref{multiple} 
contains several API invocations with compatibility issues in parameters, where the failure to repair any one of these invoked APIs could halt the entire automated repair and validation process. This is because of the sequential code execution. The execution of each API depends on its context within the code, such as the dependency of function \mintinline{python}{a.b(z)} on the value of parameter \mintinline{python}{z} and the return value of \mintinline{python}{A(x, y=1)}. 
If the incompatible invocation of \mintinline{python}{A(x, y=1)} has not been fixed, the return value would not pass to \mintinline{python}{a.b(z)}. Even if API \mintinline{python}{b} is fixed, running \mintinline{python}{a.b(z)} cannot validate the fix because the return value of \mintinline{python}{A(x, y=1)} is unavailable. Therefore, to independently repair and validate each API, it is essential to acquire the contextual dependency information of each invoked API within the code.

\begin{figure}[!t]
\centering
\begin{minipage}{3.4in}
\begin{lstlisting}[language=Python, label=multiple, caption=An example of multiple invoked APIs with compatibility issues in parameters within a single source file.]
foo1(1,x,y)   #x was removed
...
foo2(x,0.2,y) #x and y swapped positions
...
a=A(x,y=1)    #y was renamed
...
a.b(z)        #Conversion to keyword parameter 
\end{lstlisting}
\end{minipage}
\vspace{-3mm}
\end{figure}

\section{Our \tool Approach}\label{sec:pcart}

\subsection{Overview}
To address the aforementioned challenges, 
we introduce \tool (Fig.~\ref{overview}), which has the following key advantages.

\begin{itemize}
    \item \textbf{Precise Compatibility Assessment.} \tool precisely assesses the compatibility of invoked APIs based on the formulation of three information sources, i.e., parameter types (e.g., positional/keyword parameter), change types (e.g., removal/renaming), and parameter passing methods (e.g., positional/keyword passing) (Section~\ref{sec:pcart-assessment}).
    \item \textbf{Fully Automated Detection and Repair.} \tool establishes API mapping relationships automatically. To establish API signature mappings, \tool introduces a novel code instrumentation and dynamic mapping approach to precisely acquire the API definitions across the current version and the target version to be upgraded (Sections~\ref{sec:pcart-instrumentation} and~\ref{sec:pcart-mapping}). Besides, \tool establishes parameter mappings by a rule-based method (Section~\ref{sec:pcart-assessment}). Moreover, the validation of repair is also automatically performed in \tool by integrating both dynamic and static validations (Section~\ref{sec:pcart-repair}).
    \item \textbf{Diverse Parameter Changes Support.} \tool supports extracting complex forms of API calls (Section~\ref{sec:pcart-extraction}) and repairing API parameter compatibility issues raised by various types of changes, i.e., parameter addition, removal, renaming, reordering, and the conversion of positional parameters to keyword parameters (Section~\ref{sec:pcart-repair}). 
\end{itemize}

Fig.~\ref{overview} shows the overview of \tool. 
When users plan to upgrade the third-party library dependency in their project to a new version, 
initially, \tool extracts the APIs related to the upgraded library from the project's source code. Then, \tool performs code instrumentation for the invoked APIs to save their contextual dependency information by executing the project. Subsequently, it employs both dynamic and static methods to establish accurate API mappings. It then assesses the compatibility of the invoked APIs, and 
finally, if a compatibility issue is found, \tool repairs and validates the incompatible API invocation to the compatible one. 


Before running \tool, we manually prepare a compatible upgraded environment for the target third-party library by cloning a working baseline environment (e.g., via conda pack), upgrading the target library within it, and resolving any dependency conflicts reported by pip. In addition, \tool's logging mechanism records all runtime errors (including traceback information) during dynamic mapping and validation, enabling users to identify and fix any remaining inter-library compatibility issues.

\begin{figure}[!t]
\centering
\includegraphics[width=\linewidth]{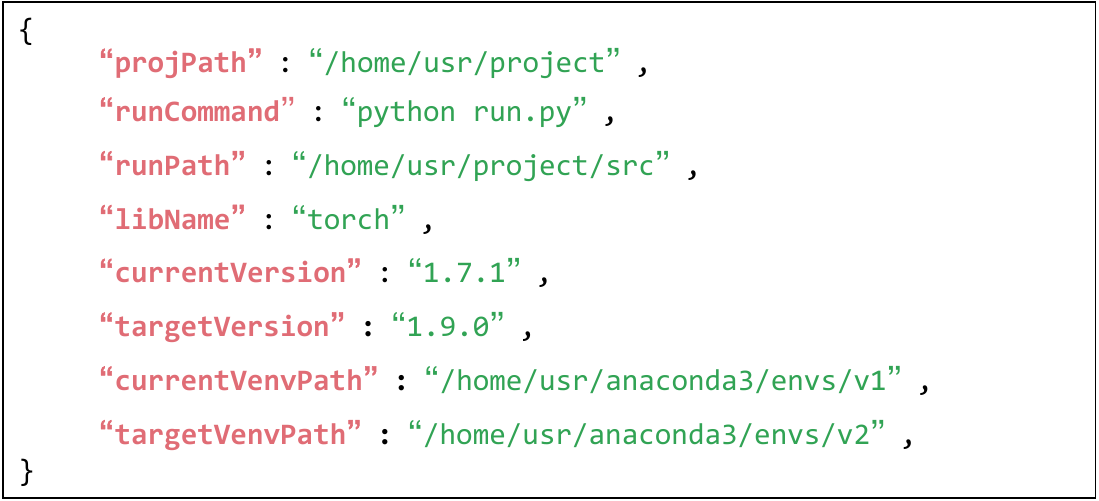} 
\vspace{-3mm}
\caption{The input configuration of \tool.}
\label{config}
\vspace{-3mm}
\end{figure}

\tool begins by accepting a configuration file as input (Fig.~\ref{config}), which contains the following information: path to the project, run command and its entry file path, library name, current version number, target version number, path to the virtual environment of the current version, and path to the virtual environment of the target version. \tool considers each source file in the project as a task to be processed and adds it to a task queue. For acceleration, 
\tool creates a pool of processes to handle these tasks concurrently. Each process executes a full set of detection and repair procedures.

\begin{figure}[!t]
\centering
\includegraphics[width=\linewidth]{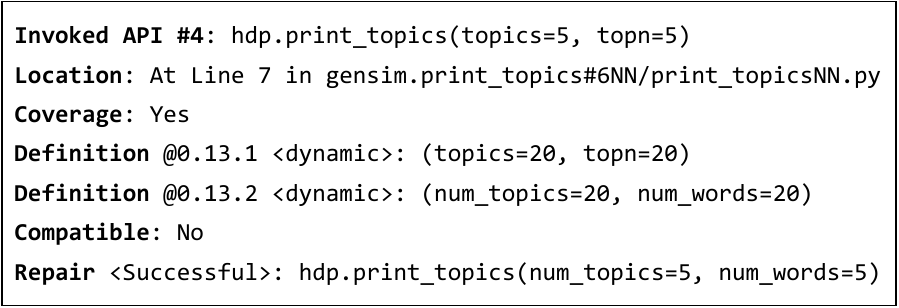} 
\vspace{-3mm}
\caption{The output repair report of \tool.}
\label{report}
\vspace{-3mm}
\end{figure}

After processing all source files, \tool outputs 
the repaired project and the repair report (Fig.~\ref{report}). The report records each API's invocation form within the project, invocation location, coverage information of dynamic execution, parameter definitions in both the current and target versions, compatibility status, and the results of the repairs.

Below, we elaborate on the design details of \tool, which consists of five main modules: \ding{182} invoked API extraction, \ding{183} instrumentation and execution, \ding{184} API mapping establishment, \ding{185} compatibility assessment, and \ding{186} repair and validation.

\subsection{Invoked API Extraction}\label{sec:pcart-extraction}
Given a Python project, which typically contains multiple \mintinline{python}{.py} source files, each file may invoke several APIs of the target third-party library. 
\tool first converts the source files into abstract syntax trees (ASTs), and then traverses the ASTs to identify all the API calls related to the specified library needing to be upgraded. The source files of the invoked APIs and the line positions within these files are also extracted. Due to the diversity in programming habits among different developers, the API invocation statements in the code vary in form, making it challenging to precisely extract all library-related API calls as strings from the source files by using text processing techniques like regular expressions. Thus, transforming the source files into a uniform format, i.e., AST, facilitates the extraction of invoked APIs. 
Details of the invoked API extraction are presented as follows.

\begin{figure}[!t]
\centering
\includegraphics[width=2in]{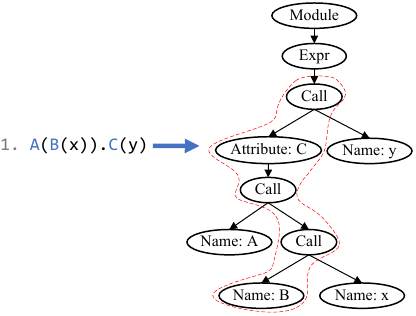}
\vspace{-3mm}
\caption{The AST structure of A(B(x)).C(y).}
\label{A(B(x)).C(y)}
\vspace{-3mm}
\end{figure}

\textit{(1) Traversing the Abstract Syntax Tree}. First, for each \mintinline{python}{.py} file, \tool uses Python's built-in AST module to parse the source code into an AST. Then, \tool employs the breadth-first search (BFS) to perform a level-order search on the AST, identifying nodes of types, i.e., \mintinline{python}{Assign}, \mintinline{python}{Import}, and \mintinline{python}{ImportFrom}, which correspond to the assignment, import, and from-import statements, 
respectively.

Second, for extracting API call statements, \tool performs the depth-first search (DFS) for branch-wise deep searches, due to the storage structure of API calls in the ASTs. As the example shown in Fig.~\ref{A(B(x)).C(y)}, 
given a complex API call statement \mintinline{python}{A(B(x)).C(y)}, API \mintinline{python}{C} is called through the return value of API \mintinline{python}{A}, while API \mintinline{python}{B} is called as an argument of API \mintinline{python}{A}. Such API invocation format is located on the same branch of the tree. Therefore, the DFS algorithm not only allows for the determination of the sequential relationship between APIs during the search process but also enables the extraction of all potential API calls within the source code. Existing tools (e.g., DLocator~\cite{wang2020exploring} and MLCatchUp~\cite{haryono2021mlcatchup}) are unable to accurately identify this type of call format. 
In contrast, \tool supports eight typical types of API calls, i.e., direct invocation, class object invocation, return value invocation, argument invocation, inheritance invocation, decorator invocation, async/await invocation, and context manager invocation (Listing~\ref{Invocation Formats}).


\textit{(2) Restoring the Conventional API Call Path.} \tool combines the \mintinline{python}{Assign} and \mintinline{python}{Import} nodes to reconstruct the call path of an API, enabling the identification of the third-party library it belongs to. First, the user explicitly specifies the target library to be upgraded in \tool's configuration file (Fig.~\ref{config}). Then, during the API extraction phase, \tool standardizes all API invocation paths in the project to the format ``\mintinline{python}{Lib.Package.Module.Class.API}''. By performing string matching on the library name ``\mintinline{python}{Lib}'', \tool accurately identifies API calls associated with the target library.

\begin{figure}[!t]
\centering
\begin{minipage}{3.4in}
\begin{lstlisting}[language=Python, label=Invocation Formats, caption=Eight typical types of API calls.]
#1. Direct Invocation
foo(x, y)

#2. Class Object Invocation
a=A(x)
a.foo(y, z)

#3. Return Value Invocation
f(x).foo(y, z)

#4. Argument Invocation
f(x, foo(y, z))

#5. Inheritance Invocation
from pkg.module import C
class Custom(C):
    def custom_method(self, x, y):
        self.foo(x, y)

#6. Decorator Invocation
@foo(param)
def bar(x, y):
    return x + y

#7. Async/Await Invocation
async def task():
    result = await foo(x, y)
    return result

#8. Context Manager Invocation
with foo(x) as r:
    r.bar(y)
\end{lstlisting}
\end{minipage}
\vspace{-3mm}
\end{figure}

\begin{figure}[!t]
\centering
\begin{minipage}{3.4in}
\begin{lstlisting}[language=Python,label=regularAPI, caption=Calling a regular API.]
from Lib.pkg.module import M as A
a=A(x)
a.b(y,z)
\end{lstlisting}
\end{minipage}
\vspace{-3mm}
\end{figure}

As demonstrated in Listing~\ref{regularAPI}, for the invocation \mintinline{python}{a.b(y, z)}, the assignment statement \mintinline{python}{a = A(x)} reveals that the variable \mintinline{python}{a} is an object instantiated from class \mintinline{python}{A}. Consequently, \mintinline{python}{a.b(y, z)} can be restored to \mintinline{python}{A(x).b(y, z)}. Further analysis of the \mintinline{python}{ImportFrom} statement shows that \mintinline{python}{A} is an alias for \mintinline{python}{M}, which is imported from \mintinline{python}{pkg.module}. Thus, the API call statement \mintinline{python}{A(x).b(y, z)} is finally restored to its fully qualified call form as \mintinline{python}{Lib.pkg.module.M(x).b(y, z)}.

\textit{(3) Restoring the Path of Inherited API Calls.} Beyond conventional API calls, users also invoke APIs from Python libraries using their custom classes through inheritance. For instance, in Listing~\ref{inheritAPI}, \mintinline{python}{self.c_method} is essentially an API from a third-party library. However, existing tools, e.g., DLocator~\cite{wang2020exploring}, PyCompat~\cite{zhang2020python}, and MLCatchUp~\cite{haryono2021mlcatchup}, fail to extract such API call formats. 

To address this issue, \tool first identifies all \mintinline{python}{ClassDef} type nodes on the AST, corresponding to class definition statements in the code, and then assesses whether each custom class has any inheritance. If so, \tool extracts all custom APIs defined within that class. Subsequently, \tool determines whether each self-invoked API belongs to the class's custom APIs. If not, it is considered an API from the inherited class. For example, \mintinline{python}{self.c_method} would be resolved back to \mintinline{python}{C.c_method}, and later further resolved to \mintinline{python}{pkg.module.C.c_method}.

\begin{figure}[!t]
\centering
\begin{minipage}{3.4in}
\begin{lstlisting}[language=Python, label=inheritAPI, caption=An example of inherited API call.]
from pkg.module import C
class Custom(C):
    def custom_method(self):
        self.c_method()
\end{lstlisting}
\end{minipage}
\vspace{-3mm}
\end{figure}

\subsection{Instrumentation and Execution}\label{sec:pcart-instrumentation}

\begin{figure}[!t]
\centering
\includegraphics[width=2in]{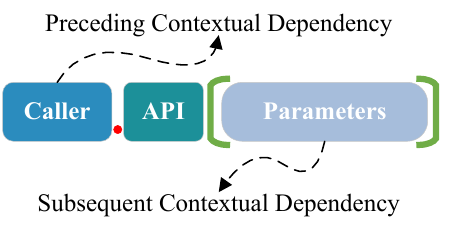}
\vspace{-3mm}
\caption{Context dependency of an invoked API.}
\label{context}
\vspace{-3mm}
\end{figure}

\rev{To achieve fully automated API mapping and enable independent repair and validation of each invoked API, we design a runtime instrumentation mechanism that preserves the complete contextual dependencies of API invocations. As shown in Fig.~\ref{context}, for each API call, \tool records two types of contextual dependencies: the \textit{preceding dependency}, capturing the caller information, and the \textit{subsequent dependency}, preserving the runtime parameter values.}

\rev{This design is crucial for dynamic API analysis in Python. In real-world projects, APIs are often invoked through complicated expressions (e.g., \mintinline{python}{a(x).b(y, z)}), where directly invoking \mintinline{python}{inspect.signature(a(x).b)} fails because the interpreter cannot resolve the runtime value of \mintinline{python}{x}. Through instrumentation, \tool serializes both the caller and its parameters, allowing reconstruction of the invocation context and safe extraction of API signatures across versions via reflection. This preserved context enables signature extraction without executing the entire project and supports fine-grained, per-API detection, repair, and validation.}

As shown in Listing~\ref{instrument}, \tool inserts corresponding assignment statements (dictionaries) into the code to obtain the contextual dependency of each invoked API. 
By running the project in the current version (the compatible one), the contextual dependency of the invoked API is recorded in the instrumented dictionaries. Later, the recorded values are serialized and stored in pickle files (\mintinline{python}{.pkl}) in binary format by utilizing Dill, a powerful library for serializing and de-serializing Python objects~\cite{Dill}. Each invoked API has one corresponding pickle file. 
The reason for choosing pickle files for storage is that they can effectively save all variable values and Python objects generated during runtime. 
Moreover, these stored values can be easily retrieved by directly loading the pickle file without the need to rerun the project. However, instrumentation for every API call may encounter several difficulties due to different coding styles and API call formats. 
Below, we discuss five typical types of processing encountered during code instrumentation.

\begin{figure}[!t]
\centering
\begin{minipage}{3.4in}
\begin{lstlisting}[language=Python, label=instrument, caption=Instrumentations for the \mintinline{python}{foo} API regarding different formats of API calls.]
#1. Direction Invocation
paraValueDict['foo(x, y)']=[x, y]
foo(x, y)

#2. Class Object Invocation
paraValueDict['A(x)']=[x]
a=A(x)
paraValueDict['@a.foo(y, z)']=a
paraValueDict['a.foo(y, z)']=[y, z]
a.foo(y, z)

#3. Return Value Invocation
paraValueDict['f(x)']=[x]
paraValueDict['@f(x).foo(y, z)']=f(x)
paraValueDict['f(x).foo(y, z)']=[y ,z]
f(x).foo(y, z)

#4. Argument Invocation
paraValueDict['foo(y, z)']=[y, z]
paraValueDict['f(x,, foo(y, z)']=[x, foo(y, z)]
f(x, foo(y, z))

#5. Inheritance Invocation
from pkg.module import C
class Custom(C):
    def custom_method(self, x, y):
        paraValueDict['@self.foo(x, y)']=self
        paraValueDict['self.foo(x, y)']=[x, y]
        self.foo(x, y)
custom=Custom()
custom.custom_method(x, y)

#6. Decorator Invocation
paraValueDict['foo(param)']=[param]
@foo(param)
def bar(x, y):
    return x + y

#7. Async/Await Invocation
async def task():
    paraValueDict['foo(x, y)']=[x, y]
    result = await foo(x, y)
    return result

#8. Context Manager Invocation
paraValueDict['foo(x)']=[x]
with foo(x) as r:
    paraValueDict['@r.bar(y)']='foo(x)'
    paraValueDict['r.bar(y)']=[y]
    r.bar(y)
\end{lstlisting}
\end{minipage}
\vspace{-3mm}
\end{figure}

\textit{(1) Handling Code Indentation.} 
Python defines the scope of different statements through the use of indentation. However, due to personal habits and differences in integrated development environments (IDEs) used by developers, code may employ various indentation styles, such as spaces and tab characters. Therefore, in the process of code instrumentation, to accurately calculate the indentation for each instrumented statement, \tool converts all tab characters in the code to four spaces by the shell command, i.e., \mintinline{shell}{expand -t 4 "$file" > "$temp" && mv "$temp" "$file"}. This ensures the instrumentation can be correctly applied across different projects. 

\begin{figure}[!t]
\centering
\begin{minipage}{3.4in}
\begin{lstlisting}[language=Python, label=location, caption=Instrumentation for API calls in \mintinline{python}{return} statements.]
def foo():
  paraValueDict["f(1,2)"]=[1, 2] #Correct Location
  return f(1,2)
  paraValueDict["f(1,2)"]=[1, 2] #Wrong Location
\end{lstlisting}
\end{minipage}
\vspace{-3mm}
\end{figure}

\textit{(2) Determining the Location for Instrumentation.} 
\tool first traverses every line of the source file and performs a string-based comparison to locate the lines of invoked APIs according to the API calls obtained from \ding{182}. Then, for each API call, 
\tool inserts the assignment statement before the line of the API call, applying the same indentation level. 
This is because when an API call occurs within a \mintinline{python}{return} statement, such as \mintinline{python}{return f(1, 2)} in Listing~\ref{location}, if the instrumentation inserted after the \mintinline{python}{return} statement, the instrumented statement would not execute during runtime, thereby failing to store the contextual dependency of the invoked API.

\textit{(3) Handling Line Breaks in Parameter Passing.} When an API call passes multiple parameters, developers may opt to write each parameter on a new line to enhance code readability. 
However, this can certainly complicate code instrumentation, as shown in Listing~\ref{breakline}. Therefore, \tool first modifies all statements in the code that pass parameters with line breaks to be on the same line before instrumentation. This ensures the correctness of the instrumentation process.

\begin{figure}[!t]
\centering
\begin{minipage}{3.4in}
\begin{lstlisting}[language=Python, label={breakline}, caption=Handling line breaks in parameter passing.]
#Before Handling
val= Foo(
  	paraValueDict["f1(x)"]=[x] #Wrong Instrumentation
  	f1(x),
  	f2(y),
  	f3(z),
	)

#After Handling
#Correct Instrumentation
paraValueDict['f1(x)']=[x]
paraValueDict['f2(y)']=[y]
paraValueDict['f3(z)']=[z]
paraValueDict['Foo(f1(x), f2(y), f3(z))']=[f1(x), f2(y), f3(z)]
val= Foo(f1(x), f2(y), f3(z))
\end{lstlisting}
\end{minipage}
\vspace{-3mm}
\end{figure}

\textit{(4) Handling List Comprehensions.} List comprehension in Python is a concise and efficient method for creating lists and dictionaries from iterable objects, structured as \mintinline{python}{[expression for item in iterable if condition]}. As depicted in Listing~\ref{comprehensions}, when the expression is a function call to \mintinline{python}{f}, since the parameters \mintinline{python}{x} and \mintinline{python}{y} that \mintinline{python}{f} depends on are located inside the list, instrumenting the line before would lead to undefined variable errors. To solve this issue, \tool parses list comprehensions using Python's built-in AST module to obtain the \mintinline{python}{item}, \mintinline{python}{iterable}, and \mintinline{python}{condition}. Then, it transforms them into the form \mintinline{python}{[item in iterable if condition]}. The first element in the list is selected as the parameter value for the function \mintinline{python}{f}.

\begin{figure}[!t]
\centering
\begin{minipage}{3.4in}
\begin{lstlisting}[language=Python, label=comprehensions, caption=Handling API calls in list comprehensions.]
#Before Handling
paraValueDict['f(x,y)']=[x,y] #NameError:'x' is not defined
a=[f(x,y) for x,y in lst if x>0 and y>0]


#After Handling
x,y=[(x,y) for x,y in lst if x>0 and y>0][0]
paraValueDict['f(x,y)']=[x,y] #Correct Instrumentation
a=[f(x,y) for x,y in lst if x>0 and y>0]
\end{lstlisting}
\end{minipage}
\vspace{-3mm}
\end{figure}

\textit{(5) Expanding if-else Statements.} Another special expression worth mentioning is the \mintinline{python}{if-else} statement (
Listing~\ref{if-else}). 
Direct instrumentation before the \mintinline{python}{if-else} statement could lead to \mintinline{python}{a(x)} receiving a parameter value less than 0, thereby causing a runtime error potentially. Therefore, \tool expands this conditional expression by modifying the \mintinline{python}{ast.IfExp} to the \mintinline{python}{ast.If} type node on the AST.

\begin{figure}[!t]
\centering
\begin{minipage}{3.4in}
\begin{lstlisting}[language=Python, label=if-else, caption=Expanding \mintinline{python}{if}-\mintinline{python}{else} statements within \mintinline{python}{return} statement for instrumentation.]
#Before Expanding
def foo():
  paraValueDict['a(x)']=[x]
  paraValueDict['@a(x).b(y, z)']=a(x) #Wrong Instrumentation
  paraValueDict['a(x).b(y, z)']=[y, z]
  return a(x).b(y, z) if x>0 else x


#After Expanding
def foo():
  if x>0:
    paraValueDict['a(x)']=x
    paraValueDict['@a(x).b(y,z)']=a(x) #Correct Instrumentation
    paraValueDict['a(x).b(y,z)']=[y, z]
    return a(x).b(y, z)
  else:
    return x
\end{lstlisting}
\end{minipage}
\vspace{-3mm}
\end{figure}

\begin{figure}[!t]
\centering
\begin{minipage}{3.4in}
\begin{lstlisting}[language=Python, label=decorator, caption=Instrumentation for API calls in consecutive decorator statements.]
paraValueDict['foo1(param1)']=[param1]
paraValueDict['foo2(param2)']=[param2] #Correct Instrumentation
@foo1(param1)
paraValueDict['foo2(param2)']=[param2] #Wrong Instrumentation
@foo2(param2)
def bar(x, y):
    return x + y
\end{lstlisting}
\end{minipage}
\vspace{-3mm}
\end{figure}

\textit{(6) Handling Consecutive Decorator Invocations.} 
When consecutive decorator API calls are present, the instrumentation stubs must not be inserted between decorators; otherwise, runtime errors may occur. For example, as shown in Listing~\ref{decorator}, if an instrumentation statement is placed between two consecutive decorators (\mintinline{python}{@foo1(param1)} and \mintinline{python}{@foo2(param2)}), the second decorator will no longer be correctly applied to the function definition, leading to execution failure. To address this, \tool collects all decorator invocations in the order they appear and arranges the corresponding instrumentation statements consecutively, placing them together before the decorated function. 

Note that when multiple API calls with the same name but different parameters appear in a code file, \tool treats each call as a separate API. During code instrumentation, \tool combines the API name and parameters to form a complete invocation statement string, which serves as the key in the instrumentation dictionary. For example, the calls \mintinline{python}{foo(1,2,3)} and \mintinline{python}{foo(4,5,6)} result in instrumentation statements: \mintinline{python}{paraValueDict["foo(1,2,3)"] = [1,2,3]} and \mintinline{python}{paraValueDict["foo(4,5,6)"]} \mintinline{python}{= [4,5,6]}, respectively. Besides, to record the file name, invocation location, and coverage information for each API call in more detail, \tool inserts coverage-collection statements in the code. For instance: \mintinline{python}{APICoveredSet}\mintinline{python}{.add(}
\mintinline{python}{f"{fileName}_}
\mintinline{python}{{lineno}_{callString}"}
\mintinline{python}{)}.

\subsection{API Mapping Establishment}\label{sec:pcart-mapping}
\textit{(1) Dynamic Mapping.} To automatically establish API signature mappings (i.e., $API_{old} \rightarrow API_{new}$), \tool first performs dynamic mapping, 
leveraging Python's dynamic reflection 
to obtain the signatures (parameter definitions) of the invoked APIs. Specifically, \tool uses Python's built-in inspect module, 
which is part of the Python standard library. The inspect module can inspect, analyze, and collect information about Python objects (e.g., modules, classes, functions, and methods) during runtime~\cite{inspect}.

Dynamic mapping enables \tool to directly observe the runtime bindings of APIs, thereby eliminating ambiguities caused by same-name APIs and import aliases. This approach is consistent with Python's execution model, in which a program is launched from a specific entry command. By changing the entry command in \tool's input configuration (Fig.~\ref{config}), developers can explore alternative execution paths. As long as the executed command covers the relevant branches, \tool can precisely detect and repair API parameter compatibility issues along those paths without requiring exhaustive static analysis of the entire codebase. In this way, dynamic mapping ensures that all API calls used during execution are accurately mapped to their true runtime definitions, substantially reducing the risk of false positives and false negatives arising from purely static analysis.

For each invoked API, \tool first generates a Python script under the project's directory structure, as shown in Listing~\ref{dynamic map}. The script imports all necessary modules required for loading the pickle files (created in \ding{183}), including user-defined modules and all third-party library modules. For example, if the project is run with the command \mintinline{shell}{python run.py}, the \mintinline{python}{import} statement in the generated script is \mintinline{python}{from run import *}. This step is crucial because if the project source files contain instances of functions or classes from specific modules, loading the pickle files depends on their definitions. 
Otherwise, Python runtime will not be able to restore these instances for the invoked APIs. 
\tool then executes this script within the project's virtual environment (current and target versions), successfully loading the previously saved contextual dependency (e.g., parameter values) from the pickle files into memory. 

After loading the pickle files, \tool uses Python's built-in inspect module to dynamically obtain the signatures of the invoked APIs. 
By loading the \mintinline{python}{a(x).b(y, z).pkl} file into memory, \tool obtains the value of \mintinline{python}{a(x)}, and 
then retrieves 
the signature of API \mintinline{python}{b} in the current library version through the \mintinline{python}{inspect.signature} function. Similarly, to obtain the API signature 
for the target version, 
\tool performs the script under the virtual environment of the new library version. Note that to obtain the signature of \mintinline{python}{b}, only the preceding contextual dependency (i.e., the value of \mintinline{python}{a(x)}) is necessary for the inspection.

\begin{figure}[!t]
\centering
\begin{minipage}{3.4in}
\begin{lstlisting}[language=Python, label=dynamic map, caption=Dynamic mapping of API signatures.]
import dill
import inspect
from run import *

with open('a(x).b(y, z).pkl','rb') as fr:
    paraValueDict=dill.load(fr)

para_def=inspect.signature(paraValueDict['@a(x).b(y, z)'].b)
print(para_def) #(x=1, y=2)
\end{lstlisting}
\end{minipage}
\vspace{-3mm}
\end{figure}

It is noted that there exists some 
API calls for which are unable to dynamically obtain their signatures.  
The primary reasons are as follows. First, when encountering built-in APIs, such as those in PyTorch and NumPy that are compiled through C/C++ extensions, it becomes impossible to use the inspect module to dynamically acquire API signatures. For example, executing \mintinline{shell}{inspect.signature(torch.nn.functional.avg_pool2d)} would raise \textit{ValueError: no signature found for builtin \textless built-in function avg\_pool2d\textgreater.}

Second, when the module that a pickle file depends on has been changed in the target version, the pickle files generated in the old library version are unable to load in the target version. For example, in Matplotlib 3.6.3, loading a class object instantiated by \mintinline{python}{matplotlib.pyplot.colorbar} in version 3.7.0 will result in an \textit{ModuleNotFoundError: No module named 'matplotlib.axes.\_subplots'}. 
In this case, \tool will attempt to regenerate the pickle files under the virtual environment of the target version (\ding{183}). However, if the invoked APIs also have compatibility issues in the target version, it leads to the inability to regenerate pickle files.

\textit{(2) Static Mapping.} When dynamic mapping fails, \tool resorts to a static mapping method, by matching the API's call path in the project with its actual path in the library source code to obtain its signature. 
The rationale for choosing the library source code over the API documentation is that the API definitions in the documentation are largely unstructured, incomplete, and outdated to a certain extent. There is significant variability in the level of detail and format regarding the recorded API 
across different library API documents. This makes it hard to implement an automated approach with high generalizability. Thus, \tool uses the library source code for the static mapping of API signatures. Details of the static mapping are provided as follows.

\textit{(2.1) Extracting APIs Defined in Libraries}.  
First, similar to the processing of project source files, \tool parses each library source file into an AST. Through traversing the AST, all \mintinline{python}{AsyncFunctionDef}, \mintinline{python}{FunctionDef}, and \mintinline{python}{ClassDef} nodes are identified, corresponding to the definition statements of functions and classes in the code, respectively.

\begin{figure}[!t]
\centering
\includegraphics[width=3in]{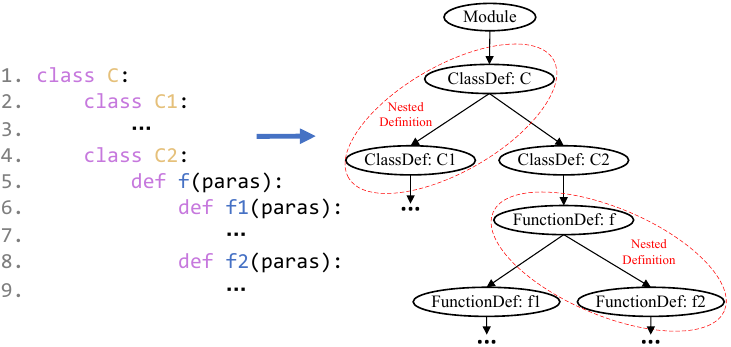} 
\vspace{-3mm}
\caption{AST structure of the nested API definitions in library source code.}
\label{nested}
\vspace{-4mm}
\end{figure}

One complicated form of API definitions is the 
nested definitions, i.e., 
classes defined within classes and functions defined within functions, as illustrated in Fig.~\ref{nested}. It is imperative to accurately discern the hierarchical relationships between classes and the affiliations among APIs to correctly construct the path of each API within the source code. Thus, 
the DFS algorithm is employed for navigation.

\begin{figure}[!t]
\centering
\begin{minipage}{3.4in}
\begin{lstlisting}[language=Python, label=overloads, caption=Examples of PyTorch buitin-in API definitions in \mintinline{python}{.pyi} file.]
class _TensorBase(object):
    @overload
    def max(self, dim: _int, keepdim: _bool=False) -> namedtuple_values_indices: ...
    
    @overload
    def max(self, dim: Union[str, ellipsis, None], keepdim: _bool=False) -> namedtuple_values_indices: ...
    
    @overload
    def max(self, other: Tensor) -> Tensor: ...
    
    @overload
    def max(self) -> Tensor: ...
\end{lstlisting}
\end{minipage}
\vspace{-5mm}
\end{figure}

In addition, regarding the built-in APIs (described in Section~\ref{sec:background-challenges}), 
developers usually declare their definitions in \mintinline{python}{.pyi} files. Listing~\ref{overloads} shows the 
declarations of the PyTorch built-in API \mintinline{python}{max} in the \mintinline{python}{torch/__init__.pyi} file (version 1.5.0). 
Hence, \tool attempts to parse \mintinline{python}{.pyi} files to acquire the definitions of built-in APIs' overloads.

Finally, by considering the class to which an API belongs, the module that contains the class, and the package that encompasses the module, \tool constructs the actual path of each API within the source code.

\textit{(2.2) Adjusting Library API Paths}. 
The actual path of library APIs in the source code often differs from the path provided to users. 
This inconsistency arises because developers leverage the \mintinline{python}{import} mechanism and \mintinline{python}{__init__.py} files to shorten the API paths for user calls (Section~\ref{sec:background-challenges}).

To address this issue, \tool recursively traverses the library's source code directory structure, focusing on \mintinline{python}{__init__.py} files at each level. By analyzing the \mintinline{python}{import} statements in these files, it simplifies the fully qualified names of the APIs in the source code to match the invocation names in the library's official API documentation. The API fully qualified name simplification process is shown in Algorithm~\ref{alg:simplify}.

\SetArgSty{}
\begin{algorithm}[!t]
\label{alg:simplify}
\footnotesize
\SetAlgoLined
\SetAlgoLined
\SetKwIF{If}{ElseIf}{Else}{if}{then}{}{}{}
\SetKwFor{ForEach}{foreach}{}{}
\SetKw{in}{in}
\SetKw{or}{or}
\KwIn{fullQualifiedName, actualPath}
\KwOut{simplifiedName}

\SetKwProg{Fn}{Function}{:}{end} \Fn{\textnormal{Simplify()}}{
currName $\gets$ fullQualifiedName \;
currPath $\gets$ actualPath \;

\While{True}{
    pos $\gets$ currPath.rfind(\texttt{'}/\texttt{'}) \;
    \If{pos == -1}{
        \textbf{break}\;
    }
    parentPath $\gets$ currPath[0 : pos]\;
    initPath $\gets$ parentPath + \texttt{"}/\_\_init\_\_.py\texttt{"}
    
    \If{isExist(initPath)}{
        root $\gets$ getAst(initPath) \;        
        importer $\gets$ ImportFrom() \;
        importDict $\gets$ importer.visit(root) \;
        repK $\gets$ \texttt{"}\texttt{"}\;
        repV $\gets$ \texttt{"}\texttt{"}\;
        
        \ForEach{key, value \in importDict}{
            \If {key.endwith(\texttt{'*'})}{
                key $\gets$ key.rstrip(\texttt{'*'}) \;
                \If{key \in currName}{
                    repK $\gets$ key \;
                    repV $\gets$ \texttt{"}\texttt{"}\;
                }
            }
            \ElseIf{\textbf{else if} key \in currName}{
                repK $\gets$ key\;
                repV $\gets$ value\;
                \textbf{break} \;
            }
        }
        \If{repK $\neq$ \texttt{"}\texttt{"}}{
            currName $\gets$ currName.replace(repK, repV)\;
        }
    }
    
    currPath $\gets$ parentPath
    
}

\Return{simplifiedName $\gets$ curreName}\;
}

\caption{Simplifying fully qualified API names.}
\end{algorithm}

The algorithm takes the fully qualified name of the API in the source code and its actual path as input, and outputs the simplified name by processing \mintinline{python}{__init__.py} files and following the \mintinline{python}{import} mechanism. Lines 5-8 check if the rightmost part of the current path contains a `/' to determine if higher-level directories exist; if not, the loop terminates. Lines 9-10 check for the presence of an \mintinline{python}{__init__.py} file in the current directory. Lines 11-13 use the \mintinline{python}{getAst} function to retrieve the AST root node of the \mintinline{python}{__init__.py} file. The \mintinline{python}{visit} function of the \mintinline{python}{ImportFrom} class then traverses the \mintinline{python}{ImportFrom} nodes in the AST, generating a dictionary that stores the \mintinline{python}{import} information. Lines 16-27 iterate through the key-value pairs in this dictionary to simplify the fully qualified API names.

For example, as illustrated in Fig.~\ref{import and init}, the import statement \mintinline{python}{from .fromnumeric import amax as max} at the \mintinline{python}{numpy/core/__init__.py} level generates a key-value pair, where the key is \mintinline{python}{fromnumeric.amax} and the value is \mintinline{python}{max}. The algorithm checks if the current API name contains the key. If it does, the key is replaced with the value, substituting \mintinline{python}{fromnumeric.amax} with \mintinline{python}{max} to obtain \mintinline{python}{numpy.core.max}. This is the first replacement. The process continues to the upper-level directory. At the \mintinline{python}{numpy/__init__.py} level, the statement \mintinline{python}{from .core import *} generates a key-value pair where the key is \mintinline{python}{core.*} and the value is an empty string. Using the same substitution method, \mintinline{python}{numpy.core.max} is further simplified to \mintinline{python}{numpy.max}. By recursively processing multi-level directories and the \mintinline{python}{import} rules in \mintinline{python}{__init__.py}, the algorithm derives the final simplified API name.

Moreover, 
although \tool has attempted to adjust the actual paths of APIs to their calling paths, static mapping may still generate multiple candidates. To resolve this uncertainty, \tool initially saves all candidates in \ding{184}. Subsequently, based on the results of the API compatibility assessment (\ding{185}), \tool eliminates the compatible candidates in the repair and validation phase (\ding{186}). 

\begin{figure*}[!t]
\centerline{\includegraphics[width=\linewidth]{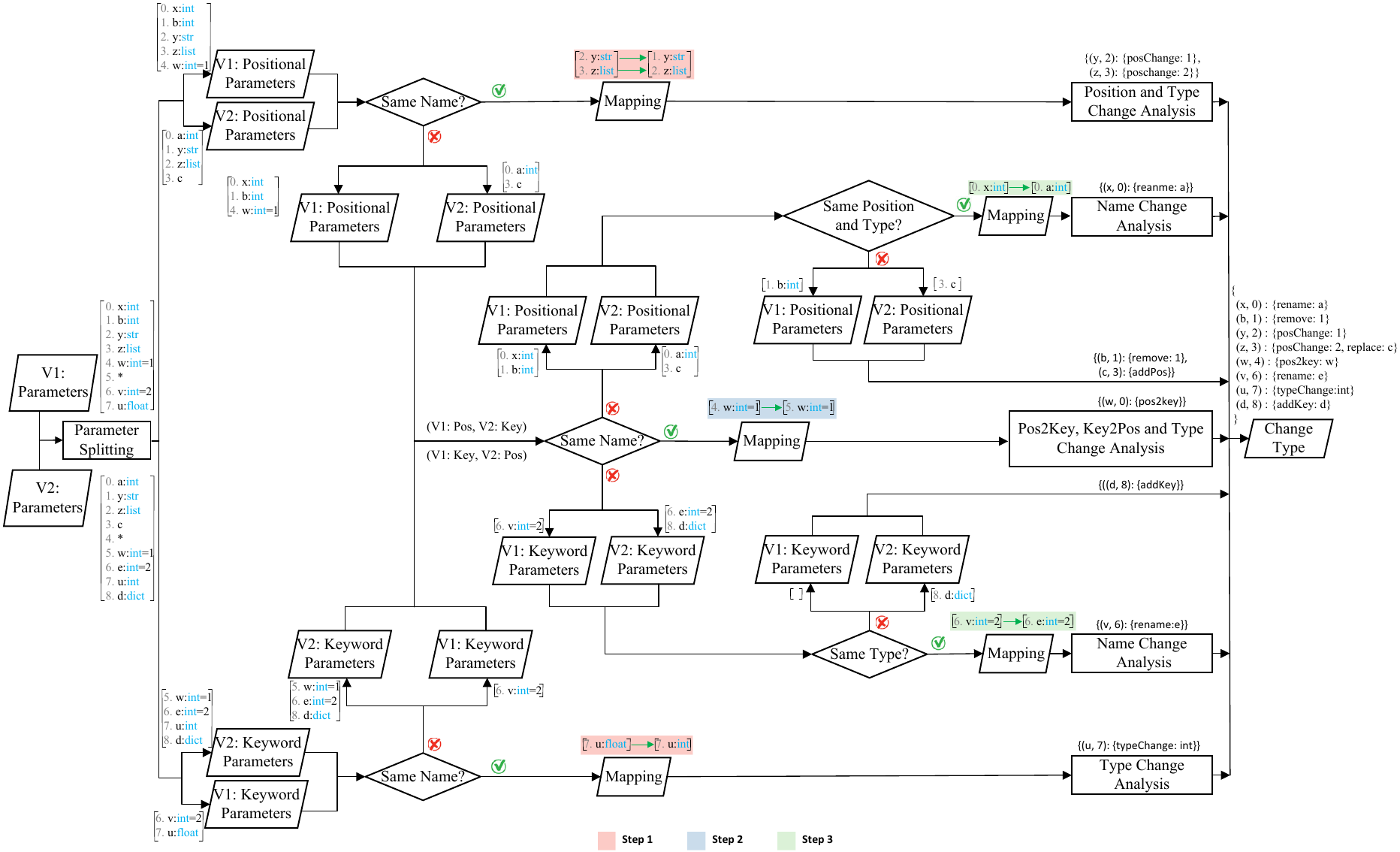}}
\vspace{-3mm}
\caption{An example of parameter mapping establishment and change analysis in the API \mintinline{python}{foo} across V1 and V2 versions.}
\label{change analysis}
\vspace{-3mm}
\end{figure*}

\subsection{Compatibility Assessment}\label{sec:pcart-assessment}
The compatibility of invoked APIs is impacted not only by the changes at the level of API parameters 
but also by the actual methods in which users pass the parameters. Therefore, \tool first analyzes the change of APIs at the parameter level. Subsequently, by integrating how parameters are passed during actual API calls, \tool precisely assesses whether an invoked API is compatible. 

\textit{(1) Analyzing API Parameter Change Types.} 
\tool begins by distinguishing between positional and keyword parameters in the API definitions using the identifier ``\mintinline{python}{*}''. Then, 
\tool establishes mappings between parameters across two library versions (i.e., the current and the target) based on attributes such as parameter name, position, and type. This process is divided into three steps. In the following, we use an example to present the procedures in detail, as depicted in Fig.~\ref{change analysis}.

\textbf{Step 1.} \tool prioritizes establishing the mapping relationship between parameters based on the consistency of parameter names. 
For positional parameters, type changes and positional changes are analyzed. For the example illustrated in Fig.~\ref{change analysis}, the positions of positional parameters \mintinline{python}{y} and \mintinline{python}{z} are changed in version V2. For keyword parameters, since their usage does not depend on position, only type changes need to be analyzed. For instance, the type of keyword parameter \mintinline{python}{u} in version V1 changes from \mintinline{python}{float} to \mintinline{python}{int} in version V2. After each round of analysis, mapped parameters are removed from the parameter list to avoid interference with subsequent mapping relationships.

\textbf{Step 2.} To determine whether there are changes from positional parameters to keyword parameters (i.e., Pos2Key) or keyword parameters to positional parameters (i.e., Key2Pos), \tool also uses parameter name consistency to establish the mapping relationship between positional parameters and keyword parameters. For example, the positional parameter \mintinline{python}{w} in version V1 becomes a keyword parameter in version V2.

\textbf{Step 3.} For the remaining positional parameters with undetermined mappings, \tool establishes mapping relationships by considering the consistency of both position and type. For each positional parameter in version V1, if a parameter with the same position and type can be found in version V2, they are considered corresponding. At this point, the renaming analysis can be conducted. For example, the positional parameter \mintinline{python}{x} in version V1 is renamed to \mintinline{python}{a} in version V2. For the remaining keyword parameters with undetermined mappings, \tool establishes mapping relationships based on type consistency. For each keyword parameter in version V1, if a parameter with the same type can be found in version V2, they are considered corresponding. For example, the keyword parameter \mintinline{python}{v} in version V1 is renamed to \mintinline{python}{e} in version V2.

After these steps, any parameter in the version V1 parameter list that still does not have a mapping is considered removed in version V2. For instance, the positional parameter \mintinline{python}{b} is removed in version V2. On the contrary, parameters in version V2 that remain unmapped are considered newly added parameters, such as the positional parameter \mintinline{python}{c} and the keyword parameter \mintinline{python}{d} in version V2.

\tool saves the changes to all parameters in a dictionary structure. The keys of the dictionary are variables of tuple type, storing the parameter name and its position. Since the position can be arbitrary when parameters are passed by keywords, the dictionary must be accessed using the parameter name. The values 
record the change types made to the parameters from versions V1 to V2 with their related changes.


\textit{(2) Analyzing Parameter Passing Methods.} In the practical usage of Python APIs, there are three typical methods of parameter passing: positional passing, keyword passing, and no passing (applicable to parameters with default values). Positional parameters can be passed either through their position or by specifying their names, while keyword parameters must be passed by specifying their names. If a positional or keyword parameter is assigned a default value, it becomes optional to pass when calling the API. \tool uses AST analysis to examine the \mintinline{python}{ast.args} (positional passing) and \mintinline{python}{ast.keywords} (keyword passing) nodes of the called API to determine how each parameter is passed during actual use. For parameters not appearing in these two nodes, they are considered not passed.

\begin{table}[!t]
\centering
\caption{Formulation of API Parameter Compatibility}
\label{compatibility}
\scalebox{0.65}{
\begin{tabular}{|l|c|c|}
\hline
Parameter Type             & $f(P,E,M)$                           & Compatibility \\ \hline
\multirow{15}{*}{Positional Parameter} & $p\wedge \Delta_d \wedge \uparrow_n$ & Compatible    \\ \cline{2-3} 
                           & $p\wedge \Delta_d \wedge \uparrow_p$ & Incompatible  \\ \cline{2-3} 
                           & $p\wedge \Delta_d \wedge \uparrow_k$ & Incompatible  \\ \cline{2-3}
                           & $p\wedge \Delta_o \wedge \uparrow_n$ & Compatible    \\ \cline{2-3} 
                           & $p\wedge \Delta_o \wedge \uparrow_p$ & Incompatible  \\ \cline{2-3} 
                           & $p\wedge \Delta_o \wedge \uparrow_k$ & Compatible    \\ \cline{2-3} 
                           & $p\wedge \Delta_r \wedge \uparrow_n$ & Compatible    \\ \cline{2-3} 
                           & $p\wedge \Delta_r \wedge \uparrow_p$ & Compatible    \\ \cline{2-3} 
                           & $p\wedge \Delta_r \wedge \uparrow_k$ & Incompatible  \\ \cline{2-3} 
                           & $p\wedge \Delta_k \wedge \uparrow_n$ & Compatible    \\ \cline{2-3} 
                           & $p\wedge \Delta_k \wedge \uparrow_p$ & Incompatible  \\ \cline{2-3} 
                           & $p\wedge \Delta_k \wedge \uparrow_k$ & Compatible    \\ \cline{2-3}
                           & $p\wedge \Delta_t \wedge \uparrow_n$ & Compatible    \\ \cline{2-3} 
                           & $p\wedge \Delta_t \wedge \uparrow_p$ & Incompatible  \\ \cline{2-3} 
                           & $p\wedge \Delta_t \wedge \uparrow_k$ & Incompatible  \\ \cline{2-3} 
                           & $p\wedge \Delta_u \wedge \uparrow_n$ & Incompatible  \\ \cline{2-3}
                           & $p\wedge \Delta_v \wedge \uparrow_n$ & Compatible    \\ \hline
\multirow{9}{*}{Keyword Parameter}   & $k\wedge \Delta_d \wedge \uparrow_n$ & Compatible    \\ \cline{2-3} 
                           & $k\wedge \Delta_d \wedge \uparrow_k$ & Incompatible  \\ \cline{2-3} 
                           & $k\wedge \Delta_r \wedge \uparrow_n$ & Compatible    \\ \cline{2-3} 
                           & $k\wedge \Delta_r \wedge \uparrow_k$ & Incompatible  \\ \cline{2-3}
                           & $k\wedge \Delta_p \wedge \uparrow_n$ & Compatible    \\ \cline{2-3}  
                           & $k\wedge \Delta_p \wedge \uparrow_k$ & Compatible    \\ \cline{2-3} 
                           & $k\wedge \Delta_t \wedge \uparrow_n$ & Compatible    \\ \cline{2-3} 
                           & $k\wedge \Delta_t \wedge \uparrow_k$ & Incompatible  \\ \cline{2-3}
                           & $k\wedge \Delta_u \wedge \uparrow_n$ & Incompatible    \\ \cline{2-3}
                           & $k\wedge \Delta_v \wedge \uparrow_n$ & Compatible    \\ \hline
\end{tabular}
}
\vspace{-3mm}
\end{table}

\textit{(3) Formulating Compatibility.} 
In \tool, we propose a model for assessing parameter compatibility based on parameter types, change types, and passing methods. The model comprehensively characterizes the compatibility of API parameters using the formula:

\begin{equation}
\centering
f(P, E, M) = P \wedge E \wedge M.
\end{equation}

Here, $P$ represents the parameter type, with a domain of 
$\{p, k\}$, where $p$ refers to the positional parameter and $k$ denotes the keyword parameter. 
$E$ denotes the parameter change type, with values from $\{\Delta_d, \Delta_r, \Delta_o, \Delta_p, \Delta_k, \Delta_t, \Delta_u, \Delta_v\}$: $\Delta_d$ indicates parameter removal, $\Delta_r$ indicates parameter renaming, $\Delta_o$ represents parameter reordering, $\Delta_p$ indicates the conversion of a keyword parameter to a positional parameter, $\Delta_k$ indicates the conversion of a positional parameter to a keyword parameter, $\Delta_t$ denotes an incompatible type change, $\Delta_u$ denotes the addition of a parameter without a default value, and $\Delta_v$ denotes the addition of a parameter with a default value. $M$ represents the parameter passing method, with values from $\{\uparrow_n, \uparrow_p, \uparrow_k\}$, where $\uparrow_n$ means the parameter is not passed, $\uparrow_p$ means it is passed positionally, and $\uparrow_k$ stands for keyword passing.

By exhaustively enumerating all valid $(P, E, M)$ combinations under Python's syntactic constraints, the model guarantees complete coverage of compatibility scenarios, as shown in Table \ref{compatibility}. 
Notably, since newly added parameters are absent in the API calls of the previous version, their passing method is limited to $\uparrow_n$. 
The model's correctness is ensured by its alignment with Python's parameter binding mechanisms. For positional parameters, the rules directly map to constraints on argument count, order, and type. For keyword parameters, the rules enforce name validity and default value handling.

\begin{figure}[!t]
\centering
\begin{minipage}{3.4in}
\begin{lstlisting}[escapeinside={(*@}{@*)}, language=Python, label=lst:positional-examples, caption=Representative examples of positional parameter changes.]
#1. Removal (p (*@$\wedge$@*) (*@$\Delta$@*)d)
Lib: Polars
API: polars.read_json
0.15.18: (file: 'str | Path | IOBase', json_lines: 'bool | None' = None) -> 'DataFrame'
0.16.1: (file: 'str | Path | IOBase') -> 'DataFrame'

#2. Reordering (p (*@$\wedge$@*) (*@$\Delta$@*)o)
Lib: Requests
API: requests.cookies.RequestsCookieJar.get
0.12.1: (self, name, domain=None, path=None, default=None)
0.13.0: (self, name, default=None, domain=None, path=None)

#3. Renaming (p (*@$\wedge$@*) (*@$\Delta$@*)r)
Lib: SymPy
API: sympy.GramSchmidt
0.7.6.1: (vlist, orthog=False)
1.0: (vlist, orthonormal=False)

#4. Conversion to keyword-only (p (*@$\wedge$@*) (*@$\Delta$@*)k)
Lib: scikit-learn
API: sklearn.cluster.SpectralCoclustering
0.22.2: (n_clusters=3, svd_method='randomized', n_svd_vecs=None, mini_batch=False, init='k-means++', n_init=10, n_jobs=None, random_state=None)
0.23.0: (n_clusters=3, *, svd_method='randomized', n_svd_vecs=None, mini_batch=False, init='k-means++', n_init=10, n_jobs='deprecated', random_state=None)

#5. Incompatible type change (p (*@$\wedge$@*) (*@$\Delta$@*)t)
Lib: scikit-learn
API: sklearn.linear_model.LogisticRegression.fit
0.24.2: (X, y, sample_weight=None) #X supports np.matrix
1.2.0: (X, y, sample_weight=None)  #np.matrix disallowed

#6. Adding required parameter (p (*@$\wedge$@*) (*@$\Delta$@*)u)
Lib: PyTorch Lightning
API: pytorch_lightning.callbacks.Callback.on_validation_batch_end
0.8.5: (trainer, pl_module)
0.9.0: (trainer, pl_module, batch, batch_idx, dataloader_idx)

#7. Adding optional parameter (p (*@$\wedge$@*) (*@$\Delta$@*)v)
Lib: NetworkX
API: networkx.laplacian
1.4: (G,nodelist=None)
1.5: (G,nodelist=None,weight='weight')
\end{lstlisting}
\end{minipage}
\vspace{-3mm}
\end{figure}

\textbf{Positional Parameters.} When a positional parameter is removed ($p \wedge \Delta_d$), both positional ($\uparrow_p$) and keyword passing ($\uparrow_k$) become invalid, since the call either exceeds the reduced argument count or references a nonexistent parameter name. Only not passing ($\uparrow_n$) remains compatible. As the example illustrated in Listing~\ref{lst:positional-examples}, in the \mintinline{python}{polars.read_json} API, the parameter \mintinline{python}{json_lines} was removed in version 0.16.1, causing calls like \mintinline{python}{pl.read_json("./output.json", None)} or \mintinline{python}{pl.read_json("./output.json", json_lines=None)} to fail, while \mintinline{python}{pl.read_json("} \mintinline{python}{./output.json")} continues to run successfully.

Reordering of positional parameters ($p \wedge \Delta_o$) disrupts argument binding only when parameters are passed positionally ($\uparrow_p$). For example, in Requests \mintinline{python}{RequestsCookieJar.get} API, the \mintinline{python}{domain} argument was moved between versions 0.12.1 and 0.13.0 (Listing~\ref{lst:positional-examples}). A call like \mintinline{python}{cookie_jar.get("cookie1", "example.com")} now misbinds its arguments, whereas keyword passing ($\uparrow_k$) such as \mintinline{python}{cookie_jar.get("cookie1", domain="example.com")} and not passing ($\uparrow_n$) such as \mintinline{python}{pl.read_json('./output.json')} remain valid.

Renaming of positional parameters ($p \wedge \Delta_r$) invalidates keyword passing ($\uparrow_k$) but leaves positional passing ($\uparrow_p$) and not passing ($\uparrow_n$) unaffected. In Sympy's \mintinline{python}{GramSchmidt} API, the keyword \mintinline{python}{orthog} was renamed to \mintinline{python}{orthonormal} in version 1.0 (Listing~\ref{lst:positional-examples}). Consequently, keyword calls like \mintinline{python}{sp.GramSchmidt(independent_vectors, orthog=True)} fail, but calls like \mintinline{python}{sp.GramSchmidt(independent_vectors,} \mintinline{python}{True)} and \mintinline{python}{sp.GramSchmidt(independent_vectors)} work.   

A different situation arises when a positional parameter is converted into a keyword-only parameter ($p \wedge \Delta_k$). Here, positional passing ($\uparrow_p$) becomes invalid, while keyword passing ($\uparrow_k$) and not passing ($\uparrow_n$) remain compatible. This is exemplified by scikit-learn's API, e.g.,  \mintinline{python}{SpectralCoclustering}, where \mintinline{python}{svd_method} became keyword-only in version 0.23.0 (Listing~\ref{lst:positional-examples}). As a result, \mintinline{python}{SpectralCoclustering(2, "randomized")} fails, while \mintinline{python}{SpectralCo} \mintinline{python}{clustering(2, svd_method='randomized')} and \mintinline{python}{SpectralCoclustering(2)} work.

Incompatible type changes ($p \wedge \Delta_t$) affect both positional ($\uparrow_p$) and keyword passing ($\uparrow_k$). A representative case occurs in scikit-learn's \mintinline{python}{LogisticRegression.fit} (Listing~\ref{lst:positional-examples}), where the input \mintinline{python}{X} deprecated \mintinline{python}{np.matrix} in version 1.0 and raises a \textit{TypeError} in version 1.2 if the old type is used.    

Adding new positional parameters without defaults ($p \wedge \Delta_u$) is also problematic: not passing ($\uparrow_n$) is no longer valid, rendering prior calls incompatible. For example, as depicted in Listing~\ref{lst:positional-examples}, \mintinline{python}{on_validation_batch_end} in PyTorch Lightning expanded three required positional parameters (i.e., \mintinline{python}{batch}, \mintinline{python}{batch_idx}, and \mintinline{python}{dataloader_idx}) in version 0.9.0, making older callbacks invalid. By contrast, new positional parameters with defaults ($p \wedge \Delta_v$) are safe. In NetworkX's \mintinline{python}{laplacian}, the parameter \mintinline{python}{weight} was added in version 1.5 with a default value, so calls like \mintinline{python}{nx.laplacian(G)} continue to work.

\textbf{Keyword Parameters.} Removal ($k \wedge \Delta_d$) breaks keyword passed ($\uparrow_k$) calls but leaves not passed ($\uparrow_n$) calls  unaffected. For example, as shown in Listing~\ref{lst:keyword-examples}, in Rich's \mintinline{python}{Layout}, the keyword \mintinline{python}{height} was dropped in version 12.6.0, so \mintinline{python}{Layout(height=None)} fails while \mintinline{python}{Layout()} remains valid.

Renaming ($k \wedge \Delta_r$) invalidates keyword passing ($\uparrow_k$) but remains compatible for not passing ($\uparrow_n$). For example, in Loguru's \mintinline{python}{logger.configure}, where \mintinline{python}{patch} was renamed to \mintinline{python}{patcher} between versions 0.3.2 and 0.4.0 (Listing~\ref{lst:keyword-examples}), breaking calls like \mintinline{python}{logger.configure(patch=None)}.

When a keyword parameter is converted to positional ($k \wedge \Delta_p$), compatibility is preserved since its name remains valid. For example, in Bleak's \mintinline{python}{BleakClient}, the \mintinline{python}{disconnected_callback} argument was previously absorbed by \mintinline{python}{**kwargs} but became a positional parameter in version 0.18.0 (Listing~\ref{lst:keyword-examples}), and calls such as \mintinline{python}{BleakClient(addr, disconnected_callback=None)} ($\uparrow_k$) and \mintinline{python}{BleakClient(addr)} ($\uparrow_n$) still work.

Although our taxonomy includes incompatible type changes for keyword parameters ($k \wedge \Delta_t$), where passing old-type values would raise a \textit{TypeError}, our empirical study did not reveal real-world instances of this pattern. Nevertheless, this rule is theoretically valid, as changing a parameter's accepted type (e.g., from \mintinline{python}{str} to \mintinline{python}{int}) would inevitably break legacy calls that pass the old type.

Finally, the addition of new required keyword-only parameters ($k \wedge \Delta_u$) is incompatible, as not passing ($\uparrow_n$) produces a runtime error. This occurred in Discord.py 2.0 (Listing~\ref{lst:keyword-examples}), where \mintinline{python}{discord.Client} introduced a mandatory \mintinline{python}{intents} parameter, making \mintinline{python}{discord.Client()} invalid. By contrast, adding optional keyword-only parameters ($k \wedge \Delta_v$) is safe. NumPy's \mintinline{python}{stack}, for example, added \mintinline{python}{dtype} and \mintinline{python}{casting} in version 1.24.0, but calls such as \mintinline{python}{np.stack((a, b, c))} continue to execute correctly.

\begin{figure}[!t]
\centering
\begin{minipage}{3.4in}
\begin{lstlisting}[escapeinside={(*@}{@*)},language=Python, label=lst:keyword-examples, caption=Representative examples of keyword parameter changes.]
#1. Removal (k (*@$\wedge$@*) (*@$\Delta$@*)d)
Lib: Rich
API: rich.Layout
12.5.1: (self,renderable:Optional[RenderableType]=None,*,name:Optional[str]=None,size:Optional[int]=None,minimum_size:int=1,ratio:int=1,visible:bool=True,height:Optional[int]=None)
12.6.0: (self,renderable:Optional[RenderableType]=None,*,name:Optional[str]=None,size:Optional[int]=None,minimum_size:int=1,ratio:int=1,visible:bool=True)

#2. Renaming (k (*@$\wedge$@*) (*@$\Delta$@*)r)
Lib: Loguru
API: loguru.logger.configure
0.3.2: (*, handlers=None, levels=None, extra=None, patch=None, activation=None)
0.4.0: (*, handlers=None, levels=None, extra=None, patcher=None, activation=None)

#3. Conversion to positional (k (*@$\wedge$@*) (*@$\Delta$@*)p)
Lib: Bleak
API: bleak.BleakClient
0.17.0: (address_or_ble_device: Union[bleak.backends.device.BLEDevice, str], **kwargs)
0.18.0: (device_or_address: 'Union[BLEDevice, str]', disconnected_callback: 'Optional[Callable[[BleakClient], None]]' = None, *, timeout: 'float' = 10.0, winrt: 'WinRTClientArgs' = {}, backend: 'Optional[Type[BaseBleakClient]]' = None, **kwargs)

#4. Adding required parameter (k (*@$\wedge$@*) (*@$\Delta$@*)u)
Lib: Discord.py
API: discord.Client
1.7.3: (*, loop=None, **options)
2.0.0: (*, intents: 'Intents', **options: 'Unpack[_ClientOptions]') -> 'None'

#5. Adding optional parameter (k (*@$\wedge$@*) (*@$\Delta$@*)v)
Lib: NumPy
API: numpy.stack
1.23.5: (arrays, axis=0, out=None)
1.24.0: (arrays, axis=0, out=None, *, dtype=None, casting='same_kind')
\end{lstlisting}
\end{minipage}
\vspace{-3mm}
\end{figure}



Thus, the overall compatibility of an API invocation is determined by the conjunction of all parameter-level results:

\begin{equation}
\centering
C_{\text{InvokedAPI}} = \bigwedge_{i=1}^{n} C_i = \bigwedge_{i=1}^{n} f(P_i, E_i, M_i),
\end{equation}

\noindent where $n$ is the total parameters in the call. This formulation ensures strict compatibility: a single incompatible parameter ($C_i=False$) invalidates the entire API invocation, mirroring Python's fail-fast semantics.

\subsection{Repair and Validation}\label{sec:pcart-repair}

\begin{figure}[!t]
\centering
\includegraphics[width=2.5in]{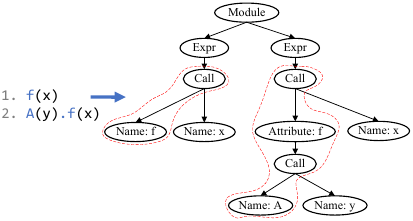} 
\vspace{-3mm}
\caption{
Comparison of AST structures between \mintinline{python}{f(x)} and \mintinline{python}{A(y).f(x)}.}
\label{locatedAPI}
\vspace{-3mm}
\end{figure}

\textit{(1) Repair.} 
\tool leverages 
the change dictionary generated during the compatibility assessment phase \ding{185}, along with the invocation and parameter passing methods of APIs in user project code, to fix the detected incompatible API calls. 
Currently, \tool supports the following repair types: parameter addition, removal, renaming, reordering, and the conversion of positional parameters to keyword parameters. The repair process involves three stages:

\textit{(1.1) Locating Incompatible API Invocations.} \tool first uses Python's built-in AST module to convert the source files of the user's project into an AST. By traversing the AST, \tool identifies all the APIs in the code that require fixing. As shown in Fig.~\ref{locatedAPI}, suppose the API needing repair is \mintinline{python}{A(y).f(x)}. However, since the BFS algorithm is insensitive to the sequence of API calls during its search process, it may mistakenly repair the wrong API when encountering another API with the same name, i.e., \mintinline{python}{f(x)}. Existing repair tools like MLCatchUp~\cite{haryono2021mlcatchup} cannot recognize this situation of API calls. \tool, on the other hand, employs the DFS algorithm, precisely resolving this problem.

\textit{(1.2) AST-based Repair.} 
\tool's repair process preserves the original invocation style whenever possible, while strictly adhering to Python's syntactic rules. Specifically, renamed positional parameters passed by position are considered \textit{Compatible} and left unchanged, while renamed parameters passed by keyword are treated as \textit{incompatible} and repaired by updating the parameter name. For reordered positional parameters, \tool adjusts their positions without adding keyword names and keeps keyword-passed parameters unchanged. 

\begin{figure}[!t]
\centering
\includegraphics[width=3.5in]{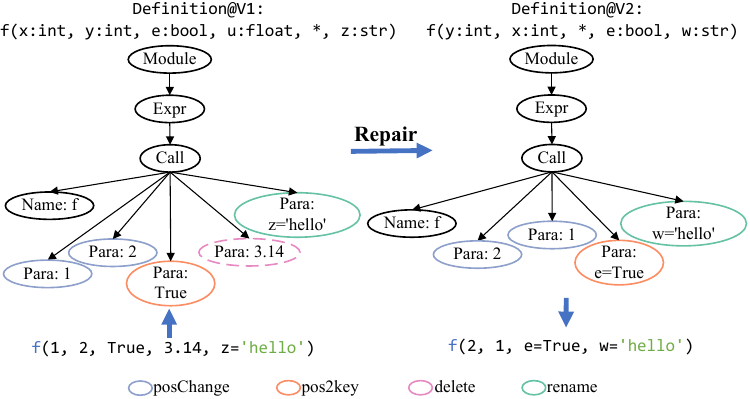} 
\vspace{-3mm}
\caption{Example of AST-based repair operations.}
\label{repair}
\vspace{-3mm}
\end{figure}

As illustrated in Fig.~\ref{repair}, \tool performs repairs at the AST level to address parameter compatibility issues between API versions. In this example, the parameter list changes from \mintinline{python}{f(x:int, y:int, e:bool, u:float, *, z:str)} in \textit{Definition@V1} to \mintinline{python}{f(y:int, x:int,} \mintinline{python}{*, e:bool, w:str)} in \textit{Definition@V2}. \tool detects and applies the following minimal-change repair operations: \textit{posChange} (reorders positional arguments so that each value remains bound to the same semantic parameter under the new signature but at its new index, e.g., \mintinline{python}{1} for \mintinline{python}{x} moves from position 1 to position 2, \mintinline{python}{2} for \mintinline{python}{y} moves from position 2 to position 1), \textit{pos2key} (converts positional to keyword when necessary, e.g., \mintinline{python}{True} $\rightarrow$ \mintinline{python}{e=True}), \textit{delete} (removes obsolete arguments, e.g., removing \mintinline{python}{3.14} after \mintinline{python}{u} is deleted), and \textit{rename} (updates keyword names, e.g., \mintinline{python}{z='hello'} $\rightarrow$ \mintinline{python}{w='hello'}). The final repaired call, \mintinline{python}{f(2, 1, e=True, w='hello')}, reflects the minimal necessary modifications to achieve compatibility with the updated API definition.

\SetArgSty{}
\begin{algorithm}[!t]
\label{alg:repair}
\footnotesize
\SetAlgoLined
\SetAlgoLined
\SetKwIF{If}{ElseIf}{Else}{if}{then}{}{}{}
\SetKwFor{ForEach}{foreach}{}{}
\SetKw{in}{in}
\SetKw{or}{or}
\KwIn{callAPI, repairDict, root}
\KwOut{root}

\caption{AST-based repair.}
\SetKwProg{Fn}{Function}{:}{end} \Fn{\textnormal{repair(callAPI, repairDict, node)}}{
    \ForEach{n \in ast.iter\_child\_nodes(node)}{
        repair(callAPI, repairDict, n)\;
    }

    \If{isFunctionCallNode(node)}{
        \If{ast.unparse(node) == callAPI}{
            posParaList $\gets$ node.args \;
            keyParaList $\gets$ node.keywords\;

            index $\gets$ 0\;
            newPosParaList $\gets$ [ ]\;
            newKeyParaList $\gets$ [ ]\;
            \ForEach{posParaNode \in posParaList}{
                opDict $\gets$ mapPos(index, repairDict)\;

                \ForEach{operation, patch \in opDict}{
                    
                    \If{operation == \texttt{"}delete\texttt{"}}{
                        \textbf{break}\;
                    }
                    \If{operation == \texttt{"}rename\texttt{"}}{
                        newPosParaList.insert(index, posParaNode)\;
                    }
                    \If{operation == \texttt{"}pos2key\texttt{"}}{
                        value $\gets$ ast.unparse(posParaNode)\;
                        newNode $\gets$ ast.keyword(arg=patch, value=value)\;
                        newKeyParaList.append(newNode)\;
                    }
                    \If{operation == \texttt{"}replace\texttt{"}}{
                        newNode $\gets$ ast.Name(id = patch)\;
                        newPosParaList.insert(index, newNode)\;
                    }
                    \If{operation == \texttt{"}posChange\texttt{"}}{
                        newPosParaList.insert(patch, posParaNode)\;
                    }
                }

                \If{isEmpty(opDict)}{
                    newPosParaList.insert(index, posParaNode)
                }
                index $\gets$ index + 1\;
            }

            node.args $\gets$ newPosParaList\;

            \ForEach{keyParaNode \in keyParaList}{
                opDict $\gets$ mapName(keyParaNode.name, repairDict)\;
                \ForEach{operation, patch \in opDict}{
                    \If{operation == \texttt{"}delete\texttt{"}}{
                        \textbf{break}\;
                    }
                    \If{operation == \texttt{"}rename\texttt{"}}{
                        keyParaNode.name $\gets$ patch\;
                        newKeyParaList.append(keyParaNode)\;
                    }
                    \If{operation == \texttt{"}posChange\texttt{"}|\texttt{"}pos2key\texttt{"}|\texttt{"}key2pos\texttt{"}}{
                        newKeyParaList.append(keyParaNode)\;
                    }
                }
                \If{isEmpty(opDict)}{
                    newKeyParaList.append(keyParaNode)
                }
            }

            node.keywords $\gets$ newKeyParaList\;
        }
    }
    \Return{node}\;
}
\end{algorithm}

As shown in Algorithm~\ref{alg:repair}, the AST-based repair takes the API call to be repaired, the change dictionary generated during compatibility assessment, and the root node of the AST for the user project code as input. The output is the repaired AST root node. 
In lines 2-3 of the algorithm, \tool recursively traverses the API call chain to the deepest level using the DFS algorithm, ensuring repairs proceed from the bottom up. In lines 4-5, it checks whether the current node is a \mintinline{python}{FunctionCallNode} to locate API call statements in the source file. If so, the node is parsed using \mintinline{python}{ast.unparse} and compared as a string with the target API (\mintinline{python}{callAPI}) to identify the specific API call to be repaired. In lines 6-10, \tool retrieves the positional and keyword parameter information from the node's \mintinline{python}{args} and \mintinline{python}{keywords} attributes, initializing two new parameter lists: \mintinline{python}{newPosParaList} for positional parameters and \mintinline{python}{newKeyParaList} for keyword parameters.

Lines 11-30 handle repairs for positional parameters by iterating over each positional parameter node in the original list and applying changes based on the parameter's index and the corresponding entry in the change dictionary. For parameters marked for removal (\mintinline{python}{delete}), the parameter is skipped and not added to the new list. If the operation is renaming (\mintinline{python}{rename}), since positional parameters are not used with explicit names, the renaming has no effect, and the original node is added to the new list at its original position. When the operation involves converting a positional parameter to a keyword parameter (\mintinline{python}{pos2key}), the patch specifies the keyword name, and \tool creates a new keyword parameter node, adding it to the keyword list. For replacement (\mintinline{python}{replace}), \tool inserts a new parameter at the specified position with the value provided in the patch. In cases where the operation is a position change (\mintinline{python}{posChange}), the patch indicates the new index, and the parameter is inserted at the corresponding position in the new list. If no changes are specified for a parameter in the dictionary, \tool assumes no modification is needed and directly adds the parameter to the new list. After all positional parameters are processed, the node's args attribute is updated with the \mintinline{python}{newPosParaList}.

Similarly, lines 31-43 handle repairs for keyword parameters. Since keyword parameters include explicit names, renaming (\mintinline{python}{rename}) updates the parameter's name with the value specified in the patch. For changes like position changes (\mintinline{python}{posChange}), conversion to positional (\mintinline{python}{key2pos}), or conversion from positional (\mintinline{python}{pos2key}), keyword parameters are unaffected, and the original parameters are added to the new list without modification. After repairs, \mintinline{python}{newKeyParaList} is used to update the node's keywords attribute.

Once the current API call is repaired, \tool moves up the call chain to assess and repair higher-level API calls. It recursively processes all incompatible API calls in the project, addressing API parameter compatibility issues one by one. If an API repair fails, \tool skips the failed API and proceeds to the next one.
\tool's repair mechanism significantly overcomes the limitations of existing tools like MLCatchUp and Relancer, offering a more accurate and automated solution to address API parameter compatibility issues.

\textit{(1.3) Repairing Incompatible Candidates.} As mentioned at the end of \ding{184} API mapping establishment, the static mapping, i.e., extracting the definition of invoked APIs through matching the API name in library source code, may generate multiple APIs with the same name or APIs with multiple overloads. This results in multiple API signature mapping candidates.

\tool first eliminates compatible mapping candidates, as they do not require fixes. This process involves iterating through all candidates and determining the correspondence of APIs with the same name across two versions based on their path names. Given two lists of definitions of an invoked API, if an API definition from the current version matches one in the target version, it is 
excluded. Additionally, if the current version's API definition has a compatible counterpart in the target version, assessed by \ding{185}, it is also excluded, as no parameter compatibility issue exists. The remaining mapping candidates are incompatible. For each incompatible candidate, \tool attempts to fix it following the aforementioned procedures (1.1) and (1.2).

\begin{figure}[!t]
\centering
\includegraphics[width=\linewidth]{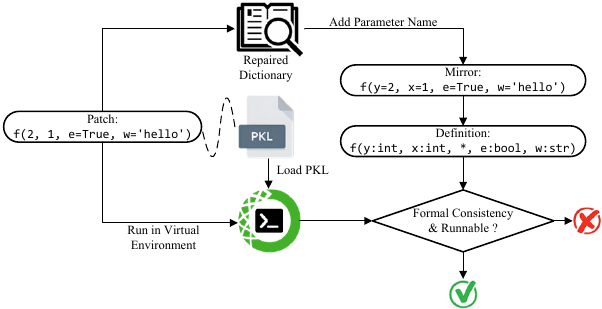}
\vspace{-3mm}
\caption{Static and dynamic validations of repairs in \tool.} 
\label{validate}
\vspace{-3mm}
\end{figure}


\textit{(2) Validation.}
\tool validates a fix via both static and dynamic approaches, as shown in Fig.~\ref{validate}. A fix is considered successful only if both validations are passed. Taking the repair example in Fig.~\ref{repair}, the API \mintinline{python}{f} undergoes several changes from version V1 to V2, i.e., \textit{posChange}, \textit{delete}, \textit{pos2key}, and \textit{rename}. As a result of these changes, the original call \mintinline{python}{f(1, 2, True, 3.14, z='hello')} becomes incompatible in the new version. \tool generates the corresponding repair patch: \mintinline{python}{f(2, 1, e=True, w='hello')}. To ensure that this repair is correct, \tool employs a combination of static and dynamic validation. Relying solely on execution success (dynamic validation) is insufficient, because Python's flexible calling conventions may allow syntactically valid but semantically incorrect bindings. For example, as shown in Listing 4, when a positional parameter (\mintinline{python}{maxcardinality}) is removed, a value passed by position might be silently shifted to the next parameter slot (\mintinline{python}{weight}). Although this does not raise a syntax error, the semantics of the call (\mintinline{python}{nx.min_weight_matching(G, None)}) have changed in the new version. To avoid such silent misbindings, \tool first performs static validation.

\textbf{Static Validation.} As shown in Fig.~\ref{validate}, \tool constructs a \textit{full parameter mirror} for the repaired API call by explicitly assigning names to all arguments. For example, the target version defines the function as \mintinline{python}{f(y:int, x:int, *, e:bool, w:str)}. The repair \mintinline{python}{f(2, 1, e=True, w='hello')} is converted into the mirror form \mintinline{python}{f(y=2, x=1, e=True, w=} \mintinline{python}{'hello')}, according to the repaired dictionary generated from parameter change analysis. \tool then compares this mirror with the definition of the target version. For keyword arguments, the repair must contain names that exactly match those in the definition. For positional arguments, both the position and the associated parameter name must be correct. For example, in the repaired call, the arguments \mintinline{python}{y=2} and \mintinline{python}{x=1} are verified to occupy the first and second positions, respectively. Therefore, this step ensures the formal correctness of the repaired API. 

\textbf{Dynamic Validation.} After passing the static check, \tool validates the patch dynamically by executing it in an isolated environment of the target library version. Specifically, \tool leverages Python's \mintinline{python}{subprocess} to spawn a child process and activate the appropriate virtual environment via \mintinline{python}{conda activate}. A dedicated validation script is then auto-generated (Listing~\ref{script}). This script imports all necessary project modules, loads the pickle file containing runtime context (e.g., \mintinline{python}{f(1, 2, True, 3.14, z='hello').pkl}, generated in \ding{183}), and reconstructs the repaired API call string (e.g., \mintinline{python}{sys.argv[2]}). The script finally executes the repaired call with \mintinline{python}{eval} to confirm its runnability. 

\tool enables independent validation of repaired APIs without running the entire project, avoiding interference between multiple incompatible APIs. Running the entire project is not only time-consuming but also risks preventing subsequent API repairs and validations if one repair fails.

Once the repair and validation are complete, 
the updated AST is converted back to code using \mintinline{python}{ast.unparse}. After processing all project files, \tool generates a report to assist users in resolving API parameter compatibility issues.

\begin{figure}[!t]
\centering
\begin{minipage}{3.4in}
\begin{lstlisting}[language=Python, label=script, caption=Dynamic validation script.]
from test_fNN import *
import sys
import dill

#sys.argv[1]: f(1, 2, True, 3.14, z='hello').pkl
pklPath=sys.argv[1]

#patch: f(2, 1, e=True, w='hello')
#sys.argv[2]: f(paraValueDict["f(1, 2, True, 3.14, z='hello')"][1], paraValueDict["f(1, 2, True, 3.14, z='hello')"][0], e=paraValueDict["f(1, 2, True, 3.14, z='hello')"][2], w=paraValueDict["f(1, 2, True, 3.14, z='hello')"][4])
fixedAPI=sys.argv[2]

with open(pklPath,'rb') as fr:
    paraValueDict=dill.load(fr)

eval(fixedAPI)
\end{lstlisting}
\end{minipage}
\vspace{-3mm}
\end{figure}

\section{\benchmark: Benchmark for Python API Parameter Compatibility Issues}\label{sec:benchmark}
To evaluate 
\tool, 
we construct a large-scale benchmark, \benchmark, providing a baseline for Python API parameter compatibility issues. In the following sections, we present the details of building \benchmark, including data collection of popular Python third-party libraries and APIs with parameter changes, test case generation, and the assignment of compatibility labels. 

\subsection{Collecting Popular Python Third-party Libraries}
To construct a representative benchmark, we need to collect popular third-party libraries in the Python ecosystem. 
The collection procedure consists of 
three rounds. The first round involves searching on GitHub with the keyword ``Python stars:\textgreater10000'', resulting in 312 GitHub projects. The second round is to filter these projects to identify popular Python third-party libraries based on two criteria: the project (a) provides relevant APIs for user calls, and (b) has an average daily download count on PyPI of over 100,000 in the most recent week (as of July 9, 2023). After the second round, 55 Python libraries were filtered from the 312 GitHub projects. The selection criteria in the third round include: the library (a) contains comprehensive API documentation and detailed version change logs, and (b) is not a command-line tool. The first criterion helps us identify and collect APIs with parameter changes, while the second criterion ensures the APIs are being called in Python projects rather than executed in the terminal (e.g., bash and Zsh). After three rounds of selection, 33 libraries were finalized, covering domains such as machine learning, natural language processing, image processing, data science, and web frameworks, as shown in Table~\ref{distribution}.

\begin{table}[!t]
\centering
\caption{Distribution of Changed APIs and Test Cases Across 33 Python Third-party Libraries}
\label{distribution}
\scalebox{0.55}{
\begin{tabular}{|l|c|c|l|c|c|l|c|c|}
\hline
Library      & \#APIs & \#Test Cases & Library      & \#APIs & \#Test Cases & Library  & \#APIs & \#Test Cases \\ \hline
PyTorch      & 4      & 91           & Redis        & 2      & 6            & HTTPX    & 8      & 191          \\ \hline
SciPy        & 193    & 5,887         & Faker        & 8      & 24           & NetworkX & 49     & 542          \\ \hline
Gensim       & 13     & 999          & LightGBM     & 1      & 10           & XGBoost  & 1      & 20           \\ \hline
TensorFlow   & 19     & 585          & Loguru       & 5      & 102          & Plotly   & 52     & 20,208        \\ \hline
Tornado      & 20     & 570          & SymPy        & 15     & 274          & Django   & 3      & 69           \\ \hline
Transformers & 1      & 44           & scikit-learn & 117    & 4,620         & Pillow   & 26     & 344          \\ \hline
Requests     & 2      & 20           & Flask        & 2      & 11           & JAX      & 1      & 28           \\ \hline
Matplotlib   & 21     & 2,027         & Click        & 4      & 241          & Polars   & 30     & 539          \\ \hline
FastAPI      & 4      & 156          & aiohttp      & 12     & 153          & pandas   & 73     & 5,103         \\ \hline
NumPy        & 85     & 2,006         & spaCy        & 2      & 19           & Rich     & 33     & 1,261         \\ \hline
Pydantic     & 1      & 16           & Keras        & 20     & 845          & Dask     & 17     & 467          \\ \hline
\end{tabular}
}
\vspace{-4mm}
\end{table}

\subsection{Collecting APIs with Parameter Changes}
\textbf{Step 1.} \textit{Analyzing Change Logs.} 
Our initial task is to manually inspect library change logs from Python 3 versions to identify APIs with parameter changes. 
These parameter changes mainly include the addition, removal, renaming, reordering of parameters, and the conversion of positional parameters to keyword parameters. Once we identified the changed APIs, we selected the version where the change occurred as the target version and its preceding version as the current version. Next, we extracted the API parameter definitions from the documentation of these two library versions. 

\textbf{Step 2.} \textit{Generating API Usage.} 
Based on the API definitions of the current version, we used ChatGPT to generate code examples that include all parameters. 
However, the code generated by ChatGPT may have missing parameters or syntax errors. Therefore, we manually examined and corrected each generated API usage. Then, we created two separate virtual environments for each API using Anaconda (23.5.2) \cite{Anaconda}. The two virtual environments install the current and target versions of the library, as well as their related third-party dependencies. This ensures that the generated API usage runs normally in the current version environment.   Steps 1 and 2 were performed independently by three authors.

\textbf{Step 3.} \textit{Cross Validation.} 
After completing the data collection, two authors with professional experience in Python project development performed a cross-check on the collected data to ensure the reliability and accuracy of the data. Specifically, the correctness of the API parameter definitions in the current and target versions was validated by using Python's inspect module, i.e., \mintinline{python}{inspect.signature}. Besides,  the change types of parameters were confirmed by comparing the API parameter definitions across the two versions. Moreover, we reviewed the API usage to confirm that all parameters were involved in the API calls.

The collection and validation process lasted six months. 
Finally, we collected 844 APIs with parameter changes from the 33 libraries. The number of collected APIs for each library is presented in Table~\ref{distribution}.

\subsection{Generating Test Cases via Parameter Mutation}
\begin{figure}[!t]
\centering
\includegraphics[width=3.75in]{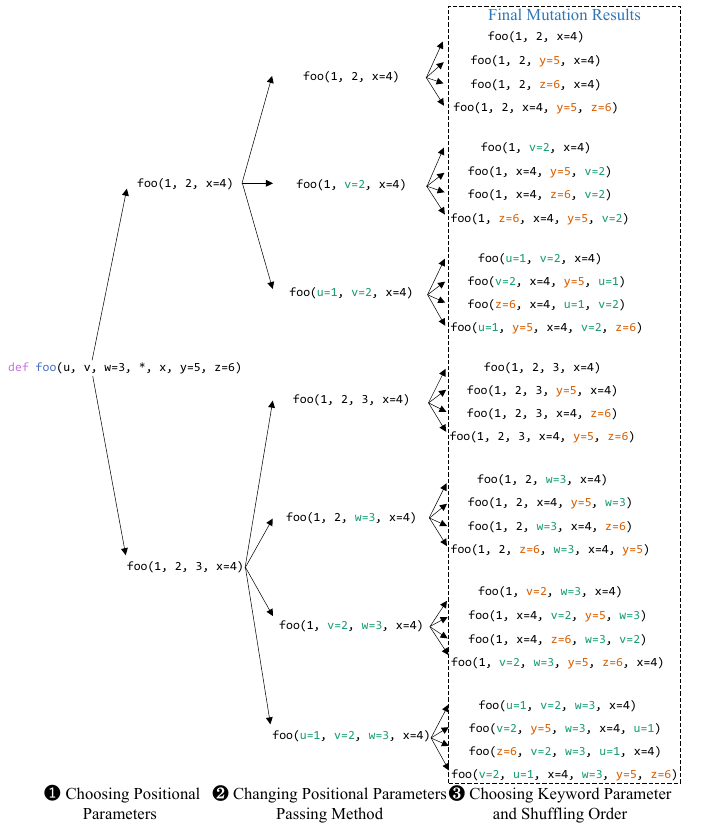}
\vspace{-4mm}
\caption{An example of parameter mutation on \mintinline{python}{foo} for generating test cases.}
\label{mutation}
\vspace{-4mm}
\end{figure}

To better simulate the diversity and flexibility of parameter passing when calling APIs in users' projects, 
we performed parameter mutation on the generated usage of the 844 APIs. 
The mutation involves changing the number of parameters, the method of parameter passing, and the order of parameters, thereby mutating a substantial number of test cases with different combinations of parameter numbers and parameter-passing methods. Fig.~\ref{mutation} illustrates the process of parameter mutation for the API \mintinline{python}{foo} with parameter definition \mintinline{python}{(u, v, w=3, *, x, y=5, z=6)} in the current version. 
Details of parameter mutation are as follows.

\textbf{Mutant Operator 1.} \textit{Choosing Positional Parameters.} We started by fixing positional and keyword parameters with no default values into combinations, where positional parameters were passed by position. Parameters with no default values must be passed when invoking APIs. Then, we added positional parameters with default values into the combination, also passed by position. For the API \mintinline{python}{foo}, this mutant operator generates two combinations, i.e., \mintinline{python}{foo(1, 2, x=4)} and \mintinline{python}{foo(1, 2, 3, x=4)}, as shown in Fig.~\ref{mutation}.

\textbf{Mutant Operator 2.} \textit{Changing Positional Parameters Passing Method.} Building on the first mutation, we changed the passing method of positional parameters from positional to keyword passing using parameter names. To ensure the syntactic correctness of Python, i.e., parameters passed by name must come after those without names, we added names to parameters from the back to the front. For example, performing this operator on \mintinline{python}{foo(1, 2, x=4)} mutates another two new combinations, i.e., \mintinline{python}{foo(1, v=2, x=4)} and \mintinline{python}{foo(u=1, v=2, x=4)}.

\textbf{Mutant Operator 3.} \textit{Choosing Keyword Parameter and Shuffling Order.} Based on the second operator, we initially selected one keyword parameter from the list containing keyword parameters with default values at a time to ensure that each keyword parameter has the possibility of being used individually. Then, in an incrementally increasing manner, we selected several keyword parameters with default values from the same list and added them to the combination. Finally, we randomly shuffled 
the order of parameters with names in the combination. 
Different orders of parameter passing align with practical usage in Python project development, which further complicates the difficulties in the detection and repair of parameter compatibility issues.

The parameter mutation was performed automatically by a script we implemented. We saved every combination generated by the three mutant operators, where the total number of combinations for each API mutation can be calculated using $\sum_{i=n}^{N}{(i+1)*2m}$, where $N$ represents the number of positional parameters, $n$ represents the number of positional parameters without default values, and $m$ represents the number of keyword parameters with default values. For the API \mintinline{python}{foo} illustrated in Fig.~\ref{mutation},  it has three positional parameters (two of them have default values) and two keyword parameters with default values. Thus, according to the formula, the parameter mutation generates 28 combinations in total. Note that, due to certain APIs having unusable default values or scenarios where only one of two parameter values can be used, some combinations were not feasible (i.e., unable to execute) and thus excluded. Finally, by mutating the parameters of 844 changed APIs, we generated a total of 47,478 test cases. The distribution of these test cases is illustrated in Table~\ref{distribution}.


\subsection{Assigning Compatibility Labels to Test Cases}
We conducted a labeling process for the 47,478 test cases containing the 844 APIs to determine their compatibility status within the target versions (i.e., the change occurred). Detailed steps are as follows:

\textit{(1) Executing Test Cases.} We executed each test case in the target version environment corresponding to the changed API. This leads to 
8,136 test cases failed to run, indicating they are incompatible with the target versions. Therefore, the label for these test cases is incompatible. Consequently, we manually analyzed these unrunnable test cases to identify the specific types of parameter changes causing the incompatibilities and categorized them based on these types. Table~\ref{Non-Runnable} shows that the majority of incompatibility is caused by position changes, followed by parameter removal. Moreover, since a test case could involve multiple types of changes, the `Total' row in Table~\ref{Non-Runnable} reflects the intersection of these test cases.

\begin{table}[!t]
\centering
\caption{Distribution of API Parameter Changes in Unrunnable Test Cases}
\label{Non-Runnable}
\scalebox{0.6}{
\begin{tabular}{|l|c|c|}
\hline
Change Type & Incompatible Cases & Incompatible APIs \\ \hline
Removal     & 3,138               & 112               \\ \hline
Rename      & 509                 & 43                \\ \hline
Pos2key     & 307                 & 13                \\ \hline
Position    & 6,013               & 103               \\ \hline
**kwargs    & 138                 & 15                \\ \hline
Total       & 8,136               & 207               \\ \hline
\end{tabular}
}
\vspace{-4mm}
\end{table}

\begin{table}[!t]
    \centering
    \caption{Distribution of API Parameter Changes in Runnable Test Cases}
    \label{runnable}
    \scalebox{0.6}{
        \begin{tabular}{|l|c|c|c|c|}
            \hline
            Change Type & Compatible Cases & Compatible APIs & Incompatible Cases & Incompatible APIs \\ \hline
            No Change   & 27,107           & 813             & 0                  & 0                 \\ \hline
            Removal     & 0                & 0               & 188                & 46                \\ \hline
            Rename      & 282              & 39              & 87                 & 13                \\ \hline
            Pos2Key     & 638              & 27              & 873                & 20                \\ \hline
            Position    & 6,668            & 86              & 3,441              & 91                \\ \hline
            **kwargs    & 334              & 15              & 0                  & 0                 \\ \hline
            Total       & 34,891           & 839             & 4,451              & 141               \\ \hline
\end{tabular}
}
\vspace{-4mm}
\end{table}

\textit{(2) Manual Labeling.} For the remaining 39,342 test cases that could run successfully, we did not immediately classify them as compatible, as successful execution does not equate to compatibility (discussed in Section~\ref{sec:background-challenges}). Therefore, we conducted further manual analysis by examining the changes in API definitions before and after the code update and their usage in test cases. We summarized the following rules to determine the compatibility of test cases in the target versions:

\textbf{Rule 1.} Test cases that do not involve changed parameters are considered compatible. They are not affected by parameter changes and are thus compatible with the target versions.

\textbf{Rule 2.} When API definitions in the target versions do not include \mintinline{python}{*args} and \mintinline{python}{**kwargs}, test cases using removed parameters are deemed incompatible; for renaming in parameter names, passing by parameter name is considered incompatible, while positional passing is considered compatible. Besides, when a parameter is converted from a positional to a keyword argument, keyword passing 
is compatible, while positional passing 
is incompatible. Furthermore, for changes in the position of parameters, again, passing by the parameter name is compatible, while those passed without the parameter names are considered incompatible. Test cases are considered incompatible if there is an incompatible parameter type change.

\textbf{Rule 3.} When API definitions in the target versions include \mintinline{python}{*args} or \mintinline{python}{**kwargs}, changes such as parameter renaming and parameter removal are considered compatible because \mintinline{python}{*args} can accept a variadic number of positional arguments, while \mintinline{python}{**kwargs} can accept a variadic number of keyword arguments, as introduced in Section~\ref{sec:background-characteristics}.

Table~\ref{runnable} presents the distribution of compatible and incompatible test cases under different types of parameter changes that are still executable. It can be observed that 27,107 test cases remain compatible because they do not utilize the changed parameters. However, 4,451 test cases are executable but incompatible. The `Total' row calculates the intersection of these test cases.

\section{Evaluation}\label{sec:evaluation}

\subsection{Research Questions}
Our work mainly focuses on answering the following five research questions (RQs):

\begin{itemize}
    \item \textbf{RQ1:} How does \tool perform in detecting API parameter compatibility issues?  
    
    \item \textbf{RQ2:} How does \tool perform in repairing API parameter compatibility issues? 

    \item \textbf{RQ3:} What is the effectiveness of \tool in real-world Python projects? 

    \item \textbf{RQ4:} How does \tool compare to ChatGPT in detecting and repairing API parameter compatibility issues? 

    \item \textbf{RQ5:} What is the time cost of \tool in detecting and repairing API parameter compatibility issues? 

\end{itemize}

\subsection{Experiment Setup}
\textit{(1) Settings of Comparison Tools.} 
The settings of comparison tools, i.e., MLCatchUp~\cite{haryono2021mlcatchup}, Relancer~\cite{zhu2021restoring}, and ChatGPT (GPT-4o), are presented as follows:

\textbf{Settings of MLCatchUp.} MLCatchUp~\cite{haryono2021mlcatchup} is an open-source tool designed to fix deprecated APIs in a single \mintinline{python}{.py} file. It requires users to manually input the signatures (parameter definitions) of APIs before and after version updates. It then performs repair through static analysis of the project code. In terms of detecting API parameter compatibility issues, MLCatchUp does not provide such functionality. 
It only outputs the repair operations and results. Therefore, we used the following settings to evaluate the detection performance of MLCatchUp. For compatible test cases, if MLCatchUp's repair operations do not affect the original compatibility of the test cases, its detection is considered correct; otherwise, it is considered incorrect. For incompatible test cases, if MLCatchUp does not provide any repair operations, it is considered an incorrect detection; if it does provide repair operations, the detection is considered correct. In terms of repairing API parameter compatibility issues, since MLCatchUp does not support automated validation of repair results, we manually reviewed and validated its repair results.

\textbf{Settings of Relancer.} Relancer~\cite{zhu2021restoring} focuses on repairing deprecated APIs in Jupyter Notebooks 
by analyzing error messages generated during the code execution. In detecting API parameter compatibility issues, Relancer simply uses whether the code can run normally without crashing as the standard to assess API compatibility. Therefore, it can only detect and repair test cases that cannot run. This means that it considers all runnable test cases as compatible. In terms of repairing API parameter compatibility issues, we determined the success of repairs by automatically parsing the information output by Relancer during the repair process. If the output information contains ``This case is fully fixed!'', the repair is considered successful; otherwise, it is regarded as a failure.

\begin{figure}[!t]
\centering
\subfloat[Prompt template without API definition.]{
    \includegraphics[width=3.25in]{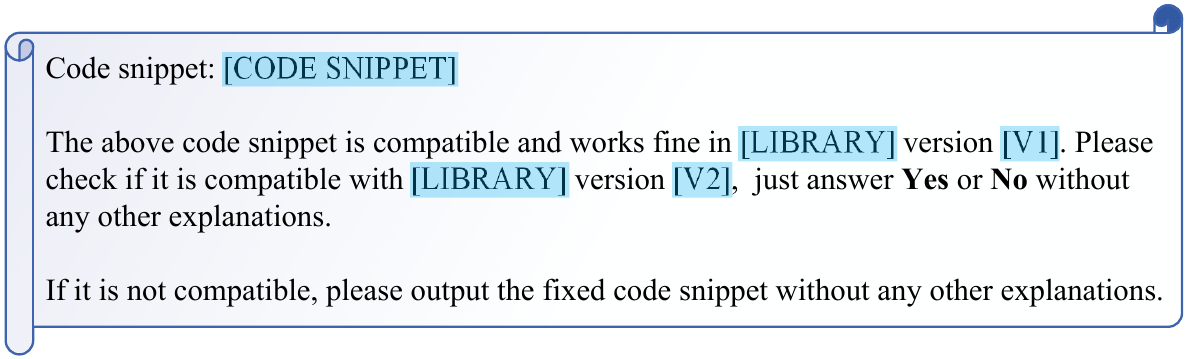}
    \label{prompt1}
}
\vspace{-3mm}
\subfloat[Prompt template with API definition.]{
    \includegraphics[width=3.25in]{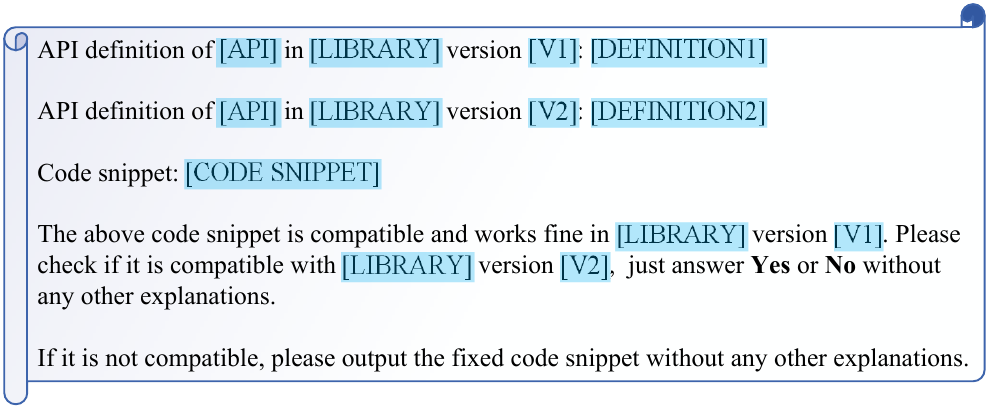}
    \label{prompt2}
}

\vspace{-3mm}
\subfloat[Prompt template with runtime message.]{
    \includegraphics[width=3.25in]{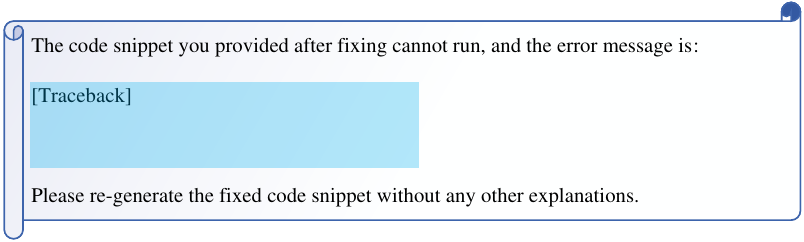}
    \label{prompt3}
}
\caption{Detection and repair prompt templates in ChatGPT (GPT-4o).}
\label{prompt}
\vspace{-4mm}
\end{figure}

\textbf{Settings of ChatGPT (GPT-4o).} 
Another solution is to directly query ChatGPT about the compatibility of APIs in test cases across different library versions and their repair results. To answer RQ 4, we used the test cases as input to query ChatGPT (GPT-4o) for detection and repair. For each test case, we conducted two experimental settings:

\begin{itemize}
    \item Setting 1: Without providing API parameter definitions (as shown in Fig.~\ref{prompt1}).
    \item Setting 2: Providing API parameter definitions for the current and target versions (as shown in Fig.~\ref{prompt2}).
\end{itemize}

To validate the fixes for incompatible test cases, we executed the repaired code locally. If error messages occurred, we used them to re-prompt ChatGPT with a revised query template (Fig.~\ref{prompt3}). Furthermore, to simulate real-world user independence and randomness, the following settings were applied:

\begin{itemize}
    \item ChatGPT's memory feature was turned off to ensure no contextual association between sessions.
    \item Tests were conducted in new session windows using both Edge and Chrome browsers.
    \item Responses from two separate attempts were manually checked for accuracy. Only if both responses were correct was ChatGPT's detection/repair deemed successful. Inconsistent results (e.g., one correct and one incorrect) were considered erroneous.
\end{itemize}

\rev{All experiments were conducted using GPT-4o (version: gpt-4o-2024-08-06, accessed on October 16, 2024, via web interface).} 

\textit{(2) Evaluation Datasets.} 
We constructed three datasets for conducting evaluation experiments.

\textbf{\benchmark.} To answer RQs 1, 2, and 5, we 
constructed a benchmark (\benchmark) mutated from 844 parameter-changed APIs across 33 popular Python libraries, 
containing a total of 47,478 test cases with compatibility labels, 
to evaluate the performance of \tool, MLCatchUp~\cite{haryono2021mlcatchup}, and Relancer~\cite{zhu2021restoring} in detecting and repairing API parameter compatibility issues. Details of \benchmark construction are described in Section~\ref{sec:benchmark}.

\textbf{MLCatchUp Dataset.} 
As shown in Table \ref{mlcatchup-dataset}, to answer RQ1 and RQ2, we extracted all the test cases related to parameter changes from the dataset provided by MLCatchUp~\cite{haryono2021mlcatchup}, obtaining 20 cases in total. These cases involve two popular libraries (scikit-learn and TensorFlow) and four different APIs. However, since MLCatchUp's dataset is designed for static analysis, some test cases cannot be executed directly, lack runtime coverage of the target library APIs, or miss specific version information. To address this, we queried the official API documentation to identify the version in which the API changes occurred (target version), and used the immediately preceding version as the current version. We then created two separate virtual environments corresponding to these versions. Finally, we added the necessary invocation statements and commented out unused or redundant import statements that could hinder execution, ensuring each test case could run successfully and cover the APIs to be tested.

\begin{table}[!t]
\centering
\caption{MLCatchUp Dataset Information}
\label{mlcatchup-dataset}
\scalebox{0.575}{
\begin{tabular}{|c|c|c|c|c|c|c|}
\hline
\multirow{2}{*}{API} & \multirow{2}{*}{Library} & \multicolumn{2}{c|}{\#Test Cases} 
& Current & Target & Change \\ \cline{3-4}
& & Compat. & Incompat. & Version & Version & Type \\ \hline
sklearn.cluster.Kmeans              & scikit-learn             & \multicolumn{1}{c|}{0}          & 8            & 0.24.2                           & 1.0                             & remove                       \\ \hline
sklearn.tree.DecisionTreeClassifier & scikit-learn             & \multicolumn{1}{c|}{0}          & 5            & 0.23.1                           & 0.24.1                          & remove                       \\ \hline
sklearn.tree.DecisionTreeRegressor  & scikit-learn             & \multicolumn{1}{c|}{0}          & 3            & 0.23.1                           & 0.24.1                          & remove                       \\ \hline
tensorflow.compat.v1.string\_split   & TensorFlow               & \multicolumn{1}{c|}{4}          & 0            & 1.13.2                           & 1.14.0                          & rename                       \\ \hline
\end{tabular}
}
\vspace{-4mm}
\end{table}

\textbf{Real-world Python Projects.} For RQ3, we selected real-world Python projects from GitHub as our test dataset. Initially, we filtered out APIs containing incompatible test cases from \benchmark. 
We then searched these APIs on GitHub, applying a filter for the Python language. 
To enhance search efficiency, we included the specific parameters that had changed in these APIs as part of our search keywords. Through this approach, we successfully collected 30 Python projects that have parameter compatibility issues.
These projects cover 9 popular Python libraries and exhibit significant diversity in code size, with lines of code ranging from tens to thousands, and the number of Python files (.py) varying between 1 and 25 (measured by cloc). 
As shown in Table~\ref{project}. For each project, we first configured the required dependencies based on the \mintinline{python}{requirements.txt} file found in the project. If the file was absent, we manually set up the virtual environment to ensure that the project could run normally. 

\begin{table}[!t]
\centering
\caption{The Collected Real-world GitHub Python Project Dataset}
\label{project}
\scalebox{0.65}{
\begin{tabular}{|l|c|c|c|c|c|}
\hline
Project                 & Files & LOC   & Library      & Current Version & Target Version \\ \hline
allnews                 & 8     & 674   & Gensim       & 3.8.3           & 4.0.0          \\ \hline
Youtube-Comedy          & 2     & 146   & Gensim       & 0.12.3          & 0.12.4         \\ \hline
recommendation-engine   & 1     & 108   & NetworkX     & 1.11            & 2              \\ \hline
political-polarisation  & 1     & 86    & NetworkX     & 2.8.8           & 3.0.0          \\ \hline
TSP                     & 1     & 36    & NetworkX     & 2.8.8           & 3.0.0          \\ \hline
CustomSamplers          & 1     & 15    & NumPy        & 1.9.3           & 1.10.1         \\ \hline
machine-learning        & 1     & 46    & NumPy        & 1.23.5          & 1.24.0         \\ \hline
gistable                & 1     & 20    & NumPy        & 1.23.5          & 1.24.0         \\ \hline
galaxiesDataScience     & 1     & 33    & NumPy        & 1.23.5          & 1.24.0         \\ \hline
fuel\_forecast\_explorer  & 1     & 32    & pandas       & 1.5.3           & 2.0.0          \\ \hline
sg-restart-regridder    & 1     & 34    & pandas       & 1.5.3           & 2.0.0          \\ \hline
MAHE\_OD\_DATASET         & 8     & 842   & pandas       & 1.5.3           & 2.0.0          \\ \hline
hfhd                    & 5     & 801   & pandas       & 1.5.3           & 2.0.0          \\ \hline
scrapping-jojo-main     & 1     & 96    & pandas       & 1.5.3           & 2.0.0          \\ \hline
Contrucao-de            & 1     & 81    & pandas       & 1.5.3           & 2.0.0          \\ \hline
polars-book-cn          & 1     & 13    & Polars       & 0.16.18         & 0.17.0         \\ \hline
EJPLab\_Computational    & 1     & 69    & Polars       & 0.16.18         & 0.17.0         \\ \hline
Deep-Graph-Kernels      & 1     & 86    & SciPy        & 0.19.1          & 1.0.0          \\ \hline
AIBO                    & 9     & 1,516 & SciPy        & 1.7.3           & 1.10.0         \\ \hline
greenbenchmark          & 10    & 620   & SciPy        & 0.19.1          & 1.0.0          \\ \hline
qho-control             & 2     & 120   & SciPy        & 1.8.1           & 1.9.0          \\ \hline
giantpopflucts          & 9     & 475   & SciPy        & 1.9.3           & 1.10.0         \\ \hline
django-selenium-testing & 1     & 94    & Tornado      & 3.1             & 5              \\ \hline
polire                  & 25    & 982   & scikit-learn & 1.1.3           & 1.2.0          \\ \hline
Python-Workshop         & 5     & 215   & scikit-learn & 1.1.3           & 1.2.0          \\ \hline
covid19-predictor       & 1     & 93    & scikit-learn & 1.1.3           & 1.2.0          \\ \hline
Final                   & 5     & 484   & scikit-learn & 1.1.3           & 1.2.0          \\ \hline
Gender-pay-gap          & 1     & 41    & scikit-learn & 1.1.3           & 1.2.0          \\ \hline
SDOML                   & 2     & 180   & Matplotlib   & 3.2.2           & 3.3.0          \\ \hline
simulations             & 8     & 523   & Matplotlib   & 3.2.2           & 3.3.0          \\ \hline
\end{tabular}
}
\vspace{-4mm}
\end{table}

\textbf{Dataset for ChatGPT (GPT-4o).} Considering the limitations on the number of queries of ChatGPT, 
it is not feasible to use all the test cases in \benchmark for evaluation. Therefore, in RQ4, we randomly selected one compatible and one incompatible test case from the 29 Python third-party libraries included in \benchmark to evaluate ChatGPT (GPT-4o). The remaining four libraries contain only compatible test cases. Thus, we did not select test cases from these libraries. 

\textit{(3) Evaluation Metrics.} The evaluation metrics for detection and repair 
are presented as follows. 

\textbf{Metrics for Detection.} In our evaluation, incompatible test cases are defined as positive instances. Accordingly, for each tool, 
we calculated the following key metrics: \textit{true positives} (TP), which are the number of incompatible cases correctly detected; \textit{false positives} (FP), which are the number of compatible test cases erroneously detected as incompatible; and \textit{false negatives} (FN), which are the number of incompatible test cases wrongly detected as compatible. Based on these metrics, we computed precision, recall, and F1-score using formulas \eqref{Precision}, \eqref{Recall}, and \eqref{F-measure}, respectively, to evaluate the detection performance of \tool and the compared tools. 

\begin{equation}
\label{Precision}
Precision=\frac{TP}{TP+FP},
\end{equation}

\begin{equation}
\label{Recall}
Recall=\frac{TP}{TP+FN},
\end{equation}

\begin{equation}
\label{F-measure}
F1-score=2 \times \frac{Precision \times Recall}{Precision+Recall}.
\end{equation}


\textbf{Metrics for Repair.} We used repair precision as the metric to evaluate the effectiveness of \tool and the baseline tools in repairing API parameter compatibility issues. Repair precision is defined as:

\begin{equation}
Precision=\frac{SR}{SR + UR},
\end{equation}

\noindent 
where $SR$ (Successful Repairs) is the number of incompatible test cases successfully repaired, while $UR$ (Unsuccessful Repairs) is the total number of unsuccessful repair attempts, including both $ICP$ (incompatible cases with failed repairs) and $CP$ (compatible cases with incorrect repairs). 
By incorporating both repair failures ($ICP$) and incorrect repairs ($CP$), 
we evaluate not only the method's ability to repair incompatible test cases but also the impact of unsuccessful and erroneous repairs. This comprehensive approach ensures an accurate reflection of the repair method's performance, considering both its strengths and weaknesses.

\textit{(4) Experiment Environment.} 
Our experiments were conducted on a server running a 64-bit Ubuntu 18.04.1 OS, equipped with two Intel Xeon Gold 6230R CPUs at 2.10GHz (26 cores with 52 threads), three Nvidia RTX 2080Ti GPUs, 160GB of RAM, 256 GB SSD, and 8 TB HDD storage. \tool is implemented using Python 3.9.

\section{Results and Analysis}\label{sec:results}

\subsection{RQ1: How does PCART Perform in Detecting API Parameter Compatibility Issues?}\label{sec:results-rq1}

\begin{table}[!t]
\centering
\caption{Comparison of Detecting API Parameter Compatibility Issues}
\label{compatibility detection}
\scalebox{0.65}{
\begin{tabular}{|lc|ccc|ccc|ccc|}
\hline
\multicolumn{1}{|l|}{\multirow{2}{*}{Library}} & \multirow{2}{*}{APIs} & \multicolumn{3}{c|}{MLCatchUp}                                              & \multicolumn{3}{c|}{RELANCER}                                                     & \multicolumn{3}{c|}{\tool}                                                            \\ \cline{3-11} 
\multicolumn{1}{|l|}{}                         &                       & \multicolumn{1}{c|}{TP}           & \multicolumn{1}{c|}{FP}    & FN         & \multicolumn{1}{c|}{TP}          & \multicolumn{1}{c|}{FP}         & FN           & \multicolumn{1}{c|}{TP}              & \multicolumn{1}{c|}{FP}         & FN           \\ \hline
\multicolumn{1}{|l|}{PyTorch}                  & 4                     & \multicolumn{1}{c|}{\textbf{7}}   & \multicolumn{1}{c|}{0}     & 0          & \multicolumn{1}{c|}{\textbf{7}}  & \multicolumn{1}{c|}{0}          & 0            & \multicolumn{1}{c|}{\textbf{7}}      & \multicolumn{1}{c|}{0}          & 0            \\ \hline
\multicolumn{1}{|l|}{SciPy}                    & 193                   & \multicolumn{1}{c|}{1}            & \multicolumn{1}{c|}{23}    & 437        & \multicolumn{1}{c|}{409}         & \multicolumn{1}{c|}{\textbf{0}} & 29           & \multicolumn{1}{c|}{\textbf{438}}    & \multicolumn{1}{c|}{6}          & \textbf{0}   \\ \hline
\multicolumn{1}{|l|}{Gensim}                   & 13                    & \multicolumn{1}{c|}{430}          & \multicolumn{1}{c|}{0}     & 24         & \multicolumn{1}{c|}{390}         & \multicolumn{1}{c|}{0}          & 64           & \multicolumn{1}{c|}{\textbf{441}}    & \multicolumn{1}{c|}{0}          & \textbf{13}  \\ \hline
\multicolumn{1}{|l|}{Tensorflow}               & 19                    & \multicolumn{1}{c|}{\textbf{57}}  & \multicolumn{1}{c|}{47}    & \textbf{0} & \multicolumn{1}{c|}{49}          & \multicolumn{1}{c|}{\textbf{0}} & 8            & \multicolumn{1}{c|}{56}              & \multicolumn{1}{c|}{\textbf{0}} & 1            \\ \hline
\multicolumn{1}{|l|}{Tornado}                  & 20                    & \multicolumn{1}{c|}{238}          & \multicolumn{1}{c|}{0}     & 52         & \multicolumn{1}{c|}{53}          & \multicolumn{1}{c|}{0}          & 237          & \multicolumn{1}{c|}{\textbf{286}}    & \multicolumn{1}{c|}{0}          & \textbf{4}   \\ \hline
\multicolumn{1}{|l|}{Transformers}             & 1                     & \multicolumn{1}{c|}{0}            & \multicolumn{1}{c|}{0}    & 0          & \multicolumn{1}{c|}{0}           & \multicolumn{1}{c|}{0} & 0            & \multicolumn{1}{c|}{0}               & \multicolumn{1}{c|}{0} & 0            \\ \hline
\multicolumn{1}{|l|}{Requests}                 & 2                     & \multicolumn{1}{c|}{0}            & \multicolumn{1}{c|}{0}     & 6          & \multicolumn{1}{c|}{2}           & \multicolumn{1}{c|}{0}          & 4            & \multicolumn{1}{c|}{\textbf{6}}      & \multicolumn{1}{c|}{0}          & \textbf{0}   \\ \hline
\multicolumn{1}{|l|}{Matplotlib}               & 21                    & \multicolumn{1}{c|}{444}          & \multicolumn{1}{c|}{0}     & 50         & \multicolumn{1}{c|}{166}         & \multicolumn{1}{c|}{0}          & 328          & \multicolumn{1}{c|}{\textbf{457}}    & \multicolumn{1}{c|}{0}          & \textbf{35}  \\ \hline
\multicolumn{1}{|l|}{FastAPI}                  & 4                     & \multicolumn{1}{c|}{\textbf{51}}  & \multicolumn{1}{c|}{0}     & \textbf{0} & \multicolumn{1}{c|}{47}          & \multicolumn{1}{c|}{0}          & 4            & \multicolumn{1}{c|}{\textbf{51}}     & \multicolumn{1}{c|}{0}          & \textbf{0}   \\ \hline
\multicolumn{1}{|l|}{NumPy}                    & 85                    & \multicolumn{1}{c|}{298}          & \multicolumn{1}{c|}{20}    & 62         & \multicolumn{1}{c|}{334}         & \multicolumn{1}{c|}{\textbf{0}} & 26           & \multicolumn{1}{c|}{\textbf{360}}    & \multicolumn{1}{c|}{\textbf{0}} & \textbf{0}   \\ \hline
\multicolumn{1}{|l|}{Pydantic}                 & 1                     & \multicolumn{1}{c|}{0}            & \multicolumn{1}{c|}{0}     & 8          & \multicolumn{1}{c|}{\textbf{8}}  & \multicolumn{1}{c|}{0}          & \textbf{0}   & \multicolumn{1}{c|}{4}               & \multicolumn{1}{c|}{0}          & 4            \\ \hline
\multicolumn{1}{|l|}{Redis}                    & 2                     & \multicolumn{1}{c|}{0}            & \multicolumn{1}{c|}{0}     & 0          & \multicolumn{1}{c|}{0}           & \multicolumn{1}{c|}{0}          & 0            & \multicolumn{1}{c|}{0}               & \multicolumn{1}{c|}{0}          & 0            \\ \hline
\multicolumn{1}{|l|}{Faker}                    & 8                     & \multicolumn{1}{c|}{0}            & \multicolumn{1}{c|}{0}     & 0          & \multicolumn{1}{c|}{0}           & \multicolumn{1}{c|}{0}          & 0            & \multicolumn{1}{c|}{0}               & \multicolumn{1}{c|}{0}          & 0            \\ \hline
\multicolumn{1}{|l|}{LightGBM}                 & 1                     & \multicolumn{1}{c|}{0}            & \multicolumn{1}{c|}{0}     & 0          & \multicolumn{1}{c|}{0}           & \multicolumn{1}{c|}{0}          & 0            & \multicolumn{1}{c|}{0}               & \multicolumn{1}{c|}{0}          & 0            \\ \hline
\multicolumn{1}{|l|}{Loguru}                   & 5                     & \multicolumn{1}{c|}{3}   & \multicolumn{1}{c|}{0}     & 0          & \multicolumn{1}{c|}{3}           & \multicolumn{1}{c|}{0}          & 0            & \multicolumn{1}{c|}{3}      & \multicolumn{1}{c|}{0}          & 0            \\ \hline
\multicolumn{1}{|l|}{SymPy}                    & 15                    & \multicolumn{1}{c|}{25}           & \multicolumn{1}{c|}{0}     & 39         & \multicolumn{1}{c|}{33}          & \multicolumn{1}{c|}{0}          & 31           & \multicolumn{1}{c|}{\textbf{63}}     & \multicolumn{1}{c|}{0}          & \textbf{1}   \\ \hline
\multicolumn{1}{|l|}{scikit-learn}             & 117                   & \multicolumn{1}{c|}{\textbf{695}} & \multicolumn{1}{c|}{623}   & \textbf{0} & \multicolumn{1}{c|}{108}         & \multicolumn{1}{c|}{\textbf{0}} & 587          & \multicolumn{1}{c|}{\textbf{695}}             & \multicolumn{1}{c|}{\textbf{0}} & \textbf{0}            \\ \hline
\multicolumn{1}{|l|}{Flask}                    & 2                     & \multicolumn{1}{c|}{\textbf{2}}   & \multicolumn{1}{c|}{0}     & \textbf{0}          & \multicolumn{1}{c|}{\textbf{2}}  & \multicolumn{1}{c|}{0}          & \textbf{0}            & \multicolumn{1}{c|}{0}               & \multicolumn{1}{c|}{0}          & 2            \\ \hline
\multicolumn{1}{|l|}{Click}                    & 4                     & \multicolumn{1}{c|}{\textbf{3}}   & \multicolumn{1}{c|}{0}     & \textbf{0} & \multicolumn{1}{c|}{0}           & \multicolumn{1}{c|}{0}          & 3            & \multicolumn{1}{c|}{\textbf{3}}      & \multicolumn{1}{c|}{0}          & \textbf{0}   \\ \hline
\multicolumn{1}{|l|}{aiohttp}                  & 12                    & \multicolumn{1}{c|}{16}  & \multicolumn{1}{c|}{0}    & 0          & \multicolumn{1}{c|}{16} & \multicolumn{1}{c|}{0} & 0           & \multicolumn{1}{c|}{16}     & \multicolumn{1}{c|}{0} & 0            \\ \hline
\multicolumn{1}{|l|}{spaCy}                    & 2                     & \multicolumn{1}{c|}{6}   & \multicolumn{1}{c|}{0}     & 0          & \multicolumn{1}{c|}{6}  & \multicolumn{1}{c|}{0}          & 0            & \multicolumn{1}{c|}{6}      & \multicolumn{1}{c|}{0}          & 0            \\ \hline
\multicolumn{1}{|l|}{Keras}                    & 20                    & \multicolumn{1}{c|}{\textbf{100}} & \multicolumn{1}{c|}{26}    & \textbf{0} & \multicolumn{1}{c|}{45}          & \multicolumn{1}{c|}{\textbf{0}} & 55           & \multicolumn{1}{c|}{\textbf{100}}    & \multicolumn{1}{c|}{\textbf{0}} & \textbf{0}   \\ \hline
\multicolumn{1}{|l|}{HTTPX}                    & 8                     & \multicolumn{1}{c|}{58}           & \multicolumn{1}{c|}{54}    & 7          & \multicolumn{1}{c|}{64}          & \multicolumn{1}{c|}{\textbf{0}} & 1            & \multicolumn{1}{c|}{\textbf{65}}     & \multicolumn{1}{c|}{\textbf{0}} & \textbf{0}   \\ \hline
\multicolumn{1}{|l|}{NetworkX}                 & 49                    & \multicolumn{1}{c|}{107}          & \multicolumn{1}{c|}{0}     & 38         & \multicolumn{1}{c|}{134}         & \multicolumn{1}{c|}{0}          & 11           & \multicolumn{1}{c|}{\textbf{142}}    & \multicolumn{1}{c|}{0}          & \textbf{3}   \\ \hline
\multicolumn{1}{|l|}{XGBoost}                  & 1                     & \multicolumn{1}{c|}{5}   & \multicolumn{1}{c|}{0}     & 0          & \multicolumn{1}{c|}{5}  & \multicolumn{1}{c|}{0}          & 0            & \multicolumn{1}{c|}{5}      & \multicolumn{1}{c|}{0}          & 0            \\ \hline
\multicolumn{1}{|l|}{Plotly}                   & 52                    & \multicolumn{1}{c|}{6,275}        & \multicolumn{1}{c|}{358}   & 45         & \multicolumn{1}{c|}{3,475}       & \multicolumn{1}{c|}{\textbf{0}} & 2,845        & \multicolumn{1}{c|}{\textbf{6,320}}  & \multicolumn{1}{c|}{\textbf{0}} & \textbf{0}   \\ \hline
\multicolumn{1}{|l|}{Django}                   & 3                     & \multicolumn{1}{c|}{3}            & \multicolumn{1}{c|}{0}     & 4          & \multicolumn{1}{c|}{\textbf{7}}           & \multicolumn{1}{c|}{0}          & \textbf{0}   & \multicolumn{1}{c|}{\textbf{7}}      & \multicolumn{1}{c|}{0}          & \textbf{0}   \\ \hline
\multicolumn{1}{|l|}{Pillow}                   & 26                    & \multicolumn{1}{c|}{0}            & \multicolumn{1}{c|}{0}     & 7          & \multicolumn{1}{c|}{\textbf{7}}  & \multicolumn{1}{c|}{0}          & \textbf{0}   & \multicolumn{1}{c|}{\textbf{7}}      & \multicolumn{1}{c|}{0}          & \textbf{0}   \\ \hline
\multicolumn{1}{|l|}{JAX}                      & 1                     & \multicolumn{1}{c|}{12}  & \multicolumn{1}{c|}{16}    & 0          & \multicolumn{1}{c|}{12} & \multicolumn{1}{c|}{\textbf{0}} & 0            & \multicolumn{1}{c|}{12}     & \multicolumn{1}{c|}{\textbf{0}} & 0            \\ \hline
\multicolumn{1}{|l|}{Polars}                   & 30                    & \multicolumn{1}{c|}{134}          & \multicolumn{1}{c|}{0}     & 3          & \multicolumn{1}{c|}{62}          & \multicolumn{1}{c|}{0}          & 75           & \multicolumn{1}{c|}{\textbf{137}}    & \multicolumn{1}{c|}{0}          & \textbf{0}   \\ \hline
\multicolumn{1}{|l|}{pandas}                   & 73                    & \multicolumn{1}{c|}{1,431}        & \multicolumn{1}{c|}{0}     & 1,170       & \multicolumn{1}{c|}{\textbf{2,485}}       & \multicolumn{1}{c|}{0}          & \textbf{116} & \multicolumn{1}{c|}{1,861}  & \multicolumn{1}{c|}{0}          & 740          \\ \hline
\multicolumn{1}{|l|}{Rich}                     & 33                    & \multicolumn{1}{c|}{137}          & \multicolumn{1}{c|}{164}   & 44         & \multicolumn{1}{c|}{159}         & \multicolumn{1}{c|}{\textbf{0}} & 22           & \multicolumn{1}{c|}{\textbf{181}}    & \multicolumn{1}{c|}{\textbf{0}} & \textbf{0}   \\ \hline
\multicolumn{1}{|l|}{Dask}                     & 17                    & \multicolumn{1}{c|}{\textbf{47}}  & \multicolumn{1}{c|}{0}     & \textbf{0} & \multicolumn{1}{c|}{42}          & \multicolumn{1}{c|}{0}          & 5            & \multicolumn{1}{c|}{8}               & \multicolumn{1}{c|}{0}          & 39           \\ \hline
\multicolumn{1}{|l|}{Total}                    & 844                   & \multicolumn{1}{c|}{10,585}       & \multicolumn{1}{c|}{1,331} & 1,996      & \multicolumn{1}{c|}{8,130}       & \multicolumn{1}{c|}{\textbf{0}} & 4,451        & \multicolumn{1}{c|}{\textbf{11,737}} & \multicolumn{1}{c|}{6}          & \textbf{842} \\ \hline
\multicolumn{2}{|l|}{Precision}                                        & \multicolumn{3}{c|}{88.83\%}                                                & \multicolumn{3}{c|}{\textbf{100.00\%}}                                            & \multicolumn{3}{c|}{99.95\%}                                                          \\ \hline
\multicolumn{2}{|l|}{Recall}                                           & \multicolumn{3}{c|}{84.13\%}                                                & \multicolumn{3}{c|}{64.62\%}                                                      & \multicolumn{3}{c|}{\textbf{93.31\%}}                                                 \\ \hline
\multicolumn{2}{|l|}{F1-score}                                         & \multicolumn{3}{c|}{86.42\%}                                                & \multicolumn{3}{c|}{78.51\%}                                                      & \multicolumn{3}{c|}{\textbf{96.51\%}}                                                 \\ \hline
\end{tabular}
}
\vspace{-4mm}
\end{table}

Table~\ref{compatibility detection} shows the comparison of MLCatchUP, Relancer, and \tool in detecting API parameter compatibility issues on \benchmark. Details of TP, FP, and FN across different libraries and the calculated precision, recall, and F1-score are presented in the table. 
MLCatchUp performs the worst in terms of FP: 1,331 compatible test cases are wrongly detected as incompatible ones. 
Relancer has the highest number of FN: 4,451 incompatible test cases are erroneously detected as compatible ones. \tool excels in detecting TP: 11,737 incompatible test cases are correctly detected. 
We constructed a contingency table containing three metrics (i.e., TP, FP, and FN) and applied the Chi-square test~\cite{pearson1900x} to assess statistical significance. With a significance level of $\alpha = 0.05$, we determined the significance 
by calculating the $p$-value. If the $p \leqslant \alpha$, we reject the null hypothesis and conclude that the differences between \tool and the other tools are statistically significant. The results ($p$-value $< 0.001$) show that \tool significantly outperforms 
MLCatchUp and Relancer in compatibility detection.

MLCatchUp assesses API compatibility solely based on API parameter definitions without adequately considering the impact of actual methods of parameter passing in the invoked APIs, 
resulting in a high number of FP. 
In contrast, Relancer 
evaluates API compatibility simply based on whether test cases can run successfully. Although this strategy effectively avoids wrongly detecting compatible test cases as incompatible ones, 
i.e., the number of FP counts to zero and achieving a precision rate of 100\%, Relancer overlooks those test cases that can run but actually have compatibility issues, leading to a significant increase in FN (i.e., 4,451 test cases).

Our tool, \tool, evaluates API compatibility by considering both the API definitions and the actual parameter-passing methods (Section~\ref{sec:pcart-assessment}). 
\tool achieves the highest TP score 
of detecting incompatible test cases, 
significantly outperforming existing tools in both the recall and F1-score metrics of 93.31\% and 96.51\%, respectively. 

The promising detection performance of \tool not only demonstrates the effectiveness of the proposed compatibility assessment (Section~\ref{sec:pcart-assessment}) but also validates the effectiveness of the automated API mapping establishment approach (Section~\ref{sec:pcart-mapping}), which is the key technique to address challenge 2 (Section~\ref{sec:background-challenges}). 
It should be noted that the API signature mappings (parameter definitions) were manually provided to MLCatchUP when performing the evaluation experiment. However, \tool 
adopts a dynamic mapping approach, which precisely and automatically obtains the signatures of APIs across different versions.

\begin{figure}[!t]
\centering
\includegraphics[width=\linewidth]{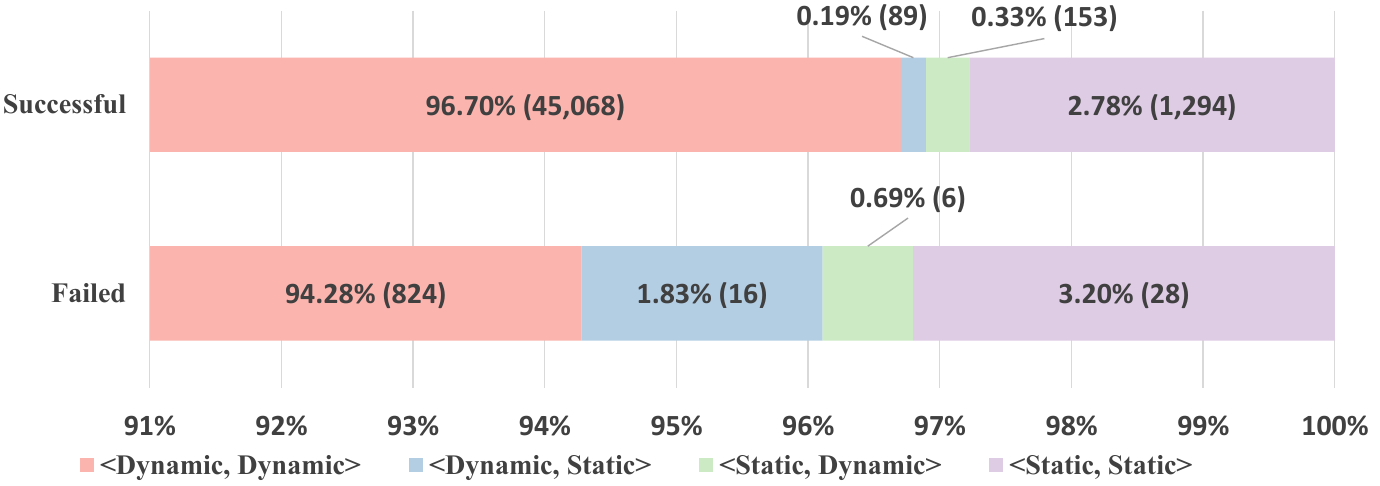}
\vspace{-4mm}
\caption{Proportion of API mapping methods employed by \tool for the current and target versions in the detection results.}
\label{mappingMethod}
\vspace{-4mm}
\end{figure}

As shown in Fig.~\ref{mappingMethod}, among the successfully detected test cases, 96.70\% utilize dynamic mapping to obtain parameter definitions in both the current and target versions, whereas this proportion is also as high as 94.28\% in the failed test cases. This implies that most API mappings are established by 
the dynamic mapping method, while only a small portion of API mappings are built through the static method. 

\begin{figure}[!t]
\centering
\begin{minipage}{3.4in}
\begin{lstlisting}[language=Python, label=kwargs, caption=Passing \mintinline{python}{**kwargs} to another API within the orignial API.]
#API Definition in Tornado 5.1.1
def fetch(self,request,callback=None,raise_error=True,**kwargs):
    ...


#API Definition in Tornado 6.0
def fetch(self,request:Union[str,'HTTPRequest'],raise_error:bool=True,**kwargs:Any):
    ...
	if not isinstance(request, HTTPRequest):
    	request = HTTPRequest(url=request, **kwargs)
    ...

import tornado.httpclient
import tornado.ioloop
async def fetch_url():
    http_client = tornado.httpclient.AsyncHTTPClient()
    response = await http_client.fetch('http://example.com', callback=None)
tornado.ioloop.IOLoop.current().run_sync(fetch_url)
\end{lstlisting}
\end{minipage}
\vspace{-4mm}
\end{figure}

Although \tool excels in accurately detecting API parameter compatibility issues, there is still room for improvement in its performance on the false negatives, with a total of 842 cases not identified correctly. 
Upon investigation, we identified four main categories of root causes that explain why \tool's parameter mapping failed for these cases. We discuss each category below, along with illustrative examples: 

\textbf{Improper Handling of Variadic Parameters.} When the invoked API in the target version contains \mintinline{python}{**kwargs} parameter, the renaming or removal of keyword arguments is supposed to be compatible, as \mintinline{python}{**kwargs} can accept a variadic number of keyword arguments. However, as the example shown in Listing~\ref{kwargs}, the \mintinline{python}{callback} parameter of the Tornado API \mintinline{python}{fetch} is removed in version 6.0. If the \mintinline{python}{callback} parameter continues to be passed in the new version, it would be automatically classified under \mintinline{python}{**kwargs}. Yet, inside the API \mintinline{python}{fetch}, the parameter \mintinline{python}{**kwargs} is passed to another API, i.e., \mintinline{python}{HTTPRequest}~\cite{HTTPRequest}, which does not include \mintinline{python}{callback} in its definitions, thus leading to a syntax error.  

\textbf{Incorrect Parameter Mapping.} In several cases, \tool established an incorrect mapping between old and new API parameters. This typically happened when parameters were renamed or removed. For example (Listing~\ref{rename and removal}), the \mintinline{python}{port} and \mintinline{python}{protocol} parameters of TensorFlow API \mintinline{python}{DispatchServer} are removed in the new version 2.4.0, and the \mintinline{python}{config} parameter is added. \tool mistakenly analyzes this change as a renaming, and because the parameter \mintinline{python}{port} is passed by position, leading to a failed detection (FN). \rev{This limitation arises because \tool primarily relies on static rules (e.g., parameter names, positions, and types) to establish mappings and lacks semantic understanding of parameter roles or meanings.}  

\textbf{Semantic Constraints Between Parameters.} Currently, \tool can not understand semantic constraints or mutual exclusivity between certain parameters. In these cases, an API call would use a combination of parameters that was valid in the older version but became invalid in the newer version, not because of a direct name change, but because the library introduced a new constraint. For example, 
in Pandas version 1.2.5, the \mintinline{python}{pandas.read_csv} API was acceptable to specify both the \mintinline{python}{named} and \mintinline{python}{prefix} parameters together, but in 1.3.0, this usage raises \textit{ValueError: Specified named and prefix; you can only specify one}. The new version enforces that only one of \mintinline{python}{named} or \mintinline{python}{prefix} can be used at a time. Currently, \tool can not catch this semantic change since its focus is on mapping parameter names, positions, and types rather than the logical interplay between parameters. This resulted in such cases being misclassified as compatible. 

\textbf{Default Value Changes.} For example, 
\mintinline{python}{dask.dataframe.} \mintinline{python}{read_parquet}: between Dask version 2022.4.1 and 2022.4.2, the default for the \mintinline{python}{split_row_groups} parameter changed from \mintinline{python}{None} (which in the older version meant an automatic or inferred behavior) to \mintinline{python}{False}. As a result, a call that explicitly passed \mintinline{python}{split_row_groups=None} would crash 
under the new version, which expects a boolean. \tool's compatibility checker did not flag this subtle change because the parameter name existed in both versions and thus appeared ``mapped'' correctly. However, the semantics of the default value changed, leading to an incorrect compatibility assessment.  

Correctly detecting parameter renaming is not trivial, especially when lacking clear semantic information, making it particularly challenging to determine whether a parameter has been renamed or deleted. Besides, due to errors in \tool's static mapping, 26 test cases in the benchmark are not correctly analyzed, whose compatibility status is labeled as unknown. 

Specifically, two primary factors contributed to the inability to determine compatibility for these cases. 
First, in some cases, static mapping yielded multiple candidate API signatures with the same name, making it difficult to identify the correct match. For example, the API \mintinline{python}{matplotlib.colorbar.Colorbar.set_ticks} in matplotlib 3.7.0 was mapped to several similarly named APIs, such as \mintinline{python}{matplotlib.axis.Axis.set_ticks}, preventing reliable compatibility assessment.  

Second, certain versions of \rev{third-party libraries} contained source code that could not be parsed due to the use of reserved keywords such as \mintinline{python}{async} in older Python versions. For example, in aiohttp version 0.17.4, the presence of the \mintinline{python}{async} keyword in \mintinline{python}{aiohttp.ClientSession}'s source file caused parsing to fail under Python 3.9, which is required by \tool's implementation. Consequently, \tool was unable to extract the necessary API signatures for analysis in such cases.

\begin{table}[!t]
\centering
\caption{\rev{Evaluation of API Compatibility Detection on Complex Invocation Patterns}}
\label{tab:new_patterns}
\scalebox{0.6}{
\begin{tabular}{|l|c|c|c|c|c|c|}
\hline
Pattern & Project & \#APIs &
\makecell{\#Covered\\APIs} &
\makecell{Current\\Version} &
\makecell{Target\\Version} &
\makecell{\#Compat.\\APIs} \\ \hline

\multirow{7}{*}{Decorator}
 & click1 & 3 & 3 & 8.0.4 & 8.1.0 & 3 \\ \cline{2-7}
 & click2 & 4 & 4 & 8.0.4 & 8.1.0 & 4 \\ \cline{2-7}
 & click3 & 3 & 2 & 8.0.4 & 8.1.0 & 2 \\ \cline{2-7}
 & flask1 & 3 & 2 & 2.2.5 & 2.3.0 & 2 \\ \cline{2-7}
 & flask2 & 25 & 24 & 2.2.5 & 2.3.0 & 24 \\ \cline{2-7}
 & tensorflow1 & 8 & 8 & 2.2.0 & 2.5.0 & 8 \\ \cline{2-7}
 & tensorflow2 & 4 & 4 & 2.7.0 & 2.9.0 & 4 \\ \hline

\multirow{2}{*}{Async/Await}
 & aiofiles1 & 4 & 4 & 22.1.0 & 23.1.0 & 3* \\ \cline{2-7}
 & aiomqtt1  & 4 & 4 & 1.2.1  & 2.0.0  & 3* \\ \hline

\multirow{3}{*}{Context Manager}
 & mysql     & 9  & 7 & 8.2.0 & 9.2.0 & 7  \\ \cline{2-7}
 & psycopg2  & 21 & 17 & 2.9.1 & 2.9.5 & 15* \\ \cline{2-7}
 & redis     & 19 & 19 & 5.2.0 & 6.0.0 & 19 \\ \hline
\end{tabular}
}
\vspace{-2mm}
\end{table}

\rev{To further evaluate \tool's capability on complex Python invocation forms, we constructed 12 representative projects covering three categories: \textit{Decorator}, \textit{Async/Await}, and \textit{Context Manager}.
As shown in Table~\ref{tab:new_patterns}, \tool successfully extracts all API invocations and accurately identifies all compatible APIs in 9 out of 12 projects. In the remaining three projects (marked with *), a few APIs (aiofiles1:1, aiomqtt1:1, psycopg2:2) have undetermined compatibility because their API signatures are unavailable in the target version. These APIs were transformed into built-in C extensions in the newer library versions, resulting in unavailable Python-level signature metadata. This situation typically occurs when previously pure-Python APIs are restructured into compiled modules. Overall, these results verify that \tool is capable of handling complex API invocation forms (e.g., decorators, asynchronous calls, and context managers), demonstrating its broad applicability across diverse Python usage patterns.}

\begin{table}[!t]
\centering
\caption{Comparison of Detecting API Parameter Compatibility Issues on the MLCatchUp Dataset}
\label{mlcatchup dataset detection}
\scalebox{0.6}{
\begin{tabular}{|l|cc|ccc|}
\hline
\multirow{2}{*}{API}                & \multicolumn{2}{c|}{\#Test Cases}               & \multicolumn{3}{c|}{Correct Compatibility Detections}                  \\ \cline{2-6} 
                                    & \multicolumn{1}{c|}{Compat.} & Incompat. & \multicolumn{1}{c|}{MLCatchUp} & \multicolumn{1}{c|}{Relancer} & \tool \\ \hline
sklearn.cluster.Kmeans              & \multicolumn{1}{c|}{0}          & 8            & \multicolumn{1}{c|}{8}         & \multicolumn{1}{c|}{8}        & 8     \\ \hline
sklearn.tree.DecisionTreeClassifier & \multicolumn{1}{c|}{0}          & 5            & \multicolumn{1}{c|}{5}         & \multicolumn{1}{c|}{5}        & 5     \\ \hline
sklearn.tree.DecisionTreeRegressor  & \multicolumn{1}{c|}{0}          & 3            & \multicolumn{1}{c|}{3}         & \multicolumn{1}{c|}{3}        & 3     \\ \hline
tensorflow.compat.v1.string\_split   & \multicolumn{1}{c|}{4}          & 0            & \multicolumn{1}{c|}{4}         & \multicolumn{1}{c|}{4}        & 4     \\ \hline
\end{tabular}
}
\vspace{-4mm}
\end{table}

Moreover, Table \ref{mlcatchup dataset detection} shows the number of compatible and incompatible test cases for each API, along with the number of cases where each method accurately detected compatibility. All three methods, i.e., MLCatchUp, Relancer, and \tool, successfully identify the compatibility issues in every test case from the MLCatchUp dataset.

\subsection{RQ2: How does PCART Perform in Repairing API Parameter Compatibility Issues?}\label{sec:results-rq2}

\begin{table}[!t]
\caption{Comparison of Repairing API Parameter Compatibility Issues}
\centering
\label{cmp repair}
\scalebox{0.55}{
\begin{tabular}{|lc|ccc|ccc|cccccc|}
\hline
\multicolumn{1}{|l|}{\multirow{3}{*}{Library}} & \multirow{3}{*}{APIs} & \multicolumn{3}{c|}{MLCatchUp}                                                   & \multicolumn{3}{c|}{Relancer}                                                      & \multicolumn{6}{c|}{\tool}                                                                                                                                                                 \\ \cline{3-14} 
\multicolumn{1}{|l|}{}                         &                       & \multicolumn{1}{c|}{\multirow{2}{*}{SR}} & \multicolumn{2}{c|}{UR}               & \multicolumn{1}{c|}{\multirow{2}{*}{SR}} & \multicolumn{2}{c|}{UR}                 & \multicolumn{1}{c|}{\multirow{2}{*}{SR}} & \multicolumn{2}{c|}{UR}                            & \multicolumn{1}{c|}{\multirow{2}{*}{SR*}} & \multicolumn{2}{c|}{UR*}                       \\ \cline{4-5} \cline{7-8} \cline{10-11} \cline{13-14} 
\multicolumn{1}{|l|}{}                         &                       & \multicolumn{1}{c|}{}                    & \multicolumn{1}{c|}{ICP}        & CP  & \multicolumn{1}{c|}{}                    & \multicolumn{1}{c|}{ICP}   & CP         & \multicolumn{1}{c|}{}                    & \multicolumn{1}{c|}{ICP} & \multicolumn{1}{c|}{CP} & \multicolumn{1}{c|}{}                     & \multicolumn{1}{c|}{ICP}          & CP         \\ \hline
\multicolumn{1}{|l|}{PyTorch}                  & 4                     & \multicolumn{1}{c|}{0}                   & \multicolumn{1}{c|}{7}          & 0   & \multicolumn{1}{c|}{0}                   & \multicolumn{1}{c|}{7}     & 0          & \multicolumn{1}{c|}{7}                   & \multicolumn{1}{c|}{0}   & \multicolumn{1}{c|}{0}  & \multicolumn{1}{c|}{\textbf{7}}           & \multicolumn{1}{c|}{\textbf{0}}   & 0          \\ \hline
\multicolumn{1}{|l|}{SciPy}                    & 193                   & \multicolumn{1}{c|}{0}                   & \multicolumn{1}{c|}{438}        & 23  & \multicolumn{1}{c|}{0}                   & \multicolumn{1}{c|}{438}   & \textbf{0} & \multicolumn{1}{c|}{415}                 & \multicolumn{1}{c|}{0}   & \multicolumn{1}{c|}{6}  & \multicolumn{1}{c|}{\textbf{415}}         & \multicolumn{1}{c|}{\textbf{23}}  & 6          \\ \hline
\multicolumn{1}{|l|}{Gensim}                   & 13                    & \multicolumn{1}{c|}{255}                 & \multicolumn{1}{c|}{199}        & 0   & \multicolumn{1}{c|}{0}                   & \multicolumn{1}{c|}{454}   & 0          & \multicolumn{1}{c|}{415}                 & \multicolumn{1}{c|}{13}  & \multicolumn{1}{c|}{0}  & \multicolumn{1}{c|}{\textbf{398}}         & \multicolumn{1}{c|}{\textbf{56}}  & 0          \\ \hline
\multicolumn{1}{|l|}{Tensorflow}               & 19                    & \multicolumn{1}{c|}{0}                   & \multicolumn{1}{c|}{57}         & 47  & \multicolumn{1}{c|}{0}                   & \multicolumn{1}{c|}{57}    & \textbf{0} & \multicolumn{1}{c|}{53}                  & \multicolumn{1}{c|}{1}   & \multicolumn{1}{c|}{0}  & \multicolumn{1}{c|}{\textbf{45}}          & \multicolumn{1}{c|}{\textbf{12}}   & \textbf{0} \\ \hline
\multicolumn{1}{|l|}{Tornado}                  & 20                    & \multicolumn{1}{c|}{0}                   & \multicolumn{1}{c|}{290}        & 0   & \multicolumn{1}{c|}{0}                   & \multicolumn{1}{c|}{290}   & 0          & \multicolumn{1}{c|}{261}                 & \multicolumn{1}{c|}{4}   & \multicolumn{1}{c|}{0}  & \multicolumn{1}{c|}{\textbf{286}}         & \multicolumn{1}{c|}{\textbf{4}}   & 0          \\ \hline
\multicolumn{1}{|l|}{Transformers}             & 1                     & \multicolumn{1}{c|}{0}                   & \multicolumn{1}{c|}{0}          & 0  & \multicolumn{1}{c|}{0}                   & \multicolumn{1}{c|}{0}     & \textbf{0} & \multicolumn{1}{c|}{0}                   & \multicolumn{1}{c|}{0}   & \multicolumn{1}{c|}{0}  & \multicolumn{1}{c|}{0}                    & \multicolumn{1}{c|}{0}            & \textbf{0} \\ \hline
\multicolumn{1}{|l|}{Requests}                 & 2                     & \multicolumn{1}{c|}{0}                   & \multicolumn{1}{c|}{6}          & 0   & \multicolumn{1}{c|}{0}                   & \multicolumn{1}{c|}{6}     & 0          & \multicolumn{1}{c|}{6}                   & \multicolumn{1}{c|}{0}   & \multicolumn{1}{c|}{0}  & \multicolumn{1}{c|}{\textbf{6}}           & \multicolumn{1}{c|}{\textbf{0}}   & 0          \\ \hline
\multicolumn{1}{|l|}{Matplotlib}               & 21                    & \multicolumn{1}{c|}{328}                 & \multicolumn{1}{c|}{166}        & 0   & \multicolumn{1}{c|}{0}                   & \multicolumn{1}{c|}{494}   & 0          & \multicolumn{1}{c|}{444}                 & \multicolumn{1}{c|}{35}  & \multicolumn{1}{c|}{0}  & \multicolumn{1}{c|}{\textbf{448}}         & \multicolumn{1}{c|}{\textbf{46}}  & 0          \\ \hline
\multicolumn{1}{|l|}{FastAPI}                  & 4                     & \multicolumn{1}{c|}{0}                   & \multicolumn{1}{c|}{51}         & 0   & \multicolumn{1}{c|}{0}                   & \multicolumn{1}{c|}{51}    & 0          & \multicolumn{1}{c|}{0}                   & \multicolumn{1}{c|}{0}   & \multicolumn{1}{c|}{0}  & \multicolumn{1}{c|}{\textbf{51}}          & \multicolumn{1}{c|}{\textbf{0}}   & 0          \\ \hline
\multicolumn{1}{|l|}{NumPy}                    & 85                    & \multicolumn{1}{c|}{10}                  & \multicolumn{1}{c|}{350}        & 20  & \multicolumn{1}{c|}{0}                   & \multicolumn{1}{c|}{360}   & \textbf{0} & \multicolumn{1}{c|}{57}                  & \multicolumn{1}{c|}{0}   & \multicolumn{1}{c|}{0}  & \multicolumn{1}{c|}{\textbf{360}}         & \multicolumn{1}{c|}{\textbf{0}}   & \textbf{0} \\ \hline
\multicolumn{1}{|l|}{Pydantic}                 & 1                     & \multicolumn{1}{c|}{0}                   & \multicolumn{1}{c|}{8}          & 0   & \multicolumn{1}{c|}{0}                   & \multicolumn{1}{c|}{8}     & 0          & \multicolumn{1}{c|}{0}                   & \multicolumn{1}{c|}{4}   & \multicolumn{1}{c|}{0}  & \multicolumn{1}{c|}{0}                    & \multicolumn{1}{c|}{8}            & 0          \\ \hline
\multicolumn{1}{|l|}{Redis}                    & 2                     & \multicolumn{1}{c|}{0}                   & \multicolumn{1}{c|}{0}          & 0   & \multicolumn{1}{c|}{0}                   & \multicolumn{1}{c|}{0}     & 0          & \multicolumn{1}{c|}{0}                   & \multicolumn{1}{c|}{0}   & \multicolumn{1}{c|}{0}  & \multicolumn{1}{c|}{0}                    & \multicolumn{1}{c|}{0}            & 0          \\ \hline
\multicolumn{1}{|l|}{Faker}                    & 8                     & \multicolumn{1}{c|}{0}                   & \multicolumn{1}{c|}{0}          & 0   & \multicolumn{1}{c|}{0}                   & \multicolumn{1}{c|}{0}     & 0          & \multicolumn{1}{c|}{0}                   & \multicolumn{1}{c|}{0}   & \multicolumn{1}{c|}{0}  & \multicolumn{1}{c|}{0}                    & \multicolumn{1}{c|}{0}            & 0          \\ \hline
\multicolumn{1}{|l|}{LightGBM}                 & 1                     & \multicolumn{1}{c|}{0}                   & \multicolumn{1}{c|}{0}          & 0   & \multicolumn{1}{c|}{0}                   & \multicolumn{1}{c|}{0}     & 0          & \multicolumn{1}{c|}{0}                   & \multicolumn{1}{c|}{0}   & \multicolumn{1}{c|}{0}  & \multicolumn{1}{c|}{0}                    & \multicolumn{1}{c|}{0}            & 0          \\ \hline
\multicolumn{1}{|l|}{Loguru}                   & 5                     & \multicolumn{1}{c|}{\textbf{3}}          & \multicolumn{1}{c|}{\textbf{0}} & 0   & \multicolumn{1}{c|}{0}                   & \multicolumn{1}{c|}{3}     & 0          & \multicolumn{1}{c|}{3}                   & \multicolumn{1}{c|}{0}   & \multicolumn{1}{c|}{0}  & \multicolumn{1}{c|}{\textbf{3}}           & \multicolumn{1}{c|}{\textbf{0}}   & 0          \\ \hline
\multicolumn{1}{|l|}{SymPy}                    & 15                    & \multicolumn{1}{c|}{2}                   & \multicolumn{1}{c|}{62}         & 0   & \multicolumn{1}{c|}{0}                   & \multicolumn{1}{c|}{64}    & 0          & \multicolumn{1}{c|}{43}                  & \multicolumn{1}{c|}{1}   & \multicolumn{1}{c|}{0}  & \multicolumn{1}{c|}{\textbf{63}}          & \multicolumn{1}{c|}{\textbf{1}}   & 0          \\ \hline
\multicolumn{1}{|l|}{scikit-learn}             & 117                   & \multicolumn{1}{c|}{411}                 & \multicolumn{1}{c|}{284}        & 623 & \multicolumn{1}{c|}{0}                   & \multicolumn{1}{c|}{695}   & \textbf{0} & \multicolumn{1}{c|}{692}                 & \multicolumn{1}{c|}{0}   & \multicolumn{1}{c|}{0}  & \multicolumn{1}{c|}{\textbf{692}}         & \multicolumn{1}{c|}{\textbf{3}}   & \textbf{0} \\ \hline
\multicolumn{1}{|l|}{Flask}                    & 2                     & \multicolumn{1}{c|}{\textbf{2}}          & \multicolumn{1}{c|}{\textbf{0}} & 0   & \multicolumn{1}{c|}{0}                   & \multicolumn{1}{c|}{2}     & 0          & \multicolumn{1}{c|}{0}                   & \multicolumn{1}{c|}{2}   & \multicolumn{1}{c|}{0}  & \multicolumn{1}{c|}{0}                    & \multicolumn{1}{c|}{2}            & 0          \\ \hline
\multicolumn{1}{|l|}{Click}                    & 4                     & \multicolumn{1}{c|}{0}                   & \multicolumn{1}{c|}{3}          & 0   & \multicolumn{1}{c|}{0}                   & \multicolumn{1}{c|}{3}     & 0          & \multicolumn{1}{c|}{3}                   & \multicolumn{1}{c|}{0}   & \multicolumn{1}{c|}{0}  & \multicolumn{1}{c|}{\textbf{3}}           & \multicolumn{1}{c|}{\textbf{0}}   & 0          \\ \hline
\multicolumn{1}{|l|}{aiohttp}                  & 12                    & \multicolumn{1}{c|}{\textbf{14}}         & \multicolumn{1}{c|}{\textbf{2}} & 0  & \multicolumn{1}{c|}{0}                   & \multicolumn{1}{c|}{16}    & 0 & \multicolumn{1}{c|}{1}                   & \multicolumn{1}{c|}{0}   & \multicolumn{1}{c|}{0}  & \multicolumn{1}{c|}{1}                    & \multicolumn{1}{c|}{15}           & 0 \\ \hline
\multicolumn{1}{|l|}{spaCy}                    & 2                     & \multicolumn{1}{c|}{0}                   & \multicolumn{1}{c|}{6}          & 0   & \multicolumn{1}{c|}{0}                   & \multicolumn{1}{c|}{6}     & 0          & \multicolumn{1}{c|}{6}                   & \multicolumn{1}{c|}{0}   & \multicolumn{1}{c|}{0}  & \multicolumn{1}{c|}{\textbf{6}}           & \multicolumn{1}{c|}{\textbf{0}}   & 0          \\ \hline
\multicolumn{1}{|l|}{Keras}                    & 20                    & \multicolumn{1}{c|}{0}                   & \multicolumn{1}{c|}{100}        & 26  & \multicolumn{1}{c|}{0}                   & \multicolumn{1}{c|}{100}   & \textbf{0} & \multicolumn{1}{c|}{100}                 & \multicolumn{1}{c|}{0}   & \multicolumn{1}{c|}{0}  & \multicolumn{1}{c|}{\textbf{100}}         & \multicolumn{1}{c|}{\textbf{0}}   & \textbf{0} \\ \hline
\multicolumn{1}{|l|}{HTTPX}                    & 8                     & \multicolumn{1}{c|}{48}                  & \multicolumn{1}{c|}{17}         & 54  & \multicolumn{1}{c|}{0}                   & \multicolumn{1}{c|}{65}    & \textbf{0} & \multicolumn{1}{c|}{57}                  & \multicolumn{1}{c|}{0}   & \multicolumn{1}{c|}{0}  & \multicolumn{1}{c|}{\textbf{65}}          & \multicolumn{1}{c|}{\textbf{0}}   & \textbf{0} \\ \hline
\multicolumn{1}{|l|}{NetworkX}                 & 49                    & \multicolumn{1}{c|}{47}                  & \multicolumn{1}{c|}{98}         & 0   & \multicolumn{1}{c|}{0}                   & \multicolumn{1}{c|}{145}   & 0          & \multicolumn{1}{c|}{123}                 & \multicolumn{1}{c|}{3}   & \multicolumn{1}{c|}{0}  & \multicolumn{1}{c|}{\textbf{118}}         & \multicolumn{1}{c|}{\textbf{27}}  & 0          \\ \hline
\multicolumn{1}{|l|}{XGBoost}                  & 1                     & \multicolumn{1}{c|}{0}                   & \multicolumn{1}{c|}{5}          & 0   & \multicolumn{1}{c|}{0}                   & \multicolumn{1}{c|}{5}     & 0          & \multicolumn{1}{c|}{5}                   & \multicolumn{1}{c|}{0}   & \multicolumn{1}{c|}{0}  & \multicolumn{1}{c|}{\textbf{5}}           & \multicolumn{1}{c|}{\textbf{0}}   & 0          \\ \hline
\multicolumn{1}{|l|}{Plotly}                   & 52                    & \multicolumn{1}{c|}{0}                   & \multicolumn{1}{c|}{6,320}       & 358 & \multicolumn{1}{c|}{0}                   & \multicolumn{1}{c|}{6,320} & \textbf{0} & \multicolumn{1}{c|}{6,320}                & \multicolumn{1}{c|}{0}   & \multicolumn{1}{c|}{0}  & \multicolumn{1}{c|}{\textbf{6,320}}       & \multicolumn{1}{c|}{\textbf{0}}   & \textbf{0} \\ \hline
\multicolumn{1}{|l|}{Django}                   & 3                     & \multicolumn{1}{c|}{3}                   & \multicolumn{1}{c|}{4}          & 0   & \multicolumn{1}{c|}{0}                   & \multicolumn{1}{c|}{7}     & 0          & \multicolumn{1}{c|}{7}                   & \multicolumn{1}{c|}{0}   & \multicolumn{1}{c|}{0}  & \multicolumn{1}{c|}{\textbf{7}}           & \multicolumn{1}{c|}{\textbf{0}}   & 0          \\ \hline
\multicolumn{1}{|l|}{Pillow}                   & 26                    & \multicolumn{1}{c|}{0}                   & \multicolumn{1}{c|}{7}          & 0   & \multicolumn{1}{c|}{0}                   & \multicolumn{1}{c|}{7}     & 0          & \multicolumn{1}{c|}{7}                   & \multicolumn{1}{c|}{0}   & \multicolumn{1}{c|}{0}  & \multicolumn{1}{c|}{\textbf{7}}           & \multicolumn{1}{c|}{\textbf{0}}   & 0          \\ \hline
\multicolumn{1}{|l|}{JAX}                      & 1                     & \multicolumn{1}{c|}{0}                   & \multicolumn{1}{c|}{12}         & 16  & \multicolumn{1}{c|}{0}                   & \multicolumn{1}{c|}{12}    & \textbf{0} & \multicolumn{1}{c|}{12}                  & \multicolumn{1}{c|}{0}   & \multicolumn{1}{c|}{0}  & \multicolumn{1}{c|}{\textbf{12}}          & \multicolumn{1}{c|}{\textbf{0}}   & \textbf{0} \\ \hline
\multicolumn{1}{|l|}{Polars}                   & 30                    & \multicolumn{1}{c|}{93}                  & \multicolumn{1}{c|}{44}         & 0   & \multicolumn{1}{c|}{0}                   & \multicolumn{1}{c|}{137}   & 0          & \multicolumn{1}{c|}{117}                 & \multicolumn{1}{c|}{0}   & \multicolumn{1}{c|}{0}  & \multicolumn{1}{c|}{\textbf{137}}         & \multicolumn{1}{c|}{\textbf{0}}   & 0          \\ \hline
\multicolumn{1}{|l|}{pandas}                   & 73                    & \multicolumn{1}{c|}{0}                   & \multicolumn{1}{c|}{2601}       & 0   & \multicolumn{1}{c|}{0}                   & \multicolumn{1}{c|}{2,601} & 0          & \multicolumn{1}{c|}{1773}                & \multicolumn{1}{c|}{740} & \multicolumn{1}{c|}{0}  & \multicolumn{1}{c|}{\textbf{1,832}}       & \multicolumn{1}{c|}{\textbf{769}} & 0          \\ \hline
\multicolumn{1}{|l|}{Rich}                     & 33                    & \multicolumn{1}{c|}{11}                  & \multicolumn{1}{c|}{170}        & 164 & \multicolumn{1}{c|}{0}                   & \multicolumn{1}{c|}{181}   & \textbf{0} & \multicolumn{1}{c|}{177}                 & \multicolumn{1}{c|}{0}   & \multicolumn{1}{c|}{0}  & \multicolumn{1}{c|}{\textbf{181}}         & \multicolumn{1}{c|}{\textbf{0}}   & \textbf{0} \\ \hline
\multicolumn{1}{|l|}{Dask}                     & 17                    & \multicolumn{1}{c|}{0}                   & \multicolumn{1}{c|}{47}         & 0   & \multicolumn{1}{c|}{0}                   & \multicolumn{1}{c|}{47}    & 0          & \multicolumn{1}{c|}{8}                   & \multicolumn{1}{c|}{39}  & \multicolumn{1}{c|}{0}  & \multicolumn{1}{c|}{\textbf{8}}           & \multicolumn{1}{c|}{\textbf{39}}  & 0          \\ \hline
\multicolumn{1}{|l|}{Total}                    & 844                   & \multicolumn{1}{c|}{1,227}               & \multicolumn{2}{c|}{12,685}           & \multicolumn{1}{c|}{0}                   & \multicolumn{2}{c|}{12,581}             & \multicolumn{1}{c|}{11,112}              & \multicolumn{2}{c|}{848}                           & \multicolumn{1}{c|}{\textbf{11,576}}      & \multicolumn{2}{c|}{\textbf{1,011}}              \\ \hline
\multicolumn{2}{|l|}{Repair Precision}                                 & \multicolumn{3}{c|}{8.82\%}                                                      & \multicolumn{3}{c|}{0.00\%}                                                        & \multicolumn{3}{c|}{92.91\%}                                                                  & \multicolumn{3}{c|}{\textbf{91.97\%}}                                                      \\ \hline
\end{tabular}%
}
\vspace{-4mm}
\end{table}

Table~\ref{cmp repair} presents the comparison results of MLCatchUp~\cite{haryono2021mlcatchup}, Relancer~\cite{zhu2021restoring}, and \tool in repairing API parameter compatibility issues on \benchmark. Details of successful and failed repairs across different libraries are given in the table. Relancer fails in all attempted repairs, achieving a repair precision of 0\%. In contrast, MLCatchUp achieves an overall precision of 8.82\%. 
Notably, \tool exhibits exceptional performance with a precision of 91.97\%, effectively fixing the majority of incompatible test cases. We constructed a contingency table with successful repairs (SR) and repair failures (UR, including ICP and CP) and used the Chi-square test to compare \tool with MLCatchUp and Relancer. With the same significance level ($\alpha = 0.05$), we calculated the $p$-value and found that \tool also significantly outperforms the comparison tools ($p$-value $< 0.001$). 

The limited repair precision of MLCatchUp is primarily due to its constraints in handling repair operations. 
MLCatchUP does not support the repair of removal and reordering of positional parameters and is only capable of recognizing simple API calls within user code. It fails to address more complex invocation scenarios such as \mintinline{python}{a().b()} where an API call is made through another API's return value, or \mintinline{python}{a(b())} where an API call is made using another API as an argument. 

For example, the TensorFlow API \mintinline{python}{random} in Listing~\ref{a.b} has a new parameter (i.e., \mintinline{python}{rerandomize_each_iteration}) in the target version, but MLCatchUp incorrectly recognizes \mintinline{python}{random}, and thus applied the new parameter to the API \mintinline{python}{take}, leading to a repair failure. Similarly, as shown in Listing~\ref{a(b())}, the TensorFlow API \mintinline{python}{Embedding} has a new parameter \mintinline{python}{sparse} in the target version, but MLCatchUp mistakenly applies this parameter to the API \mintinline{python}{add}, resulting in a repair failure, too.

\begin{figure}[!t]
\centering
\begin{minipage}{3.35in}
\begin{lstlisting}[language=Python, label=a.b, caption=Calling with a function's return value.]
import tensorflow as tf
#Before MLCatchUp repair
ds1 = tf.data.Dataset.random().take(10)
#After MLCatchUp Repair
ds1 = tf.data.Dataset.random().take(10, rerandomize_each_iteration=None)
\end{lstlisting}
\end{minipage}
\vspace{-4mm}
\end{figure}

\begin{figure}[!t]
\centering
\begin{minipage}{3.35in}
\begin{lstlisting}[language=Python, label=a(b()), caption=Calling with a function's parameter.]
import tensorflow as tf
model = tf.keras.Sequential()
#Before MLCatchUp repair
model.add(tf.keras.layers.Embedding(1000,  64))
#After MLCatchUp repair
model.add(tf.keras.layers.Embedding(1000, 64), sparse=False)
\end{lstlisting}
\end{minipage}
\vspace{-4mm}
\end{figure}

Relancer supports modifications to parameter names and values, but since its repair knowledge base is built upon a predefined dataset extracted from GitHub and API documentation, its repair strategies and capabilities are limited to the known API deprecation patterns. Thus, when faced with new or unrecorded code snippets such as those in \benchmark, Relancer fails to generate effective repair solutions. 

\tool's repair operations are deduced in real-time based on API parameter definitions 
and the actual parameter-passing methods of the invoked APIs, 
without the need for a pre-established repair knowledge base. Hence, it has a broader applicability and a higher repair precision than existing tools. Among the 12,581 incompatible test cases in \benchmark, 11,112 are confirmed as successfully repaired through automated validation, while 842 failed, and 627 test cases remained unknown. The repair precision in the automated validation phase reaches 92.91\%. The remaining 627 test cases with unknown repair status are primarily due to failures in pickle file creation or loading, which prevent automated validation. For these unknown cases, manual confirmation later changes the repair precision to 91.97\%, noted by ``*'' in the last two columns of Table~\ref{cmp repair}.

For the repair failures, we manually analyzed the repair reports by first checking whether the compatibility detection was correct. If the detection was correct, we then manually inspected the repaired results to identify the errors. On the other hand, if the compatibility detection was incorrect, we manually compared the API parameters before and after the version update to determine the cause of the detection failure.

Regarding the 842 incompatible test cases that failed to repair, 
we find that they are due to incorrect parameter mapping relations, which result in failure detection, i.e., detecting incompatible test cases as compatible (Table~\ref{compatibility detection}). Therefore, these mistakenly detected cases would not undergo a repair. 

For the remaining 627 test cases with unknown repair status, among which 122 cases are repair failures, the main reasons for repair failure are due to parameter default values and improper handling of \mintinline{python}{**kwargs}. For instance, in Listing~\ref{default value}, the parameters \mintinline{python}{include_start} and \mintinline{python}{include_end} of the Pandas API \mintinline{python}{between_time} are removed in version 2.0.0. 
Although \tool properly removes the deprecated \mintinline{python}{include_start} and \mintinline{python}{include_end} parameters, the repair is incomplete, i.e., the change of \mintinline{python}{inclusive} parameter's default value to \mintinline{python}{both} causes \textit{ValueError: Inclusive has to be either `both', `neither', `left' or `right'}.  
\tool technically supports the modification of parameter values, but it does not proceed with this step because it is uncertain whether the default values in the new version would align with users' intentions. 

\begin{figure}[!t]
\centering
\begin{minipage}{3.4in}
\begin{lstlisting}[language=Python, label=default value, caption=Change in the default value of a parameter.]
#API Definition in pandas1.2
def between_time(start_time, end_time, include_start: 'bool_t | lib.NoDefault' = <no_default>, include_end: 'bool_t | lib.NoDefault' = <no_default>, inclusive: 'IntervalClosedType | None' = None, axis=None) 

#API Definition in pandas2.0.0
def between_time(start_time, end_time, inclusive: 'IntervalClosedType' = 'both', axis: 'Axis | None' = None) 

import pandas as pd
i = pd.date_range('2018-04-09', periods=4, freq='1D20min')
ts = pd.DataFrame({'A': [1, 2, 3, 4]}, index=i)
#Before repair
ts.between_time('0:15',  '0:45',  True,  True,  None) 
#After repair
ts.between_time('0:15',  '0:45',  None) 
\end{lstlisting}
\end{minipage}
\vspace{-4mm}
\end{figure}



\begin{table}[!t]
\centering
\caption{Comparison of Repairing API Parameter Compatibility Issues on the MLCatchUp Dataset}
\label{mlcatchup dataset repair}
\scalebox{0.6}{
\begin{tabular}{|l|cc|ccc|}
\hline
\multirow{2}{*}{API}                & \multicolumn{2}{c|}{\#Test Cases}               & \multicolumn{3}{c|}{Correct Compatibility Repairs}                     \\ \cline{2-6} 
                                    & \multicolumn{1}{c|}{Compat.} & Incompat. & \multicolumn{1}{c|}{MLCatchUp} & \multicolumn{1}{c|}{Relancer} & \tool \\ \hline
sklearn.cluster.Kmeans              & \multicolumn{1}{c|}{0}          & 8            & \multicolumn{1}{c|}{7}         & \multicolumn{1}{c|}{0}        & 8     \\ \hline
sklearn.tree.DecisionTreeClassifier & \multicolumn{1}{c|}{0}          & 5            & \multicolumn{1}{c|}{4}         & \multicolumn{1}{c|}{0}        & 5     \\ \hline
sklearn.tree.DecisionTreeRegressor  & \multicolumn{1}{c|}{0}          & 3            & \multicolumn{1}{c|}{3}         & \multicolumn{1}{c|}{0}        & 3     \\ \hline
tensorflow.compat.v1.string\_split   & \multicolumn{1}{c|}{4}          & 0            & \multicolumn{1}{c|}{0}         & \multicolumn{1}{c|}{0}        & 0     \\ \hline
\end{tabular}
}
\vspace{-4mm}
\end{table}

According to Table \ref{mlcatchup dataset repair}, the MLCatchUp dataset contains 16 incompatible test cases. MLCatchUp successfully repairs 14 of these, while two fail due to its inability to 
correctly identify APIs with breaking changes, particularly those with complex invocation forms. Relancer, relying on pre-constructed repair strategies, cannot repair all incompatible test cases when encountering new code snippets in the MLCatchUp dataset. In contrast, \tool performs exceptionally well, accurately identifying and successfully repairing all incompatible APIs.

\subsection{RQ3: What is the Effectiveness of PCART in Real-world Python Projects?}\label{sec:results-rq3}

The evaluation of \tool on the collected 30 real-world projects is presented in Table~\ref{project test}. 
After manual confirmation, \tool correctly identifies all the target library APIs invoked in each project and whether they are covered during execution. 
In these projects, \tool successfully detects all API parameter compatibility issues, as listed in the TP (incompatible) and TN (compatible) columns of Table~\ref{project test}. \tool further repairs the detected compatibility issues in 25 projects via automated validation, while providing repairs for the remaining five projects (noted by ``*''), in which 
it could not complete the automated validation due to the failure of loading 
pickle files. For these projects, the manual validation confirms that \tool's repairs are all correct. Notably, two projects, ``polars-book-cn'' and ``EJPLab\_Computational'', can run but actually have underlying compatibility issues. \tool not only correctly detects these issues but also successfully repairs them. The evaluation demonstrates that \tool has good practicality in detecting and repairing API parameter compatibility issues in practical Python project development.


\begin{table}[!t]
\centering
\caption{Evaluation of \tool on Real-world GitHub Python Projects}
\label{project test}
\scalebox{0.625}{
\begin{tabular}{|l|c|c|cc|c|}
\hline
\multirow{2}{*}{Project} & \multirow{2}{*}{\#APIs} & \multirow{2}{*}{\#Covered APIs} & \multicolumn{2}{c|}{Detection} & Repair \\ \cline{4-6} 
                         &                        &                                & \multicolumn{1}{c|}{TP}  & TN  & SR     \\ \hline
allnews                  & 4                      & 4                              & \multicolumn{1}{c|}{1}   & 3   & 1      \\ \hline
Youtube-Comedy           & 1                      & 1                              & \multicolumn{1}{c|}{1}   & 0   & 1      \\ \hline
recommendation-engine    & 21                     & 21                             & \multicolumn{1}{c|}{3}   & 18  & 3*     \\ \hline
political-polarisation   & 8                      & 8                              & \multicolumn{1}{c|}{1}   & 7   & 1      \\ \hline
TSP                      & 7                      & 7                              & \multicolumn{1}{c|}{1}   & 6   & 1      \\ \hline
CustomSamplers           & 1                      & 1                              & \multicolumn{1}{c|}{1}   & 0   & 1      \\ \hline
machine-learning         & 17                     & 17                             & \multicolumn{1}{c|}{1}   & 16  & 1      \\ \hline
gistable                 & 2                      & 2                              & \multicolumn{1}{c|}{1}   & 1   & 1      \\ \hline
galaxiesDataScience      & 16                     & 16                             & \multicolumn{1}{c|}{2}   & 14  & 2     \\ \hline
fuel\_forecast\_explorer   & 16                     & 16                             & \multicolumn{1}{c|}{4}   & 12   & 4*     \\ \hline
sg-restart-regridder     & 3                      & 3                              & \multicolumn{1}{c|}{1}   & 2   & 1      \\ \hline
MAHE\_OD\_DATASET          & 5                      & 5                              & \multicolumn{1}{c|}{1}   & 4   & 1*      \\ \hline
hfhd                     & 25                     & 20                             & \multicolumn{1}{c|}{5}   & 15  & 5      \\ \hline
scrapping-jojo-main      & 3                      & 3                              & \multicolumn{1}{c|}{1}   & 2   & 1      \\ \hline
Contrucao-de             & 4                      & 4                              & \multicolumn{1}{c|}{1}   & 3   & 1      \\ \hline
polars-book-cn           & 10                     & 10                             & \multicolumn{1}{c|}{1}   & 9   & 1      \\ \hline
EJPLab\_Computational     & 12                     & 12                             & \multicolumn{1}{c|}{1}   & 11  & 1      \\ \hline
Deep-Graph-Kernels       & 21                     & 21                             & \multicolumn{1}{c|}{1}   & 20  & 1      \\ \hline
AIBO                     & 4                      & 1                              & \multicolumn{1}{c|}{1}   & 0   & 1      \\ \hline
greenbenchmark           & 3                      & 3                              & \multicolumn{1}{c|}{1}   & 2   & 1      \\ \hline
qho-control              & 5                      & 4                              & \multicolumn{1}{c|}{2}   & 2   & 2      \\ \hline
giantpopflucts           & 3                      & 1                              & \multicolumn{1}{c|}{1}   & 0   & 1*     \\ \hline
django-selenium-testing  & 7                      & 3                              & \multicolumn{1}{c|}{1}   & 2   & 1*     \\ \hline
polire                   & 6                      & 5                              & \multicolumn{1}{c|}{1}   & 4   & 1      \\ \hline
Python-Workshop          & 3                      & 3                              & \multicolumn{1}{c|}{3}   & 0   & 3      \\ \hline
covid19-predictor        & 35                     & 35                             & \multicolumn{1}{c|}{1}   & 34  & 1      \\ \hline
Final                    & 109                    & 52                             & \multicolumn{1}{c|}{2}   & 50  & 2      \\ \hline
Gender-pay-gap           & 8                      & 8                              & \multicolumn{1}{c|}{1}   & 7   & 1      \\ \hline
SDOML                    & 2                      & 1                              & \multicolumn{1}{c|}{1}   & 0   & 1      \\ \hline
simulations              & 26                     & 26                             & \multicolumn{1}{c|}{1}   & 25  & 1      \\ \hline
\end{tabular}
}
\vspace{-4mm}
\end{table}

\subsection{RQ4: How does PCART Compare to ChatGPT in Detecting and Repairing API Parameter Compatibility Issues?}\label{sec:results-rq4}

Table~\ref{chatgpt test} and Table~\ref{chatgpt repair} compare the detection and repair performance
between ChatGPT (GPT-4o) and \tool on the 58 test cases, i.e., 29 compatible and 29 incompatible test cases, randomly selected in \benchmark.

\begin{table}[!t]
\centering
\caption{Comparison of \tool and ChatGPT (GPT-4o) in Detecting API Parameter Compatibility Issues}
\label{chatgpt test}
\scalebox{0.625}{
\begin{tabular}{|ccc|ccc|ccc|}
\hline
\multicolumn{3}{|c|}{\makecell{ChatGPT (GPT-4o) \\ W./O. Definition}} 
& \multicolumn{3}{c|}{\makecell{ChatGPT (GPT-4o) \\ W. Definition}}      
& \multicolumn{3}{c|}{\tool} \\ \hline
\multicolumn{1}{|c|}{TP} & \multicolumn{1}{c|}{FP} & FN & \multicolumn{1}{c|}{TP} & \multicolumn{1}{c|}{FP} & FN & \multicolumn{1}{c|}{TP} & \multicolumn{1}{c|}{FP} & FN \\ \hline
\multicolumn{1}{|c|}{27} & \multicolumn{1}{c|}{25} & 2  & \multicolumn{1}{c|}{28} & \multicolumn{1}{c|}{23} & 1  & \multicolumn{1}{c|}{25} & \multicolumn{1}{c|}{1}  & 4  \\ \hline
\multicolumn{3}{|c|}{Precision: 51.92\%}       & \multicolumn{3}{c|}{Precision: 54.90\%}       & \multicolumn{3}{c|}{Precision: \textbf{96.15\%}}       \\ \hline
\multicolumn{3}{|c|}{Recall: 93.10\%}          & \multicolumn{3}{c|}{Recall: \textbf{96.55\%}}          & \multicolumn{3}{c|}{Recall: 86.21\%}          \\ \hline
\multicolumn{3}{|c|}{F1-score: 66.67\%}       & \multicolumn{3}{c|}{F1-score: 70.0\%}        & \multicolumn{3}{c|}{F1-score: \textbf{90.91\%}}       \\ \hline
\end{tabular}
}
\vspace{-4mm}
\end{table}

\begin{table}[!t]
\centering
\caption{Comparison of \tool and ChatGPT (GPT-4o) in Repairing API Parameter Compatibility Issues}
\label{chatgpt repair}
\scalebox{0.625}{
\begin{tabular}{|cccc|cccc|cc|}
\hline
\multicolumn{4}{|c|}{\makecell{ChatGPT (GPT-4o) \\ W./O. Definition}} 
& \multicolumn{4}{c|}{\makecell{ChatGPT (GPT-4o) \\ W. Definition}} 
& \multicolumn{2}{c|}{\tool} \\ \hline
\multicolumn{2}{|c|}{W./O. Error Retry}            & \multicolumn{2}{c|}{W. Error Retry} & \multicolumn{2}{c|}{W./O. Error Retry}            & \multicolumn{2}{c|}{W. Error Retry} & \multicolumn{1}{c|}{\multirow{2}{*}{SR}} & \multirow{2}{*}{UR} \\ \cline{1-8}
\multicolumn{1}{|c|}{SR} & \multicolumn{1}{c|}{UR} & \multicolumn{1}{c|}{SR}     & UR    & \multicolumn{1}{c|}{SR} & \multicolumn{1}{c|}{UR} & \multicolumn{1}{c|}{SR}     & UR    & \multicolumn{1}{c|}{}                    &                     \\ \hline
\multicolumn{1}{|c|}{11} & \multicolumn{1}{c|}{43} & \multicolumn{1}{c|}{11}     & 41    & \multicolumn{1}{c|}{21} & \multicolumn{1}{c|}{31} & \multicolumn{1}{c|}{24}     & 28    & \multicolumn{1}{c|}{22}                  & 8          \\ \hline
\multicolumn{4}{|c|}{\makecell{Repair Precision: \\ 21.15\%}}                                  & \multicolumn{4}{c|}{\makecell{Repair Precision: \\ 46.15\%}}                                  & \multicolumn{2}{c|}{\makecell{Repair Precision: \\ \textbf{73.33\%}}}         \\ \hline
\end{tabular}
}
\vspace{-4mm}
\end{table}

As shown in Table~\ref{chatgpt test}, in test cases without parameter definitions, ChatGPT achieves a detection precision of 51.92\%, a recall of 93.10\%, and an F1-score of 66.67\%. Among the errors in compatibility detection, five cases result from inconsistent responses across the two attempts (e.g., one correct and one incorrect). When API parameter definitions are provided, precision, recall, and F1-scores improve to 54.90\%, 96.55\%, and 70.00\%, respectively, though three cases still show inconsistent responses across attempts. Compared to ChatGPT, \tool outperforms in both precision and F1-score, achieving 96.15\% and 90.91\%, respectively. Additionally, ChatGPT exhibits a significantly higher number of false positives (FP), indicating a tendency to classify samples as incompatible. This is because ChatGPT's judgments rely more on natural language interpretation, where minor string differences in parameter definitions can lead to misclassification. 

For repair, as shown in Table~\ref{chatgpt repair}, in test cases without parameter definitions, ChatGPT successfully fixes 11 cases on the first attempt. Among the 43 failed fixes, 25 involve unnecessary modifications to compatible test cases. For the remaining 18 failed fixes involving incompatible cases, 2 failures are due to incorrect compatibility detection, 4 result from inconsistent responses across two attempts, and the rest are consistently incorrect. 
Among these failures, 4 cases produced error messages during execution. Although we re-prompted ChatGPT using these error messages, none of the fixes were successful. Ultimately, 
ChatGPT's repair precision in this setting is 21.15\%.

In test cases with parameter definitions, ChatGPT successfully fixes 21 cases on the first attempt. Among the 31 failed fixes, 23 involve unnecessary modifications to compatible test cases. Of the 8 failed fixes for incompatible cases, 1 failure is due to an incorrect compatibility judgment, 1 results from inconsistent fixes across two attempts, and the remaining cases are consistently incorrect. Among these failures, 5 cases produced error messages during execution, and re-prompting ChatGPT with these error messages led to successful fixes in 3 cases. Overall, 
ChatGPT's repair precision in this setting improves to 46.15\%. 
In contrast, \tool achieves the best performance, successfully repairing 22 incompatible cases. 

Although ChatGPT demonstrates potential for addressing API compatibility issues, its performance is limited by the model's inherent hallucination knowledge, response inconsistencies, and randomness. Among the 59 API test cases, 13 exhibit inconsistent responses across two attempts. In comparison, \tool consistently demonstrates superior and more reliable performance in both detection and repair of API compatibility issues.

\subsection{RQ5: What is the Time Cost of PCART in Detecting and Repairing API Parameter Compatibility Issues?}\label{sec:results-rq5}

\begin{figure}[!t]
\centering
\subfloat[Cumulative count plot.]{
    \includegraphics[width=3in]{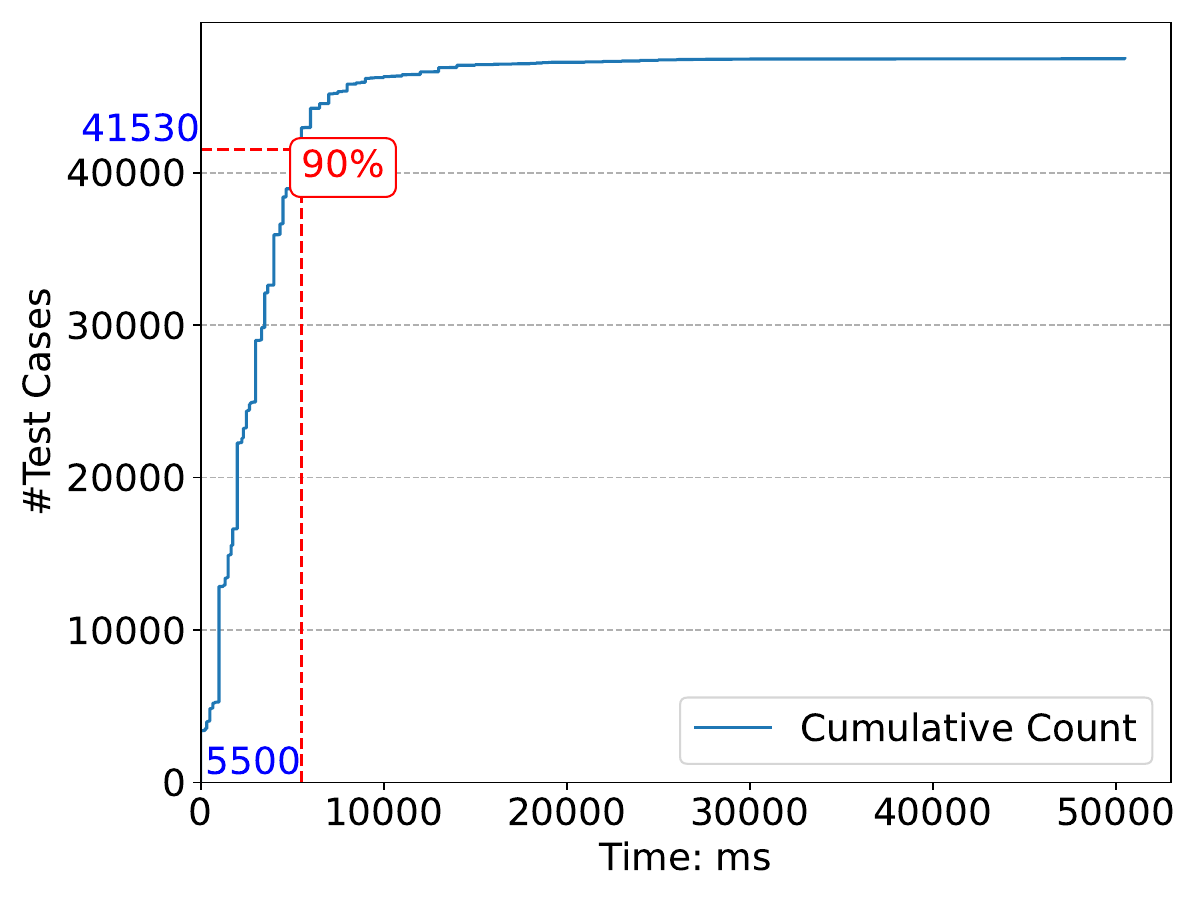}
    \label{cumulative}
}
\vspace{-4mm}
\subfloat[Box plot.]{
    \includegraphics[width=3in]{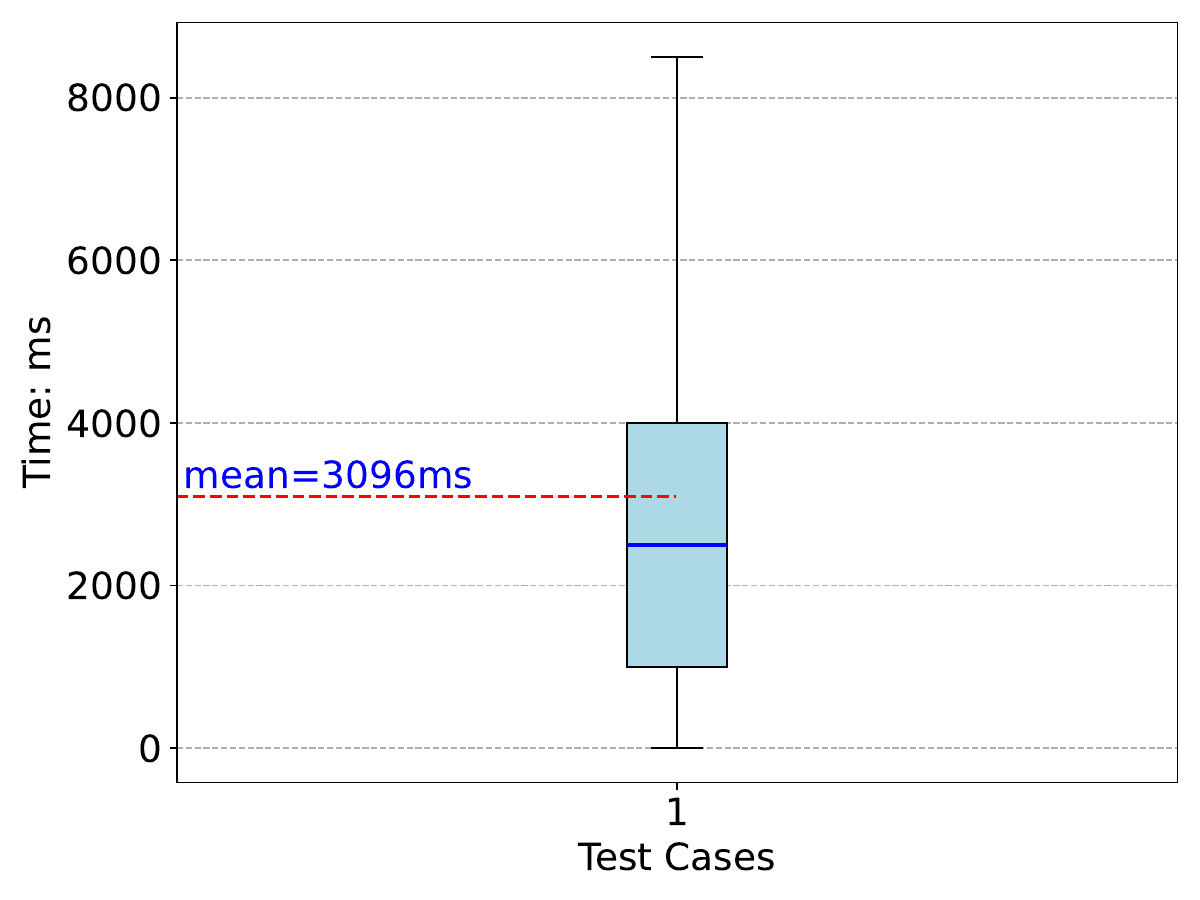}
    \label{boxplot}
}

\caption{Time spent by \tool on processing an invoked API in \benchmark.}
\label{Cumulative}
\vspace{-4mm}
\end{figure}

To answer RQ5, we measure the runtime of each test case and divide it by the number of target library APIs invoked to calculate the average runtime for processing one API. Fig.~\ref{cumulative} shows that in 90\% of the test cases, the average processing time for an API is within 5,500 ms, indicating that \tool is efficient in detecting and repairing API parameter compatibility issues for most test cases. Besides, as depicted in Fig.~\ref{boxplot}, after removing outliers (i.e., data points exceeding the upper quartile plus 1.5 times the interquartile range or below the lower quartile minus 1.5 times the interquartile range), the average processing time for one API per test case is 3,096 ms.

Note that some test cases have average processing times significantly exceeding the norm, primarily due to APIs related to model training, such as \mintinline{python}{gensim.models.fasttext.FastText} and \mintinline{python}{sklearn.manifold.TSNE}. 
These APIs substantially increase the overall processing time. This is because, during the code instrumentation phase, \tool runs the instrumented project code to record and save the context information of each API call. Since this process requires executing the project's code, the time taken is directly related to the project's runtime.

To fully automate the process from detection to repair of API parameter compatibility issues, \tool employs both dynamic and static methods. Dynamic processes, such as instrumentation and execution to obtain contextual dependencies of the invoked APIs, as well as loading contextual dependencies to establish API mappings and validate repairs, significantly increase \tool's runtime, especially for large Python projects. However, compared to manual API mapping, repair, and validation efforts, we believe \tool is both efficient and effective.

When using \tool in large Python projects, it is recommended that developers apply strategies such as reducing the dataset size or setting fewer execution epochs (particularly for deep learning projects) to minimize execution time. For example, the runtime of \mintinline{python}{gensim.models.fasttext.FastText} is affected by parameters such as \mintinline{python}{iter} (training iterations), \mintinline{python}{size} (vector dimensions), and \mintinline{python}{corpus_file} (training corpus size). Since \tool dynamically executes these APIs during the repair process to validate fixes, time-consuming APIs directly impact the overall efficiency. Therefore, reducing the dataset size or setting fewer training epochs can significantly decrease \tool's execution time.

\subsection{Limitations}
Although \tool successfully addresses the limitations of existing tools in detecting and repairing API parameter compatibility issues, it still has some shortcomings. In the following, we identify and discuss the limitations of \tool. 

\textit{(1) API Context Serialization.} The automated API mapping establishment in \tool mainly relies on the dynamic mapping, which requires capturing the contextual dependencies of invoked APIs. During instrumentation and execution, \tool captures these contextual dependencies by running the instrumented project and serializing the context information of each API call 
using the Dill library. This serialized context is saved in binary format to pickle files for later use in dynamic mapping and automated validation. However, certain variable types, such as file descriptors, sockets, and other OS resources, are generally not serializable. For example, for the aiohttp library, an object instantiated with \mintinline{python}{aiohttp.ClientSession()} will throw a \textit{TypeError: Cannot serialize socket object} when attempting serialization. 
Besides, even if some variables are serializable, they may fail to load correctly if the internal modules they depend on have changed in the new version. These factors can cause failures in dynamic mapping and automated validation.

\textit{(2) Mapping Relationship Establishment.} On one hand, when dynamic mapping fails, \tool transitions to the static mapping phase, which involves extracting the parameter definitions of the invoked APIs from the library source code. However, as mentioned in Section~\ref{sec:background-challenges}, this task becomes challenging when the fully qualified call path of an API does not match its real path in the source code.
Although \tool has converted some API paths in library source code to standard call paths provided by official documentation, a portion of APIs remains affected by issues such as APIs with the same name, API aliases, and API overloadings, which can still lead to ambiguous mappings. Therefore, \tool may mark the compatibility status of certain cases as ``unknown'' when encountering ambiguous scenarios, such as multiple candidate APIs during static mapping, necessitating manual inspection for these cases. 
On the other hand, in the compatibility assessment phase, when determining the mapping relationships of API parameters between two versions, situations where the ratio of remaining unmapped parameters from the current to the target version is $1:N$ or $N:M$ ($N>1$, representing the number of parameters) make it challenging to accurately determine the mapping relationships, especially in cases involving renaming or removal. \rev{Currently, \tool relies on syntactic and structural information, which limits its ability to capture semantic correspondences. In future work, we plan to integrate LLM-based parameter change analysis into \tool to enhance semantic reasoning and improve the accuracy of parameter mapping.}

\textit{(3) Parameter Type Analysis.} \tool primarily relies on parameter annotations to analyze parameter types, comparing the literal values of type annotations to determine if the parameter types have changed between the current and target versions. However, the style of type annotations varies among developers of different libraries; some use the Python standard type format, while others employ descriptive statements, which complicates the analysis of type changes. As a result, \tool does not support the repair of type changes and only uses limited type annotation information to assist in establishing parameter mapping relationships.

\textit{(4) Parameter Value Handling.} Changes in default parameter values in new versions can affect API compatibility. 
While \tool's architecture permits modifications to default values, we intentionally disable this functionality in the current implementation because we lack a mechanism to verify that the modified values truly meet users' needs.  
Furthermore, new parameters without default values can cause compatibility issues. \tool does not repair such issues, as generating parameter values that satisfy type requirements is challenging.

\textit{(5) Variadic Parameter Handling.} When the target API adopts variadic parameters (\mintinline{python}{*args}, \mintinline{python}{**kwargs}), \tool currently assumes that parameter renaming or removal remains compatible. However, this assumption can cause false negatives, i.e., semantically incompatible API calls may be incorrectly classified as compatible. This misclassification occurs when the variadic arguments are internally forwarded to another API that does not accept them, potentially causing runtime errors. We refer to this issue as the variadic parameter pitfalls (VPPs)~\cite{zhang2024coding}. We have developed a prototype tool, VPPDetector, to detect such VPPs, and plan to integrate it into \tool in the future to enhance its ability to avoid false negatives in variadic-parameter-related compatibility detection.

\textit{(6) Semantic Constraints Between Parameters.} \tool analyzes API parameters independently and does not account for semantic constraints or interdependencies among them. In practice, certain parameters are mutually exclusive or subject to conditional usage rules, which the current design overlooks. As a consequence, \tool may generate repairs that match the syntactic requirements of the updated API but violate its semantic expectations, resulting in incorrect behavior.



\rev{\textit{(7) Scalability of Dynamic Analysis.} The current implementation of \tool executes three main stages in its workflow (\mintinline{python}{main.py}): (i) static code preprocessing for instrumentation, (ii) project execution to collect runtime API contexts and generate \mintinline{python}{.pkl} files, and (iii) repair tasks, which are processed in parallel using Python's \mintinline{python}{multiprocessing.Pool}. Each process handles one source file for API extraction, signature retrieval, and repair validation. Although the project execution step is required only once, both the dynamic analysis for API mapping and the repair validation operate on individual API invocations, without executing the entire project. We observed that the overall speed-up from multiprocessing is limited, since files containing more API invocations dominate total execution time. For large projects, particularly deep learning systems with long-running computations, developers may reduce the dataset size or configure fewer training epochs to mitigate execution overhead. In future work, we plan to support API-level parallelism and adopt more advanced concurrency mechanisms (e.g., the free-threaded build introduced in Python 3.14) to further enhance scalability and efficiency. }

\section{Threats to Validity}\label{sec:threats}
\textit{Threats to Internal Validity.} The main threat to internal validity arises from potential implementation flaws in \tool. To mitigate this threat, we thoroughly examined the implementation logic of our code and used the test results from both the benchmark and real-world projects as feedback to continuously modify and refine \tool. Moreover, the process of manually labeling the compatibility of the test cases in \benchmark and manually checking some experimental results may introduce subjective biases. Therefore, we mitigated this type of threat through independent checks and cross-validation of all results by the authors, and have made our dataset publicly available for review and reproduction.

\textit{Threats to External Validity.} 
The primary threat to external validity comes from the selection of datasets used to evaluate the performance of \tool. To mitigate this threat, we constructed a benchmark comprising 844 APIs with parameter changes, covering 33 popular Python libraries. We further performed 
three mutant operators on the number of parameters, the method of parameter passing, and the sequence of parameter transmission, to generate \benchmark (i.e., 47,478 test cases). We believe \benchmark represents the diversity of user calls of parameters in practice. We also incorporated the MLCatchUp dataset for evaluation. Additionally, we collected 30 real-world projects from GitHub to assess \tool's effectiveness and practicality in actual environments. Finally, we compared \tool with existing tools, i.e., MLCatchUP~\cite{haryono2021mlcatchup}, Relancer~\cite{zhu2021restoring}, and ChatGPT (GPT-4o)~\cite{ChatGPT}, which are all representative. The experimental results demonstrated that \tool performs best in detecting and repairing API parameter compatibility issues.


\textit{Threats to Construct Validity.} The primary threat to construct validity lies in the possibility that the performance metrics used to evaluate \tool might not be comprehensive enough. To address this threat, we introduced incompatible test cases as positive cases and separately counted false positives (FP), true positives (TP), and false negatives (FN) in the detection of API parameter compatibility issues, calculating precision, recall, and F1-score. In terms of repairing API parameter compatibility issues, we tallied the successfully and unsuccessfully repaired test cases and calculated the repair precision, thus comprehensively evaluating \tool's performance through these multidimensional assessment metrics.

\section{Related Work}\label{sec:relatedwork}
In this section, we introduce the related work in two areas: API evolution analysis and compatibility issues repair techniques in Python, and API migration techniques in other programming languages and systems.

\subsection{API Evolution Analysis}
Many studies have summarized the characteristics of API evolution in Python third-party libraries and conducted various analyses to guide developers and practitioners \cite{zhang2020python, zhang2021unveiling, wang2020exploring}. 
Zhang \textit{et al.} \cite{zhang2020python} presented the first comprehensive analysis of API evolution patterns within Python frameworks. They analyzed six popular Python frameworks and 5,538 open-source projects built on these frameworks. Their research identified five distinct API evolution patterns that are absent within Java libraries. Zhang \textit{et al.}~\cite{zhang2021unveiling} delved into TensorFlow 2's API evolution trends by mining relevant API documentation and mapping API changes to functional categories. They determined that efficiency and compatibility stand out as the primary reasons for API changes within TensorFlow 2, constituting 54\% of the observed variations.

Du \textit{et al.} \cite{du2022aexpy} presented a system-level method based on an API model to detect breaking changes in Python third-party libraries. Building upon this, they designed and implemented a prototype tool, AexPy, for detecting documented and undocumented breaking changes in real-world libraries.
Montandon \textit{et al.} \cite{montandonunboxing} found that 79\% of the breaking changes in default parameter values in scikit-learn impact machine learning model training and evaluation, leading to unexpected results in client programs reliant on these APIs.

The study of API deprecation has become increasingly prevalent. Wang \textit{et al.} \cite{wang2020exploring} investigated how Python library developers maintain deprecated APIs. They found that inadequate documentation and declarations for deprecated APIs pose obstacles for Python API users. Vadlamani \textit{et al.}~\cite{vadlamani2021apiscanner} implemented an extension (APIScanner) that issues warnings when developers use deprecated APIs from 
Python libraries. 

Compared to existing tools that analyze API definition changes in Python libraries, our work focuses on designing an automated approach (\tool) for detecting and repairing parameter compatibility issues within the invoked APIs in user projects. When evaluating API parameter compatibility, \tool comprehensively analyzes both the changes in the API definitions 
and the actual usage of parameter-passing methods within the invoked APIs. This significantly improves the detection performance. 

\subsection{Compatibility Issues Repair Techniques}
Zhu \textit{et al.} proposed Relancer \cite{zhu2021restoring}, an iterative runtime error-driven method with a combined search space derived from API migration examples and API documentation. It combined machine learning models to predict the API repair types required for correct execution, automating the upgrading of deprecated APIs to restore the functionality of breaking Jupyter Notebook runtime environments. Haryono \textit{et al.} \cite{haryono2021characterization} initiated an empirical study to learn how to update deprecated APIs in Python libraries. Subsequently, they introduced MLCatchUp \cite{haryono2021mlcatchup}, which automatically infers the transformations necessary to migrate deprecated APIs to updated ones based on the differences mined from the manually provided signatures. Recently, Navarro \textit{et al.}~\cite{navarro2023automated} presented a closed-source tool to automatically update deprecated APIs 
in Python projects. 
The tool demonstrates limited effectiveness in handling API parameter compatibility issues due to its reliance on prebuilt deprecation knowledge from library change logs and its exclusive dependence on explicit keyword argument matching.

Compared to existing repair tools, \tool mainly fixes compatibility issues caused by API parameter changes (i.e.,  addition, removal, renaming, reordering of parameters, and the conversion of positional parameters to keyword parameters) in Python libraries. To the best of our knowledge, \tool is the first to implement a fully automated process from API extraction, code instrumentation, API mapping establishment, to compatibility assessment, and finally to repair and validation, achieving outstanding detection and repair performance. 

\subsection{API Migration Techniques}
Over the past two decades, API migration has been an active research area across multiple ecosystems~\cite{lamothe2021systematic}. Much of this work has focused on statically typed languages, particularly Java and Android~\cite{wu2010aura, dagenais2011recommending, yu2017api, zhang2022has, apiwattanapong2007jdiff, foo2018efficient, jan2013testing, yamaoka2024comparing, wu2014acua, ochoa2022breakbot, xi2019migrating, huang2021repfinder, hora2014apievolutionminer, brito2018apidiff, li2020cda, liu2023automatically, silva2022saintdroid, mahmud2023detecting, li2018cid, xing2007api, kapur2010refactoring, nguyen2010graph, yamaguchi2022two, henkel2005catchup, tronivcek2012refactoringng, vstrobl2013migration, li2015swin, kang2019semantic, gharaibeh2007coping, zhong2024compiler, fazzini2019automated, scalabrino2019data, xu2019meditor, haryono2022androevolve, lamothe2020a3, wang2022augraft, zhao4254659autopatch, zhao2022towards, perkins2005automatically, csavga2008practical, sun2022mining, mobilio2024filo, mahmud2024apicia}. Beyond these ecosystems, smaller bodies of work have addressed C++~\cite{ponomarenko2012backward, kalra2016pollux, collie2020m3}, C\#~\cite{gao2021apifix}, JavaScript~\cite{moller2020detecting}, Pharo~\cite{zaitsev2022depminer, leuenberger2019exploring}, Web frameworks~\cite{schmiedmayer2023reducing, Bonorden2024Detecting}, and HarmonyOS~\cite{ma2023cid4hmos}.

Most detection approaches analyze library-defined APIs to identify incompatibilities. Techniques range from call dependency analysis~\cite{wu2010aura, dagenais2011recommending, yu2017api, zhang2022has}, control flow analysis~\cite{apiwattanapong2007jdiff, foo2018efficient}, bytecode analysis~\cite{jan2013testing, yamaoka2024comparing}, binary code analysis~\cite{wu2014acua, ochoa2022breakbot}, and mining Javadoc annotations~\cite{xi2019migrating, huang2021repfinder}. While effective in statically typed ecosystems, these methods face limitations: documentation often lags behind actual code, bytecode approaches assume compilation artifacts, and binary analysis cannot be directly transferred to dynamically typed languages. Alternative strategies include mining revision histories~\cite{hora2014apievolutionminer} or applying similarity heuristics to detect breaking changes~\cite{brito2018apidiff}.

In mobile frameworks like Android, specialized tools leverage \mintinline{java}{@deprecated} annotations and framework metadata to detect removed or replaced APIs~\cite{li2020cda, liu2023automatically}. Other works combine lifecycle modeling with application bytecode analysis to detect compatibility issues in actual client projects~\cite{li2018cid, silva2022saintdroid, mahmud2023detecting}. Collectively, these approaches highlight the importance of analyzing not just library definitions but also client invocations, a principle that becomes even more critical in dynamic languages. 

Repair methods in Java and Android generally follow two paradigms: (1) learning-based techniques that infer update patterns from repository histories~\cite{xing2007api, kapur2010refactoring, nguyen2010graph, yamaguchi2022two}, and (2) rule/template-based techniques that rely on developer-provided mappings or documentation~\cite{henkel2005catchup, tronivcek2012refactoringng, vstrobl2013migration, li2015swin, kang2019semantic}. Tools like LibSync~\cite{nguyen2010graph} recommend adaptations based on mined migration patterns, whereas RefactoringNG~\cite{tronivcek2012refactoringng} applies author-provided refactoring rules. Android-focused systems such as AppEvolve~\cite{fazzini2019automated} and AUGraft~\cite{wang2022augraft} mine migration instances from GitHub to generate automated patches. While these methods offer automation, they often require non-trivial preprocessing (e.g., mining snippet pairs) and rely heavily on the availability of well-documented or frequent migrations. Template-based repair tools (e.g., RepairDroid~\cite{zhao2022towards}) highlight another limitation: the need for human-crafted rules, which is labor-intensive and not easily scalable. Moreover, even when repair suggestions are generated automatically, validation typically depends on manual review or limited test execution, reducing confidence in correctness. 

Despite progress, two critical gaps remain:
(1) End-to-end automation is rare. Most methods stop short of full repair pipelines, either requiring manual mappings or manual validation.
(2) Dynamic language challenges remain unresolved. Techniques built for statically typed languages rely on compiler feedback, static signatures, or type information. These assumptions break down in Python, where parameter passing can vary between positional and keyword styles, and where incompatibilities often manifest only at runtime. 

In contrast, \tool is designed specifically for Python's dynamic environment and addresses these gaps directly. First, it automates the inference of repair actions without requiring developers to predefine change mappings or templates. Second, it combines static and dynamic analyses to establish API mappings and capture runtime parameter contexts: capabilities that prior static-only approaches cannot provide. Third, \tool integrates validation into its repair process: static validation ensures formal consistency with new API signatures, while dynamic validation confirms runtime executability in isolated environments. Together, these features enable end-to-end automated migration, which to our knowledge has not been achieved in Python or other dynamic languages at the parameter level.

Moreover, by constructing the first large-scale benchmark for Python parameter compatibility, \tool not only advances the state of the art in Python but also highlights migration challenges broadly relevant to other dynamic ecosystems such as JavaScript, Ruby, and R. This positions \tool as both a practical repair tool and a research contribution to the general field of automated API migration.

\section{Conclusion}\label{sec:conclusion}

In this paper, we introduced \tool, an open-source tool that combines static and dynamic approaches to achieve end-to-end automation for API extraction, code instrumentation, API mapping establishment, compatibility assessment, repair, and validation, precisely addressing Python API parameter compatibility issues. To comprehensively 
evaluate \tool's detection and repair performance, we constructed \benchmark, a large-scale benchmark consisting of 844 parameter-changed APIs from 33 popular Python libraries and a total of 47,478 test cases. 
Experimental results show that \tool outperforms existing tools (MLCatchUP and Relancer) and ChatGPT (GPT-4o) in both detecting and repairing API parameter compatibility issues. 
Furthermore, we evaluated \tool on 30 real-world Python projects, demonstrating its ability to accurately detect and successfully repair all incompatible APIs. Finally, we evaluated \tool's efficiency, and the results highlight its strong performance.

\tool presents an exploratory step towards fully automated Python API compatibility issues repair. In the future, we plan to address its limitations and further improve its practicality and effectiveness by testing it on more projects. 


\section*{Acknowledgments}
The authors would like to thank the anonymous reviewers for their valuable comments and suggestions for improving this paper.

\bibliographystyle{IEEEtran}
\bibliography{main}

\begin{IEEEbiography}[{\includegraphics[width=1in,height=1.25in,clip]{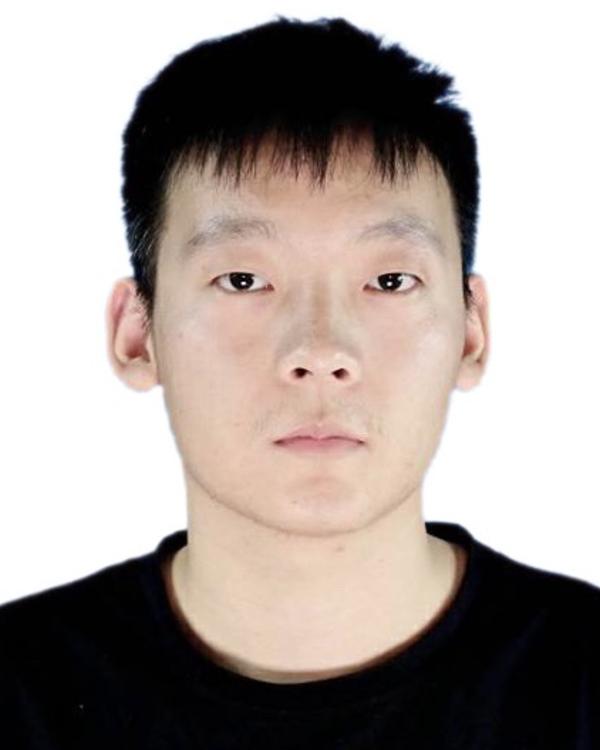}}]{Shuai Zhang} is working toward the master's degree with the College of Computer Science and Technology, Nanjing University of Aeronautics and Astronautics, China. His current research interest is software compatibility detection and repair.
\end{IEEEbiography}


\begin{IEEEbiography}[{\includegraphics[width=1in,height=1.25in,clip]{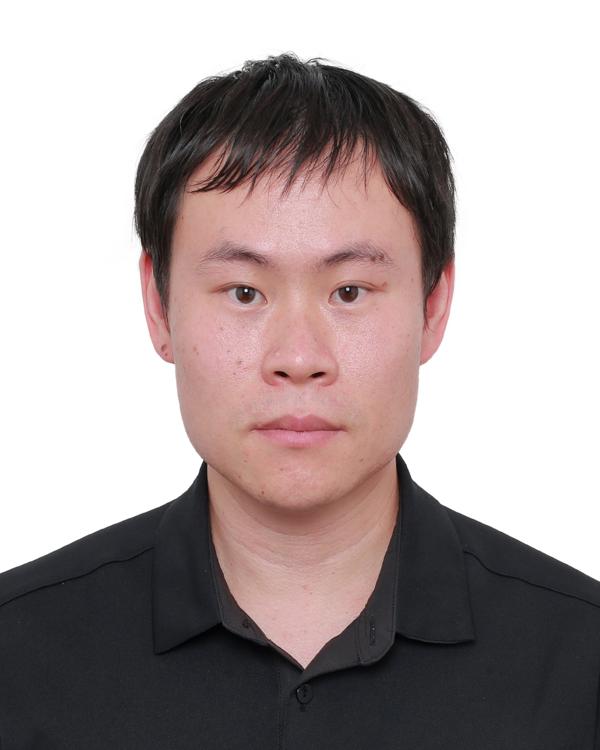}}]{Guanping Xiao} is a tenured associate professor at the College of Computer Science and Technology, Nanjing University of Aeronautics and Astronautics, China. He received his Ph.D. degree in software engineering from Beihang University, Beijing, China. His research interests include software reliability, software evolution, and program analysis.
\end{IEEEbiography}


\begin{IEEEbiography}[{\includegraphics[width=1in,height=1.25in,clip]{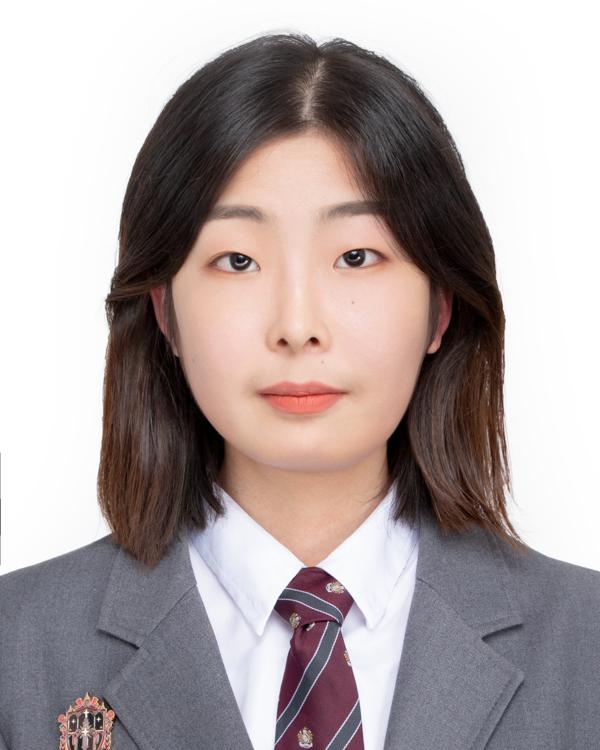}}]{Jun Wang} is working toward the master's degree with the College of Computer Science and Technology, Nanjing University of Aeronautics and Astronautics, China. Her current research interest is in detecting compatibility issues in deep learning systems.
\end{IEEEbiography}


\begin{IEEEbiography}[{\includegraphics[width=1in,height=1.25in,clip]{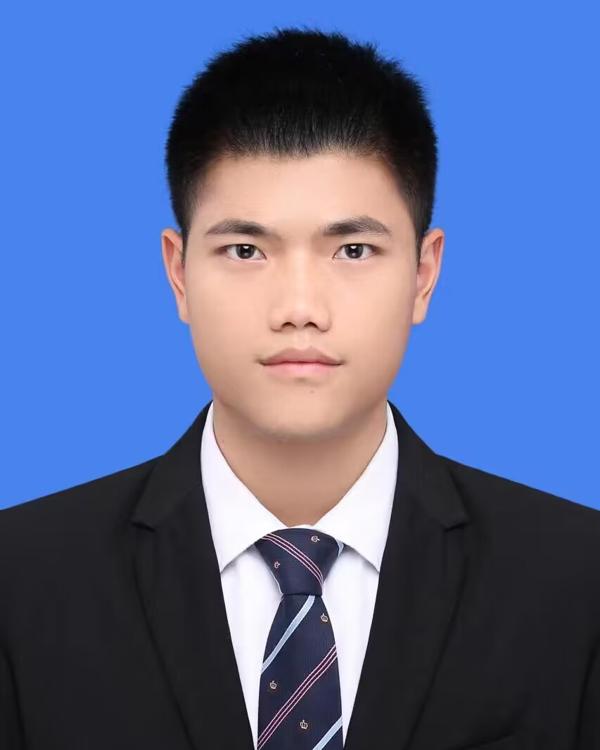}}]{Huashan Lei} is working toward the master's degree with the College of Computer Science and Technology, Nanjing University of Aeronautics and Astronautics, China. His current research interest is software configuration management.
\end{IEEEbiography}


\begin{IEEEbiography}[{\includegraphics[width=1in,height=1.25in,clip]{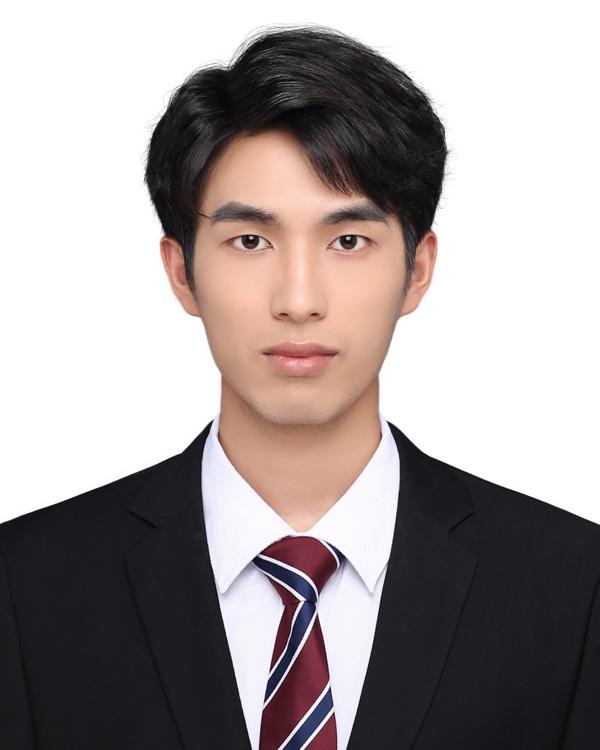}}]{Gangqiang He} is working toward the master's degree with the College of Computer Science and Technology, Nanjing University of Aeronautics and Astronautics, China. His current research interest is software compatibility detection and repair.
\end{IEEEbiography}


\begin{IEEEbiography}[{\includegraphics[width=1in,height=1.25in,clip]{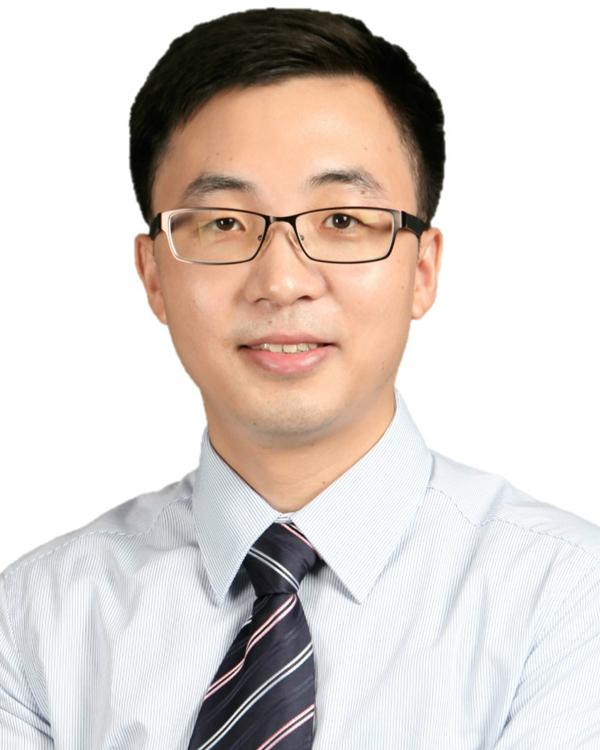}}]{Yepang Liu} is a tenured associate professor with the CSE Department of SUSTech and leads the Software Quality Lab. He is also the director of the Trustworthy Software Research Center at the Research Institute of Trustworthy Autonomous Systems. His research interests mainly include software testing and analysis, empirical software engineering, cyber-physical systems, software security, and trustworthy AI.
\end{IEEEbiography}


\begin{IEEEbiography}[{\includegraphics[width=1in,height=1.25in,clip]{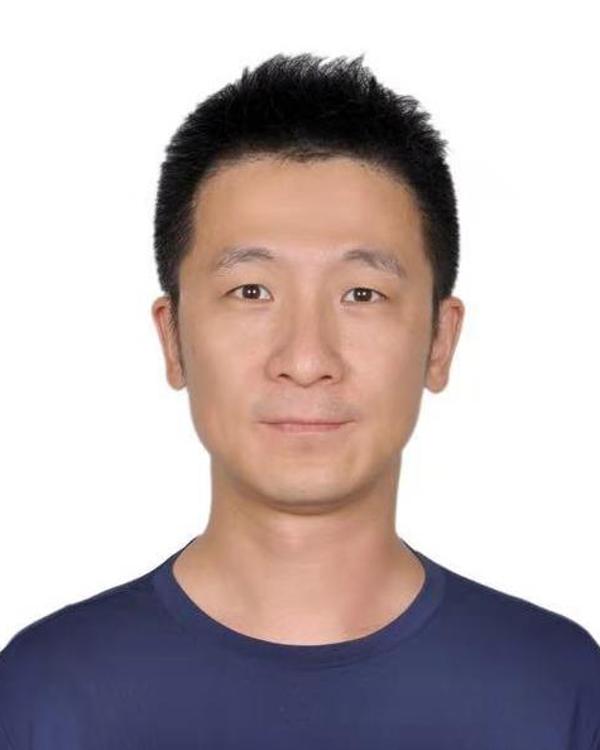}}]{Zheng Zheng} is a professor with Beihang University and the deputy dean of the School of Automation Science and Electrical Engineering. His research focuses primarily on software reliability and testing. Recently, He paid more attention to intelligent software reliability engineering. He has co-authored more than 100 journal and conference publications, including IEEE Transactions on Dependable and Secure Computing, IEEE Transactions on Information Forensics and Security, IEEE Transactions on Software Engineering, IEEE Transactions on Reliability, IEEE Transactions on Services Computing, and JSS, among others. He serves for IEEE PRDC2019, IEEE DASC 2019, IEEE ISSRE 2020, and IEEE QRS 2021 as PC Co-Chairs, as well as WoSAR 2019, DeIS 2020, and DeIS 2021 as General Co-Chairs. He is the editor-in-chief of Atlantis Highlights in Engineering (Springer Nature), associate editor of IEEE Transactions on Reliability (2021-), Elsevier KBS (2018-), Springer IJCIS (2012-), and guest editor of IEEE Transactions on Dependable and Secure Computing (2021). He is an IEEE CIS Emerging Technologies TC member.
\end{IEEEbiography}


\end{document}